\documentclass[twocolumn]{aastex63}
\usepackage{float}
\usepackage{amsmath}

\newcommand{\Uphi}{U$_\phi$ }
\newcommand{\Qphi}{Q$_\phi$ }
\newcommand{\Uphip}{U$_\phi$}
\newcommand{\Qphip}{Q$_\phi$}
\def\Msun   {\ifmmode{{\rm ~M}_\odot}\else{~M$_\odot$}\fi}

\received{April 8th, 2022}
\revised{May 26th, 2022}
\accepted{May 27th, 2022}

\submitjournal{AJ}

\shorttitle{Gemini-LIGHTS}
\shortauthors{Rich et al.}

\graphicspath{{./}{figures/}}

\begin{document}

\title{Gemini-LIGHTS: Herbig Ae/Be and massive T-Tauri protoplanetary disks imaged with Gemini Planet Imager}

\correspondingauthor{Evan A. Rich}
\email{earich@umich.edu}

\author[0000-0002-1779-8181]{Evan A. Rich}
\affiliation{Department of Astronomy, University of Michigan, West Hall, 1085 South University Avenue, Ann Arbor, MI 48109-1090, USA}

\author[0000-0002-3380-3307]{John D. Monnier}
\affiliation{Department of Astronomy, University of Michigan, West Hall, 1085 South University Avenue, Ann Arbor, MI 48109-1090, USA}

\author[0000-0002-1327-9659]{Alicia Aarnio}
\affiliation{The University of North Carolina at Greensboro, USA}

\author[0000-0002-2145-0487]{Anna S. E. Laws}
\affiliation{Astrophysics Group, University of Exeter, Stocker Road, Exeter, EX4 4QL, UK}

\author[0000-0001-5980-0246]{Benjamin R. Setterholm}
\affiliation{Department of Astronomy, University of Michigan, West Hall, 1085 South University Avenue, Ann Arbor, MI 48109-1090, USA}

\author[0000-0003-1526-7587]{David J. Wilner}
\affiliation{Center for Astrophysics $|$ Harvard \& Smithsonian, 60 Garden Street, Cambridge, MA 02138, USA}

\author[0000-0002-3950-5386]{Nuria Calvet}
\affiliation{Department of Astronomy, University of Michigan, West Hall, 1085 South University Avenue, Ann Arbor, MI 48109-1090, USA}

\author[0000-0001-8228-9503]{Tim Harries}
\affiliation{Astrophysics Group, University of Exeter, Stocker Road, Exeter, EX4 4QL, UK}

\author[0000-0002-1365-0841]{Chris Miller}
\affiliation{Department of Astronomy, University of Michigan, West Hall, 1085 South University Avenue, Ann Arbor, MI 48109-1090, USA}

\author[0000-0001-9764-2357]{Claire L. Davies}
\affiliation{Astrophysics Group, University of Exeter, Stocker Road, Exeter, EX4 4QL, UK}

\author[0000-0002-8167-1767]{Fred C. Adams}
\affiliation{Department of Astronomy, University of Michigan, West Hall, 1085 South University Avenue, Ann Arbor, MI 48109-1090, USA}
\affiliation{Physics Department, University of Michigan, Randall Lab, 450 Church Street, Ann Arbor, MI 48109-1090, USA}

\author[0000-0003-2253-2270]{Sean M. Andrews}
\affiliation{Center for Astrophysics $|$ Harvard \& Smithsonian, 60 Garden Street, Cambridge, MA 02138, USA}

\author[0000-0001-7258-770X]{Jaehan Bae}
\affiliation{Department of Astronomy, University of Florida, Gainesville, FL 32611, USA}

\author[0000-0001-9227-5949]{Catherine Espaillat}
\affiliation{Department of Astronomy \& Institute for Astrophysical Research, Boston University, 725 Commonwealth Avenue, Boston, MA 02215, USA}

\author[0000-0002-7162-8036]{Alexandra Z. Greenbaum}
\affiliation{IPAC, Caltech, 1200 E California Blvd, Pasadena, CA 91125, USA}

\author[0000-0001-8074-2562]{Sasha Hinkley}
\affiliation{Astrophysics Group, University of Exeter, Stocker Road, Exeter, EX4 4QL, UK}

\author[0000-0001-6017-8773]{Stefan Kraus}
\affiliation{Astrophysics Group, University of Exeter, Stocker Road, Exeter, EX4 4QL, UK}

\author[0000-0003-1430-8519]{Lee Hartmann}
\affiliation{Department of Astronomy, University of Michigan, West Hall, 1085 South University Avenue, Ann Arbor, MI 48109-1090, USA}

\author[0000-0001-8061-2207]{Andrea Isella}
\affiliation{Department of Physics and Astronomy, Rice University, 6100 Main Street, MS-108, Houston, TX 77005, USA}

\author[0000-0003-1878-327X]{Melissa McClure}
\affiliation{University of Amsterdam, Netherlands}

\author[0000-0001-7130-7681]{Rebecca Oppenheimer}
\affiliation{Astrophysics Department, American Museum of Natural History, Central Park West at 79th Street, New York, NY 10024}

\author[0000-0002-1199-9564]{Laura M. P\'erez}
\affiliation{Departamento de Astronom\'ia, Universidad de Chile, Camino El Observatorio 1515, Las Condes, Santiago, Chile}

\author[0000-0003-3616-6822]{Zhaohuan Zhu}
\affiliation{Department of Physics and Astronomy, University of Nevada, Las Vegas, 4505 S. Maryland Parkway, Box 454002, Las Vegas, NV 89154-4002, USA}

\begin{abstract}

We present the complete sample of protoplanetary disks from the Gemini- Large Imaging with GPI Herbig/T-tauri Survey (Gemini-LIGHTS) which observed bright Herbig Ae/Be stars and T-Tauri stars in near-infrared polarized light to search for signatures of disk evolution and ongoing planet formation. The 44 targets were chosen based on their near- and mid-infrared colors, with roughly equal numbers of transitional, pre-transitional, and full disks. Our approach explicitly did not favor well-known, ``famous'' disks or those observed by ALMA, resulting in a less-biased sample suitable to probe the major stages of disk evolution during planet formation. Our optimized data reduction allowed polarized flux as low as 0.002\% of the stellar light to be detected, and we report polarized scattered light around 80\% of our targets. We detected point-like companions for 47\% of the targets, including 3 brown dwarfs (2 confirmed, 1 new), and a new super-Jupiter mass candidate around V1295 Aql. We searched for correlations between the polarized flux and system parameters, finding a few clear trends: presence of a companion drastically reduces the polarized flux levels, far-IR excess correlates with polarized flux for non-binary systems, and systems hosting disks with ring structures have stellar masses $<$ 3 \Msun{.} Our sample also included four hot, dusty ``FS CMa" systems and we detected large-scale ($>100$ au) scattered light around each, signs of extreme youth for these enigmatic systems. Science-ready images are publicly available through multiple distribution channels using a new FITS file standard jointly developed with members of the VLT/SPHERE team.

\end{abstract}

\section{Introduction}

Orbiting reservoirs of gas and dust are often observed around young stars and referred to as ``protoplanetary'' disks.  As the name suggests, it has long been thought \citep[e.g.,][]{Kant1755} that planets form within such disks, though definitive evidence had been lacking until quite recently (PDS 70b,c: \citealt{Keppler2018, Haffert2019}).  While the location for planet formation is not in doubt, the critical physical mechanisms at play are still hotly debated for a predictive theory of planet formation that can robustly explain the wild diversity seen in exoplanet demographics.

Protoplanetary disks were initially classified based on the shapes of their Spectral Energy Distributions (SEDs).  ``Full disks'' show a continuous spectrum resulting from thermal emission from $\sim 1500$\,K to 10s of K.  ``Transition disks'' contrast in the SED shape of ``Full disks'' and lack near-infrared emissions due to a large gap or cavity close to the star \citep{Strom1989, Calvet2002, Espaillat2014}, possibly indicating planet formation or merely disk dissipation. With the advent of high angular resolution imaging by ALMA and with extreme adaptive optics systems on 8m-class telescopes, a more complex picture emerges.   Recent surveys such as DSHARP and DARTT-S have revealed a host of substructure including spiral arms, rings, gaps, and non-azimuth asymmetries \citep{Andrews2018,Avenhaus2018,Garufi2018}. These features can be interpreted as signposts of forming planet, however exoplanets is not directly detected with the exception of PDS 70 b,c \citep{Keppler2018}.

Here, we bring the power of ``Polarized Differential Imaging (PDI)'' to the study of planet formation. PDI reveals faint light scattered off the disk surfaces, cavity walls,  and other dust structures in the circumstellar environment, revealing the 3-dimensional distributions of small dust grains \citep{Avenhaus2018,Rich2021}. When coupled with ALMA imaging, which is sensitive to large dust grains settled into the disk mid-plane, PDI can monitor how dust grains grow and evolve with time and search for differences for systems with different stellar masses. Also, since PDI is dependent on illumination, the polarized flux will be influenced by shadowing of the inner disk onto the outer disk \citep{Debes2017,Rich2019,Labdon2019,Muro2020} and the flaring angle of the outer protoplanetary disk.

In this work, we define the Gemini- Large Imaging with GPI Herbig/T-tauri Survey (Gemini-LIGHTS) sample of 44 Herbig Ae/Be and T-Tauri proto-stars imaged in near-infrared scattered light with the Gemini Planet Imager (GPI; Section \ref{sec:sample}). Our new survey complements existing surveys (e.g., SEEDS, \citealt{Tamura2009}; DARTT-S \citealt{Avenhaus2018}) by better populating the high-mass range (Herbig Ae/Be stars; $>$ 3 \Msun).
We describe the reduction techniques utilized in our sample (Section \ref{sec:pipeline}) then present our calibrated images along with a descriptive analysis (Section \ref{sec:results}). Next we limit our sample to targets with stellar masses between 1.4 and 8 \Msun{} to search for trends between polarized flux and system characteristics (Section \ref{sec:trends}).
Finally we discuss what trends we observe in the sample to help explain Herbig Ae/Be and T-Tauri evolution (Section \ref{sec:discussion}).

\section{Gemini-LIGHTS Sample}\label{sec:sample}

\begin{figure*}[!ht]
    \centering
    \includegraphics[width=0.99\linewidth,trim=5 5 0 5 ,clip]{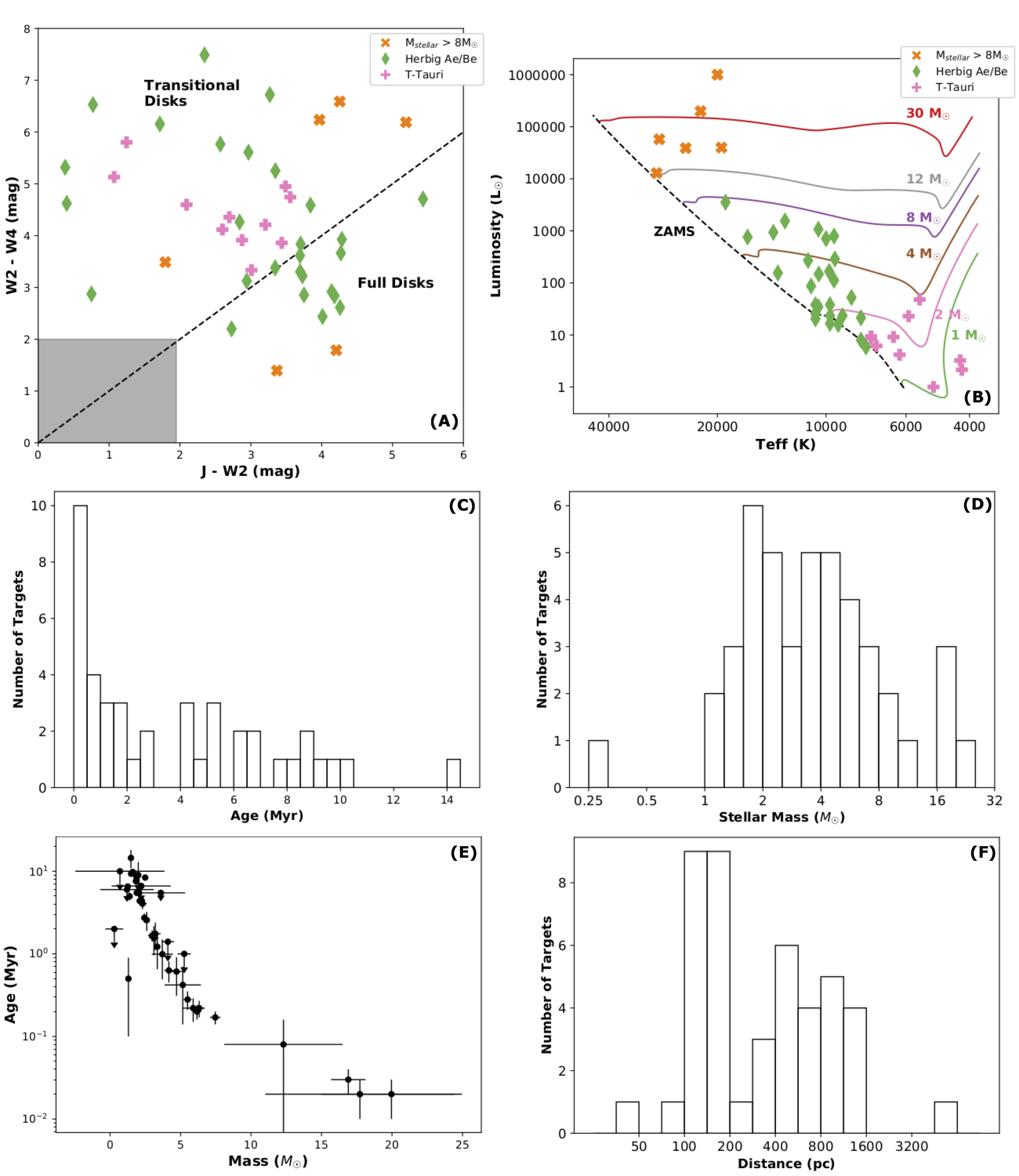}
    \caption{Sample parameters of the 44 targets in the Gemini-LIGHTS survey. (A) is an infrared color-color diagram using WISE and 2MASS colors. The dashed line represents a flat spectrum SED and the grey shaded region are objects with no near-infrared or mid-infrared excess. (B) HR diagram with pre-main-sequence mass tracks (colored lines) and the zero age main sequence (ZAMS) assuming solar metallicity mass tracks from \citet{Bressan2012}. Targets are classified as: stars with mass $>$ 8 \Msun{} (orange x's), T-Tauri stars (pink plus), and Herbig Ae/Be stars (green diamonds). (C) Histogram of estimated age. (D) Histogram of estimated stellar mass. (E) Age versus stellar mass. (F) Histogram of target distances. Specific target parameter values can be found in the appendix in Tables \ref{tbl:target_properties} and \ref{tbl:target_SED_table}. \label{fig:sample_GLITHS}. Note that FU Ori is not plotted on the HR diagram as the T$_{eff}$ temperature is unknown.}
    
\end{figure*}

The Gemini-Lights sample was chosen to represent a broad range of T-Tauri and Herbig Ae/Be stars with different protoplanetary disk structures including transition, pre-transition, and full disks. 
First, the sample consists of objects that were R $<$ 9 mag (GPI wavefront sensor limit), declination between +20$^\circ$ to -80$^\circ$ (lower airmass).
Next, we identified these targets with significant infrared excess based on a color-color diagram of WISE and 2MASS colors as shown in Figure \ref{fig:sample_GLITHS}. Additionally, we chose targets over a range of colors to achieve a mixture of transitional and full disks. 
Transitional disk targets host less near-infrared flux than far-infrared flux associated with a gap in the inner disk. This upper left portion of the diagram coincides with the location of the full disks shown in Figure \ref{fig:sample_GLITHS}. 
Full disk targets host equal near-infrared flux to far-infrared flux associated with no gaps in the disk. We note that a systems inclination can effect their broad SED categorization.
Known equal brightness binaries were not selected with a separation between 0$\farcs$05 - 2$\farcs$0 because they would inhibit the performance of the adaptive optics (AO) system. Compact binaries (e.g., HD 34700A) and unequal brightness binaries (e.g., FU Ori) were not selected against. Observations were taken in the J- and H-band, however not every object has both bands observed. H-band observations were prioritized for dim R-band stars for better AO performance, otherwise J-band observations were prioritized.

We also added archival observations of targets to our sample that have previously been observed by GPI. First we include an early science GPI project \citep{Monnier2017} which include data from MWC 275, HD 144432, MWC 863, and HD 169142. We have also included data found on the Gemini archive including targets PDS 66, HD 100546 , HD 101412
, HD 100453, HD 142527, and AK Sco in which some have previously been published \citep{Rodigas2014,Wagner2015,Follette2017}.

The Gemini-LIGHTS sample includes 44 targets which are summarized in Table \ref{tbl:target_coords}, along with their stellar properties in Table \ref{tbl:target_properties}, and photometry used for the sample selection in Table \ref{tbl:target_SED_table}.
Stellar properties are primarily taken from \citet{Vioque2018} to create a uniform sample.
An HR diagram and histograms of the systems' properties are shown in Figure \ref{fig:sample_GLITHS} where we also label which targets have masses $>$ 8 \Msun, Herbig Ae/Be stars ($>$ 7600 K), and T-Tauri stars ($<$ 7600 K). 
Targets have distances between 40-5000 pc with the median distance of 350 pc, target age estimates span from 0.02 -15 Myrs and a median age of 2.3 Myr, and target stellar masses range from 0.3-20\Msun with a median value of 2.5\Msun. The majority of our sample have masses consistent with Herbig Ae/Be stars. Also, all but FU Ori ($\sim$0.3 \Msun) have central star masses larger than 1 \Msun.

We note a limitation in our sample.
We plot the estimated age versus stellar mass in Figure \ref{fig:sample_GLITHS} and find that the two properties are highly correlated. This is likely due to our sample having few intermediate mass T-Tauri stars. This effect was noted by \citet{Guzman2021} where their recent survey of Herbig Ae/Be stars was also missing young intermediate mass stars (see Figure 8 by \citet{Guzman2021}).

\subsection{Observations}

Observations of the targets were taken at Gemini South Observatory using the Gemini Planet Imager (GPI; \citealt{Macintosh2014,Poyneer2014,Larkin2014}) in J- and H-band filters. Observations were in polarimetry mode using coronagraphic spots of sizes 184.7 mas and 246.7 mas for J- and H-bands respectively and a pixel scale of 14.14 mas. Observations of the targets HD 100453 and HD 142527 were the only observations taken without a coronagraphic mask. Each image of a target measures the orthogonal polarization state of the 2$\farcs$0 x 2$\farcs$0 field of view (FOV). Between each image of a target, the half wave plate is rotated 22.5$^\circ$. This creates a polarization set of 4 images at 4 different waveplate positions (0$^\circ$, 22.5$^\circ$, 45$^\circ$, 67.5$^\circ$). A typical observational epoch of a target observed 8 sets of 4 polarization images producing 32 images per epoch. Exposure of the image for each target was adjusted for the targets brightness such that the PSF of the star does not saturate the images.
Observations of the targets occurred between 2014 April and 2019 May with the majority of observations occurring between 2017-2019 and a list of the observations can be found in Table \ref{tbl:obs_log}. Calibration files such as lamps and dark images where taken by the Gemini observatory staff every 2-4 weeks and accessed through the Gemini archive.

\section{Data Reduction} \label{sec:pipeline}

The data was primarily reduced using the IDL based Data Reduction Pipeline (DRP) \citep{Maire2010,Perrin2014} using version 1.5.0 (rev c0cad3f), written to reduce GPI data. 
We also employed our own python based wrapper which automated the execution of the DRP IDL pipeline to batch process our data to allow for consistent and reproducible reductions.
The python wrapper edits the DRP parameter files needed and executes the IDL commands. 
Changes to the standard reduction were based on previous work by \citet{Monnier2019} and \citet{Laws2020}. The python wrapper is available on Github \footnote{\url{https://github.com/earich/Gemini_LIGHTS_pipeline}}.

In order to reduce our raw observational data, we inspected the individual raw frames and removed any frames if there was poor Adaptive Optics (AO) performance or if the target star slips to the edge of the coronographic mask.
If a single frame was removed or missing from a polarization set, a frame from an adjacent set was used in its place. If more than one frame from a polarization set was removed or missing, the entire polarization set is not used. From Raw images to Polarization Data Cubes (PODC) files, we utilized the IDL DRP without modifications as described by \citet{Maire2010}, \citet{Perrin2014}, \citet{Perrin2015}, \citet{Millar2016}, and \citet{DeRosa2020}. Each PODC file contains two images of the orthogonal polarization states. After PODC files were created, each observation was inspected by eye to verify if the flexure solution was correct.

The PODC cubes are centered using the IDL DRP code which utilizes a Radon transformation of the satellite spots to calculate the location of the star behind the coronagraph. We found that this transformation could fail when there was a bright point source in the field near the satellite spots. We found that applying a circular mask of radius=10 pixels over the point source were sufficient to correct for this issue. We tested the centering function on point sources in the FOV around HD 50138 and found a centroid accuracy of  0.22 pix (3.1 mas).
This method was applied to all targets with a point source in the FOV. All centering solutions were inspected by eye to ensure the method worked properly. There were two instances where the Radon transformation within IDL DRP failed to find the correct center with the companion massking improvements.
The coronographic mask for HD 98800 was placed over the A component of the system rather than the disk hosting B component of the system \citep{Kennedy2019}. We centered the images on the disk hosting B component using an interactive Radon transformation algorithm described by \citet{Monnier2017}. Secondly, the centering uncertainty for MWC 863 was large (1.9 pixels) using the IDL DRP Radon transformation. Using the interactive Radon transformation used from \citet{Monnier2017} improved the centering measured by the companions center ($>$ 1.2 pixels). We found the best solution was taking the average centering position from the IDL DRP Radon transformation and correcting for relative offsets of the PODC frames by measuring the center of the companion resulting in a centroid accuracy of (0.4 pixels). The relative centering method with the companion comes at the cost of greater uncertainty for the absolute position of the star behind the coronographic mask.

We next create Stokes cubes (I,Q,U,V) from the derotated (north-up) PODC files.
Each set of half-waveplate positions (4 PODC files) are double differenced into Stokes I, Q and U, creating a set of approximately 8 Stokes cubes per observing epoch. We note that while a circular polarization (V) image is created in the DRP pipeline, we do not use it in our analysis and will not refer to V further in this work. 

We removed Stellar and Instrumental Polarization (SIP) from the Q and U images. Measuring the amount of SIP is typically done by picking a region in the image and measuring the ratio of Q$_{\textrm{SIP}}$/I and U$_{\textrm{SIP}}$/I which we will refer to as $f_Q$ and $f_U$ respectively. 
The shape of the SIP should be similar to the I image, when dominated by the PSF, thus we can remove the SIP by multiplying the fraction of $f_Q$ and $f_U$ polarization to the intensity image and subtracting it from the Q and U images creating a set of corrected $Q^*$ and $U^*$

\begin{align}
    \label{eqn:remove_stellar_polQ}
    Q^* = Q - I \times f_Q \\
    U^* = U - I \times f_U  \label{eqn:remove_stellar_polU}
\end{align}
where images that have been corrected for SIP. 

Next we picked a region in the image to best measure the SIP for each Q and U frame. As noted by \citet{Laws2020}, using the region inside the coronagraphic mask region to estimate $f_Q$ and $f_U$ does not reliably remove the SIP for all targets. \citet{Laws2020} choose to use the region between 70-80 pixels away from the central star. This method was effective as long as the disk does not extend into this region or if there is a bright point source in this region. We chose to use a new method in which the entire FOV (0-140 pix) was used and mask out any regions where the Q/I or U/I ratio is larger than 0.05 or where there is known to be a point source within 10 pixels.

We tested if the masked method effectively measured $f_Q$ and $f_U$ by
rotating the set of Stokes Q and U cubes into \Qphi and \Uphi frames where:
\begin{align}
    Q_\phi =- Q^* \textrm{cos}(2\phi) -U^* \textrm{sin}(2\phi)  \\
    U_\phi = +Q^* \textrm{sin}(2\phi) - U^* \textrm{cos}(2\phi)
\end{align}
as defined by \citet{Monnier2019}.
This rotation will result in a distinctive quadrupole structure in \Qphi and \Uphip. If the masked method left a quadrupole structure, we utilized annulus regions of 10 pixels wide that minimized the quadrupole structure in the \Qphi and \Uphi images and recalculated Q$^*$ and U$^*$ described in Equations \ref{eqn:remove_stellar_polQ} and \ref{eqn:remove_stellar_polU}. These methods are compared in the appendix of \citet{Davies2022}. Systems hosting bright point sources in the FOV occasionally exhibited some quadrupole structure in the \Qphi and \Uphi that could not be removed with the above methods (e.g. HD 98800 B and HD 144432). Future work is needed to better remove SIP from the GPI polarimetric data. For those wishing to replicate our reductions, the regions used to calculate $f_Q$ and $f_U$ can be found in the headers of our reduced FITS images along with the average $f_Q$ and $f_U$ removed. The average for a given epoch $f_Q$ and $f_U$ are listed in Table \ref{tbl:obs_log} as the polarization angle (PA) and \% polarization. We note that this information can be used to investigate the unresolved polarization of the inner disk region but caution that targets with low \% polarization ($<$ 0.6\%) will be dominated by instrumental polarization \citep{Millar2016}.

Once the SIP is finally removed, the final \Qphi and \Uphi images are calculated, and the set of \Qphi and \Uphi are median combined to produce the final I, \Qphip, and \Uphi images as shown in Section \ref{sec:results}. All I, \Qphip, and \Uphi images for every epoch can be found as a figure set appearing on the online version of this work (Figure \ref{fig:images_example}) \footnote{arXiv version has these figures can be found in appendix \ref{sec:IQU_appendix}}. An example of the I, \Qphip, and \Uphi images can be seen in Figure \ref{fig:images_example}. The fully reduced data can be found on Vizier 
and Data Behind the Figures (DBF). We also note that we have adopted a FITS header standard, in collaboration with VLT/SPHERE team members, that is presented in appendix \ref{sec:header} to aid in easy comparison of data and result replication.

\begin{figure*}[h]
    \centering
    \includegraphics[width=0.99\linewidth]{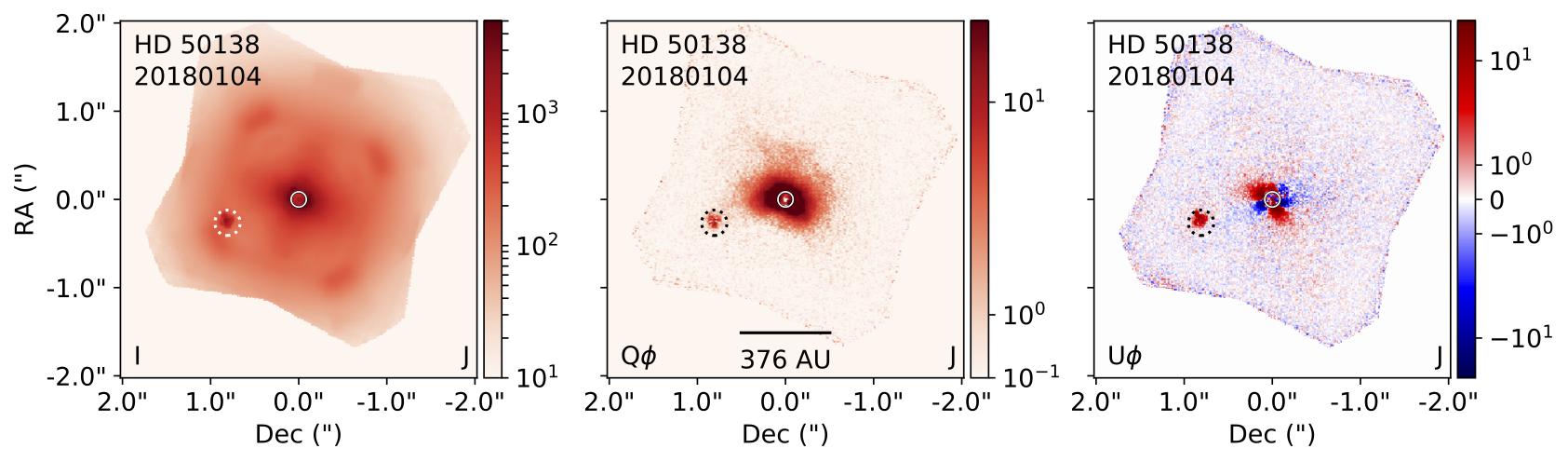}
    \caption{I, \Qphi, and \Uphi images for the Gemini-LIGHTS sample. Coronagraphic mask region marked with black circle. Location of likely binaries indicated in dashed black circles. Target and epoch date (YYYYMMDD) indicated in upper left of each panel. The complete figure set (44 figures) is available in the online journal (AJ). For arXiv version, see appendix \ref{sec:IQU_appendix}. 
    \label{fig:images_example}}
\end{figure*}
\subsection{Flux Calibration}\label{sec:flux_calibration}

The images are flux calibrated using the four satellite spots in each of the PODC images. To increase the signal to noise of satellite spots, the PODC files are mean combined. For H-band images, we used the first order satellite spots. For J-band images, we used the second order spots to ensure the spots were further away from the central PSF core. We note that the conversion factor between the satellite spots was updated  since version 1.4 of DRP that was utilized by \citet{Laws2020} and \citet{Monnier2019}. We used the known flux of the star from the 2MASS catalogue to calculate the flux conversion factor and apply it to the \Qphi and \Uphi images as shown in Figure \ref{fig:qphi_Uphi}. In our sample, we measure an average scale factor of 3.05 $\pm$ 0.57 for J-band and 3.09 $\pm$ 0.51 for H-band mJy\,$\mathrm{\prime\prime}^{-2}$/(ADU/s). 
We note that time (s) in the above scale factor is the total exposure time (ITIME $\times$ coadds). 

\begin{figure}[h]
    \centering
    \includegraphics[width=0.99\linewidth]{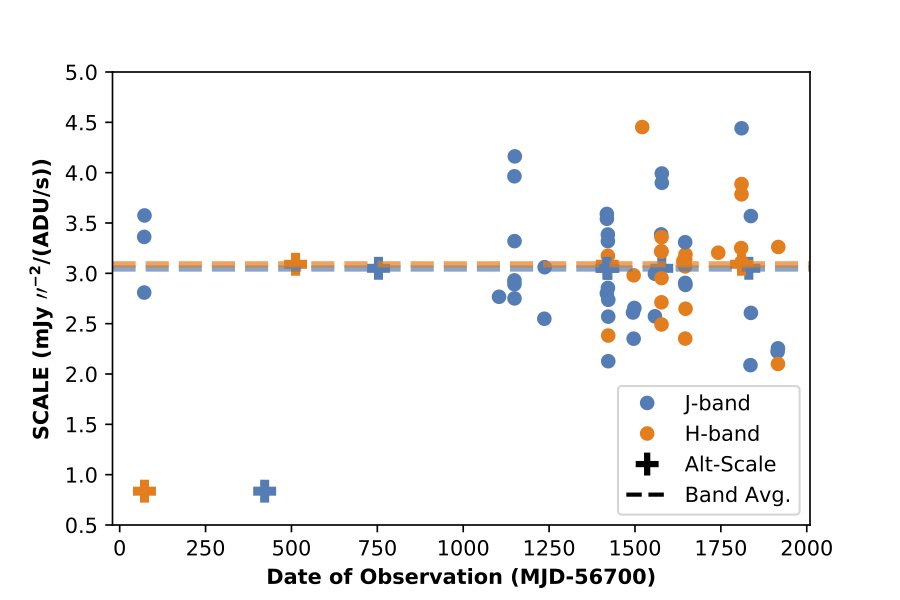}
    \caption{This figure shows the flux calibration scale for each of the targets and epochs in the sample. Blue dots are J-band observations and orange dots are H-band observations. Observations that have their scale factors corrected (see section \ref{sec:flux_calibration}) are noted with plus signs with their corrected values. Average J and H - band scale factors are shown as dashed lines. \label{fig:flux_claibration}}
\end{figure}

Central stars flux can be variable with time, thus without contemporaneous photometry we have taken effort to investigate and correct target epochs with discrepant flux calibration values. 
These different scale values could also be due to seeing conditions, and AO performance. We investigated 14 observations that had a scale value 1$\sigma$ (0.57 for J-band, 0.51 for H-band) away from the average scale value of 3.05 in J-band (3.09 in H-band). Four of these epochs had similar values to observations taken on the same night suggesting that the flux scale deviation is a correction for poor seeing or AO performance. Another four of these epochs had similar scale values to previous epochs suggesting that the divergent high or low scale value might be due to an over or under estimate of the flux of the target. The other six targets: FU Ori, HD 104237, HD 142666, HD 45677, HD 144432, and PDS 66 are likely to be due to variability thus we use the average flux scale of other targets in that epoch.
Finally, HD 100453 and HD 142527 were observed without coronographic spots thus we choose to use flux calibration scale factors from \citet{Long2017} 0.836 mJy\,$\prime\prime^{-2}$/(ADU/s) for both observations.

\subsection{Flat field accuracy} \label{sec:flatfield}

\begin{figure}[!ht]
    \centering
    \includegraphics[width=0.95\linewidth,trim=80 20 40 10,clip]{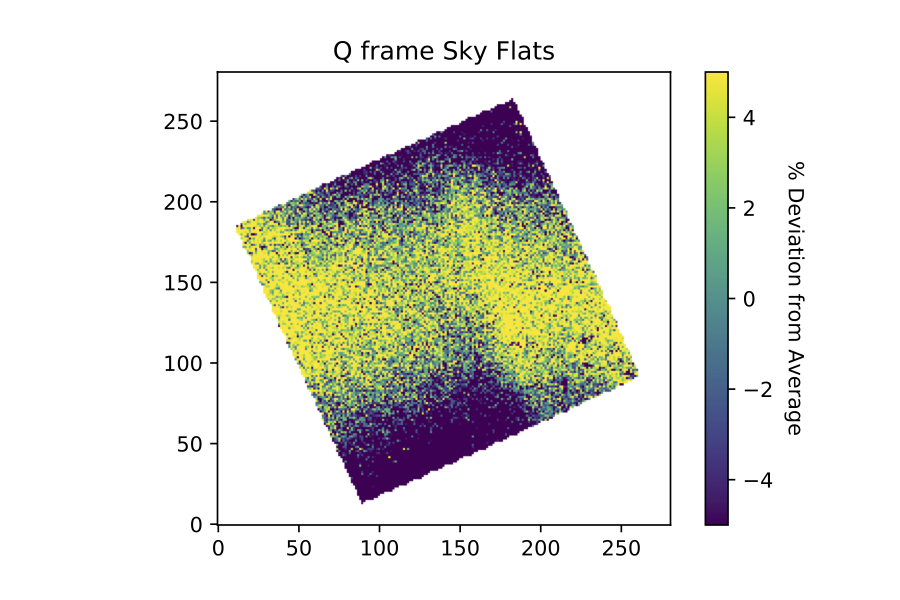} \\
    \includegraphics[width=0.95\linewidth,trim=80 20 40 10,clip]{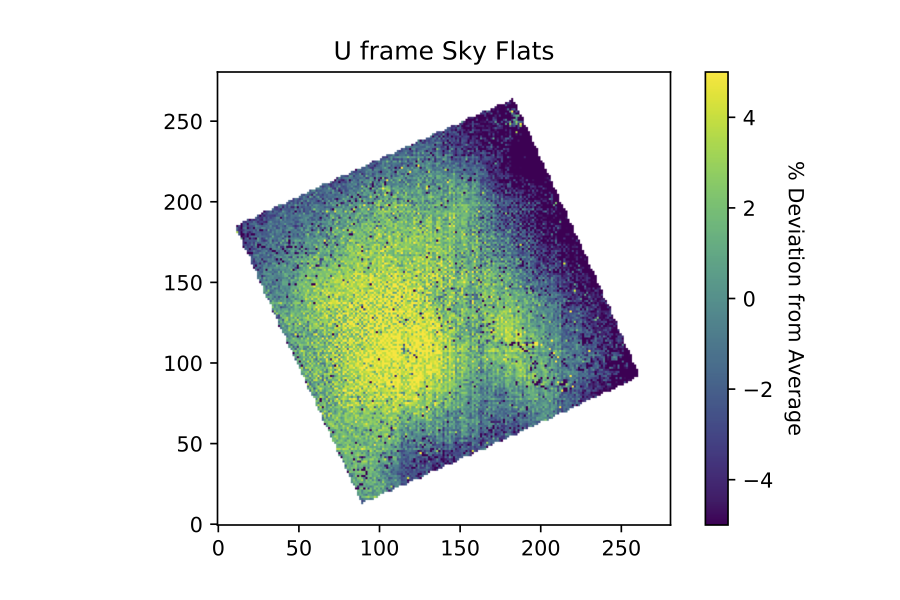}
    \caption{Showing the Q (top) and U (bottom) example sky flats taken on 20180103. Flux plotted as the percent deviation from the flux average of the frame.
    \label{fig:sky_flats}}
\end{figure}

Due to the design of the GPI instrument, flat field corrections of the pixel-to-pixel variations in the raw frame are not currently possible. The current correction uses lamp flat observations to correct for low-frequency variations across the FOV. This issue was first investigated by \citet{Millar2016}.
We observed twilight sky to obtain sky flats in order to independently estimate the effects of the flat field correction onto our data. 
We reduced the data using the same parameters as described above in section \ref{sec:pipeline}.
We plot an example of an individual Q and U frame as a \% flux deviation from the average of the Q and U frames as shown in Figure \ref{fig:sky_flats}. We show that the low-frequency flat field fails to correct for all large scale flat fielding variations, especially towards the edge of the detector.
However, there are still sizable flux variations between 2-4\% at the center of the image. Thus we conclude that any azimuthal flux variations observed in the disk that are on the scale of 2-4\% may not be astrophysical but instrumental due to poor flat fielding of the image.

\subsection{Uncertainty Propagation}\label{sec:bootstrap}

We estimate the uncertainty of our images by bootstrapping images from the fully reduced Stokes \Qphi and \Uphi images.
We performed this bootstrapping 100 times and these sets of bootstrapped images are used for all analysis and error propagation in section \ref{sec:results} and utilized in calculating the uncertainties of the polarization statistics for each observed epoch as shown in Table \ref{tbl:epoch_parameters}.

\subsection{Point Source Reduction}\label{sec:binaries}

We identified point sources in the FOV by looking for bright companions in the intensity image, and identifying dimmer point sources using an Angular Differential Imaging (ADI) reduction technique. For the ADI reduction, we used the Karhunen-Lo\'eve Image Projection (KLIP) algorithm for epochs that had FOV rotation that was sufficiently large to take advantage of an ADI reduction.
Using the centered PODC files from the standard reduction discussed above, we utilized the pyKLIP \footnote{https://pyklip.readthedocs.io/en/latest/index.html} package output \citep{Wang2015}. In order to avoid self-subtraction issues, we used parameters of subsection=1 and annuli$=1$ which minimizes the number of individual segments in the FOV used to calculate the PSF. We found that 16 systems have point sources identified by the pyKLIP reduction. Upon close examination, all of these point sources are visible in the intensity images. A further 6 targets have point sources that were identified in the total intensity image but did not have sufficient FOV rotation for the pyKLIP analysis. 

We measured the flux and position of each point source by fitting a Gaussian to each point source in the intensity frame. 
The location of these point sources can be found in \Qphi and \Uphi images in the online figure sets (e.g. Figure \ref{fig:images_example}) and are marked by dotted black circles. The point source location, flux, and estimated mass are shown in Table \ref{tbl:companion_results}. For the companion mass estimates we used models from \citet{Baraffe2015} for masses $>$ 0.01 \Msun{} and the chemical equilibrium models from \citet{Phillips2020} for masses $<$ 0.01 \Msun{}.

Using the pyKLIP routine, we also placed limits on the point source companions we are sensitive to detecting in our sample. We calculated the 5-$\sigma$ flux limit of point sources at angular separation of 0$\farcs$2 and 0$\farcs$5 for all targets with sufficient field rotation. We define sufficient rotation where a point source would move $>$ 1 pixel due to field rotation, corresponding to a total field rotation of 4$^\circ$ at 0\farcs2 and 1.6$^\circ$ at 0\farcs5. We note that 6 systems, HD 98800 B, HD 100453, HD 142527, HD 142666, HD 158643, and WRAY 15-535 did not have sufficient field rotation at 0$\farcs$2 to determine the contrast at a separation of 0$\farcs$2. Based on the contrast and the systems age, we estimated the upper limit mass that would be detectable at 0$\farcs$2 and 0$\farcs$5. For angular separations of 0$\farcs$2, in 9 systems we can detect down to Jupiter mass companions, for 26 we can detect down to brown dwarf mass companions, and in 3 systems we can detect some stellar mass companions.  For angular separations of 0$\farcs$5, in all systems we can detect companions down to 1 M$_{jup}$ mass companions. The results for each epoch can be found in the appendix in Table \ref{tbl:epoch_parameters}. For the companion mass estimate limits we used models from \citet{Baraffe2015} for masses $>$ 0.01 \Msun{} and the chemical equilibrium models from \citet{Phillips2020} for masses $<$ 0.01 \Msun{}.

\section{Results}\label{sec:results}

We now present the complete sample of our results from the Gemini-LIGHTS survey. All I, \Qphi, and \Uphi images for all 71 epochs of the 44 targets can be found as a figure set appearing on the online version of this work (Figure \ref{fig:images_example}). We will first identify targets in which we do not detect \Qphi flux. Next we will group the remaining targets by their disk morphological properties. We note that some targets can be found in multiple categories.

Some of the observations from this long and large observational campaign have already been published as part of the Gemini-LIGHTS campaign. \citet{Laws2020} published observations of 4 large disks with irregular features (FU Ori, MWC 789, HD 45677, Hen 3-365). \citet{Monnier2019} investigated the J- and H-band observations of HD 34700A which exhibit strong spiral structures. \citet{Davies2022} published the J- and H-band observations of HD 145718 that compared the scattered light images to photometry and near-IR interferometry and constrained by radiative transfer modeling. Finally, \citet{Kraus2020} investigated the triple star system of GW Ori finding evidence of disk tearing.

\subsection{Non-detections} 
\label{sec:nondetections}
\begin{figure}[!ht]
    \centering
    \includegraphics[width=1.0\linewidth]{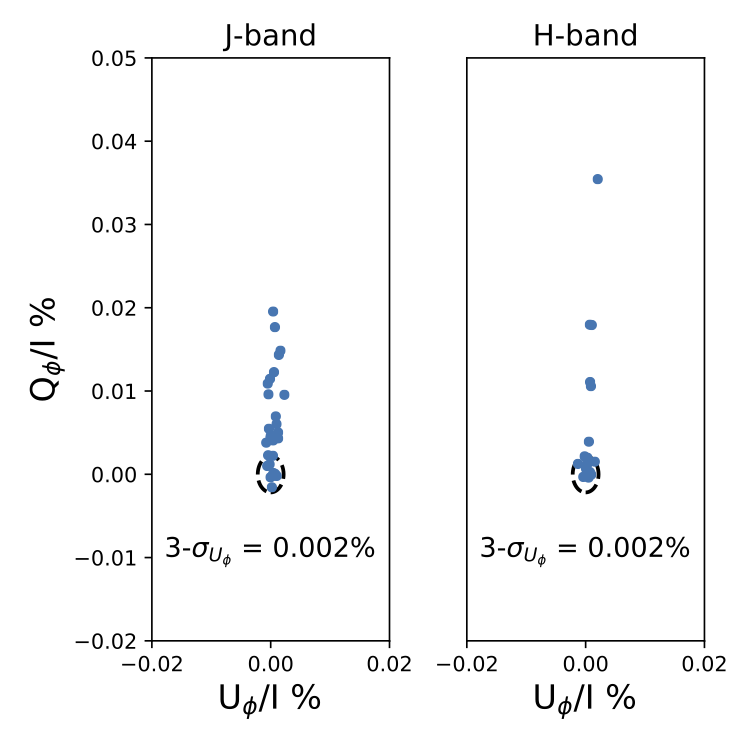}
    \caption{The \% of Q$_\phi$/I flux versus U$_\phi$/I in a 0$\farcs$4 annulus (30 pixels) for J-band (left) and H-band (right) observations. The dashed circle represents the 3-$\sigma$ significance.
    \label{fig:qphi_Uphi}}
\end{figure}

Here we establish which targets have detected scattered light and which targets are non-detections. One advantage in rotating the polarization I, Q and U into \Qphi and \Uphi is the ease of interpretation. \Qphi flux should be dominated by photons that scattered only once and from one source while \Uphi flux could be dominated by photons that scattered multiple times, or scattering that is not azimuthally symmetric with the assumed central star (i.e. presence of a binary). We compute the amount of \Qphi and \Uphi flux within 0$\farcs$4 (30 pixels) divided by the stellar flux for that given band. For sources with no close-in point source (e.g. within 0$\farcs$4 of central star), for disks with very low \Qphip, the amount of \Uphi will go to zero before \Qphi as shown in Figure \ref{fig:qphi_Uphi}. Since \Qphi and \Uphi are drawn from the same observables, the noise in \Qphi is the same as \Uphip. Thus we look at the standard deviation of \Uphi for epochs $<$ 0.01\% \Qphi and \Uphi and removed binaries by removing epochs $>$ 3-$\sigma$. We then took the standard deviation of the summed \Uphi flux which is also the uncertainty of the \Qphi flux. Using this metric, we establish that we do not detect polarized light around 18 of the 77 epochs using a 3-$\sigma$ deviation of \Uphi of 0.002\% in J-band and 0.002\% in H-band. Targets that are non-detections include: HD 36917, HD 101412, HD 144432, HD 158643, HD 176386, Hen 2-225, Hen 3-1330, MWC 147, MWC 166. Two targets, V1295 Aql (20180608 H-band, 20180816 H-band) and TY CrA (20180608 H-band) where not detected in some epochs, but were detected in other epochs (V1295 Aql: 20180816 J-band, TY CrA: 20180817 H-band). Thus, we do not detect polarization light around 9 of the 44 targets and additionally do not always detect polarization light in 2 of the targets.  

\subsection{Disk Structure Categorization}\label{sec:catagories}

We are categorizing our disks only by the observed disk structure in the \Qphi images. The categories include spiral armed disks, ringed disks, continuous disks, irregular disks, and undetermined disks structure. Our chosen categories are inspired by the disk structure categories defined by \citet{Garufi2018}. We have deviated when appropriate to better describe our sample. Results of our classification are tabulate in the appendix in Table \ref{tbl:target_properties}.

\subsubsection{Spiral Armed Disks}
\begin{figure*}[!ht]
    \centering
    \includegraphics[width=0.9\linewidth]{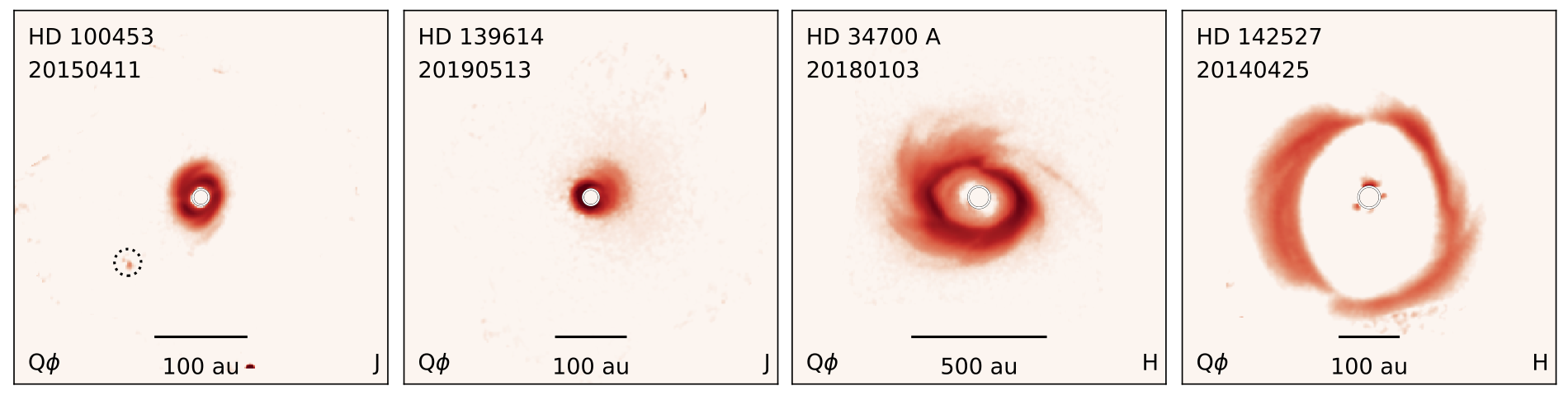}
    \caption{Spiraled Disks: \Qphi images of disks classified as hosting spiral arms including targets: HD 100453, HD 139614, HD 34700 A, and HD 142527. See online figure sets (e.g. Figure \ref{fig:images_example} for target specific color bars. Image flux scale is log and chosen to highlight target structure. Point sources in the FOV are labeled with a dashed black circle. The name of the target and the epoch in YYYYMMDD format can be found in the upper left of each image. The type of polarized image (\Qphi), the scale of the image (in au), and the photometry band (J or H) can be found along the bottom of each sub-image.
    \label{fig:giant_spirals}}
\end{figure*}

We find there are four targets that host one or more spiral arms including HD 100453, HD 139614, HD 34700 A, and HD 142527 as shown in Figure \ref{fig:giant_spirals}. HD 100453 hosts symmetric spiral arms while HD 34700A and HD 142527 most multiple arms that are not symmetric. HD 139614 has a arm structure on one side of the disk. All four of these systems have had the origin of their spiral arms investigated previously (HD 100453: \citealt{Wagner2015}, HD 139614: \citealt{Laws2020}, HD 34700 A: \citealt{Monnier2019, Laws2022}, HD 142527: \citealt{Long2017}). 

\subsubsection{Ringed Disks} 
\begin{figure*}[!ht]
    \centering
    \includegraphics[width=0.9\linewidth]{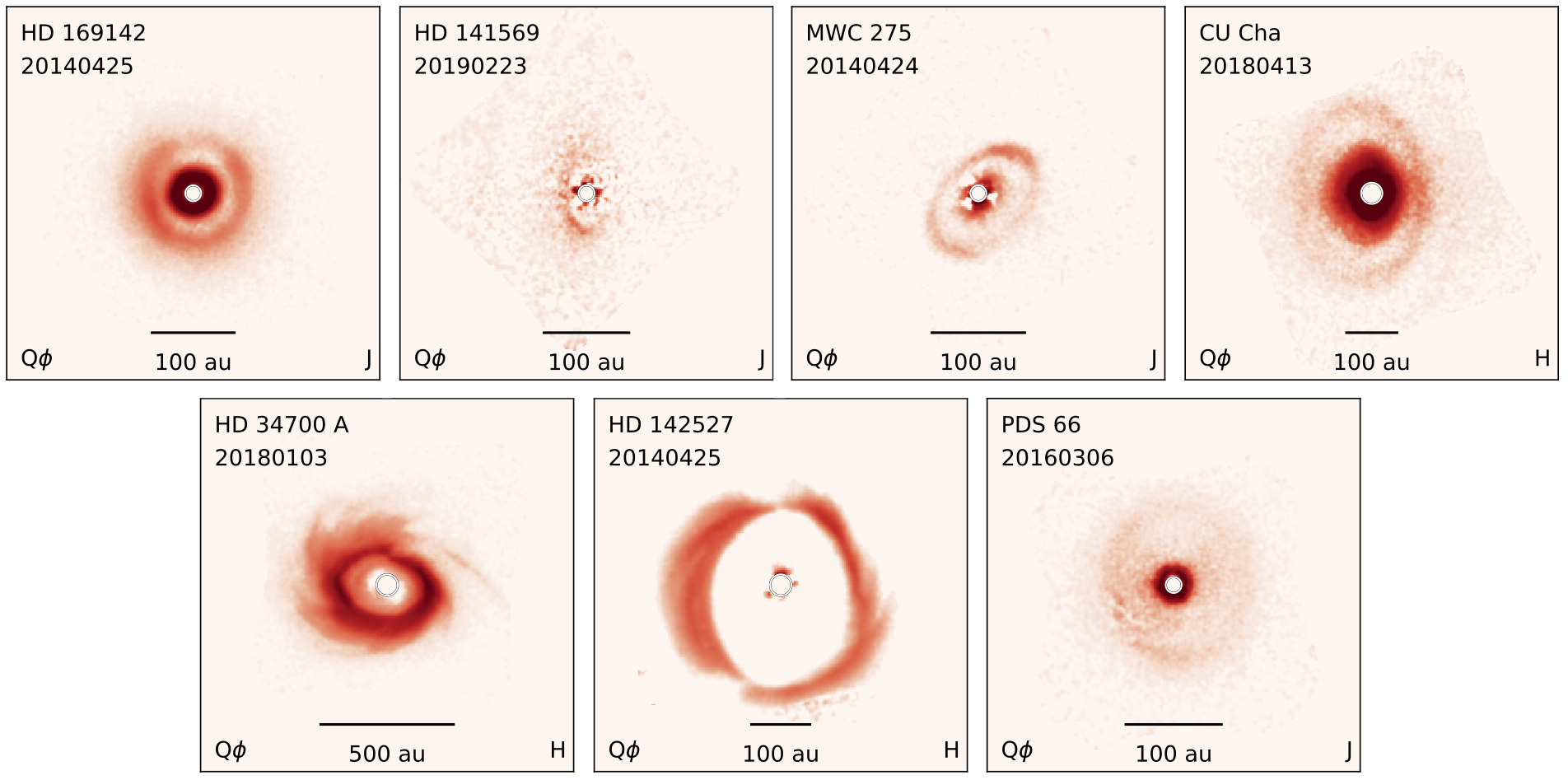}
    \caption{Ringed Disks: \Qphi images of disks classified as hosting rings including targets: HD 169142, HD 141569, MWC 275, CU Cha, HD 34700 A, HD 142527, PDS 66. See online figure sets (e.g. Figure \ref{fig:images_example} for target specific color bars. See Figure \ref{fig:giant_spirals} caption for figure details.
    \label{fig:ringed}}
\end{figure*}

We define disks that have one or more concentric polarized light dust rings in their \Qphi image as shown in Figure \ref{fig:ringed}. We note that we deviate from the definition of \citet{Garufi2018} as they make a distinction between ringed and rimmed systems, however scattered light imaging lacks the information to robustly make the distinction due to the inner working angle. We find that 7 targets show signatures of rings: HD 169142, HD 141569, MWC 275, CU Cha, HD 34700 A, HD 142527, and PDS 66. We note that 2 of the 4 spiral armed disks are also ringed disks (HD 142527, HD 34700 A).

\subsubsection{Continuous Disks} 
\begin{figure*}[!ht]
    \centering
    \includegraphics[width=0.99\linewidth]{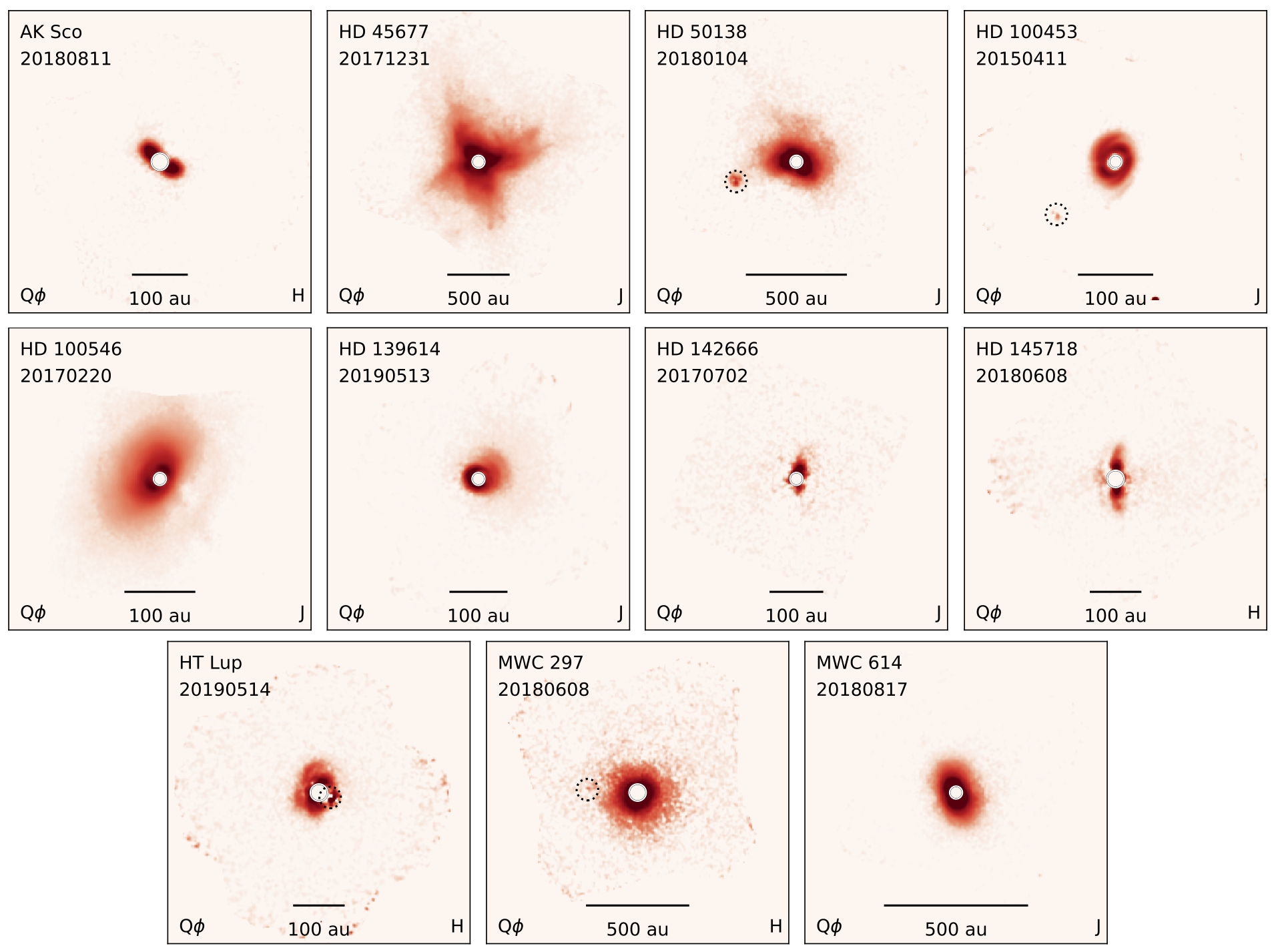}
    \caption{Continuous Disks: \Qphi images of disks classified as continuous including targets: AK Sco, HD 45677, HD 50138, HD 100453, HD 100546, HD 139614, HD 142666, HD 145718, HT Lup, MWC 297, and MWC 614. See online figure sets (e.g. Figure \ref{fig:images_example} for target specific color bars. See Figure \ref{fig:giant_spirals} caption for figure details.
    \label{fig:continuous}}
\end{figure*}

We define continuous disks as disks that are resolved and do not appear to have a gap or hole in their PDI or show strong non-azimuthal structure as shown in Figure \ref{fig:continuous}. Some of these disks may have gap or ringed structure but is not shown in the polarized light imagery due to the inclination (HD 145718: \citealt{Davies2022}) or Inner Working Angle (IWA) (MWC 614: \citealt{Kluska2018}).  
We note that this does not imply that these disks do not have gaps at all, but those gaps are not visible in our polarized light imaging. We find that 11 targets show signatures of rings: AK Sco, HD 45677, HD 50138, HD 100453, HD 100546, HD 139614, HD 142666, HD 145718, HT Lup, MWC 297, and MWC 614. We note that one of the spiral armed disks are also continuous disks (HD 139614).

\subsubsection{Irregulars} 
\begin{figure*}[!ht]
    \centering
    \includegraphics[width=0.99\linewidth]{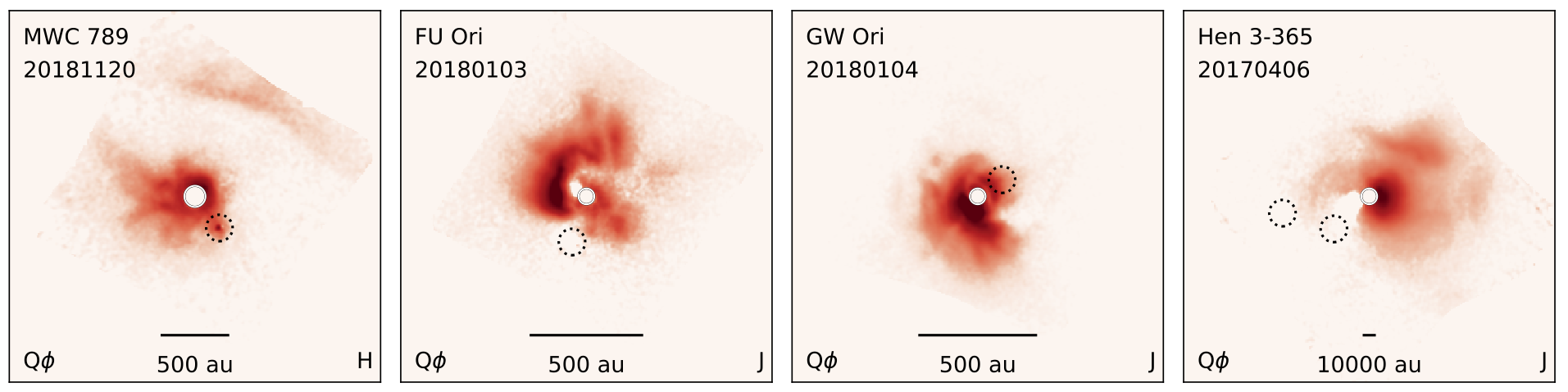}
    \caption{Irregular Disks: \Qphi images of disks classified including targets: MWC 789, FU Ori, GW Ori, and Hen 3-365.See online figure sets (e.g. Figure \ref{fig:images_example} for target specific color bars. See Figure \ref{fig:giant_spirals} caption for figure details.
    \label{fig:irregulars}}
\end{figure*}

We define irregular disks as disks that host structures that have strong non-azimuthal features and the disks are very large in size ($>$ 300 au). There are four disks in our sample that exhibit these features: MWC 789, FU Ori, GW Ori, and Hen 3-365 which are shown in Figure \ref{fig:irregulars}. GW Ori irregularity is thought to be due to disk tearing as investigated by \citet{Kraus2020}.

\subsubsection{Undetermined} 
\begin{figure*}[!ht]
    \centering
    \includegraphics[width=0.99\linewidth]{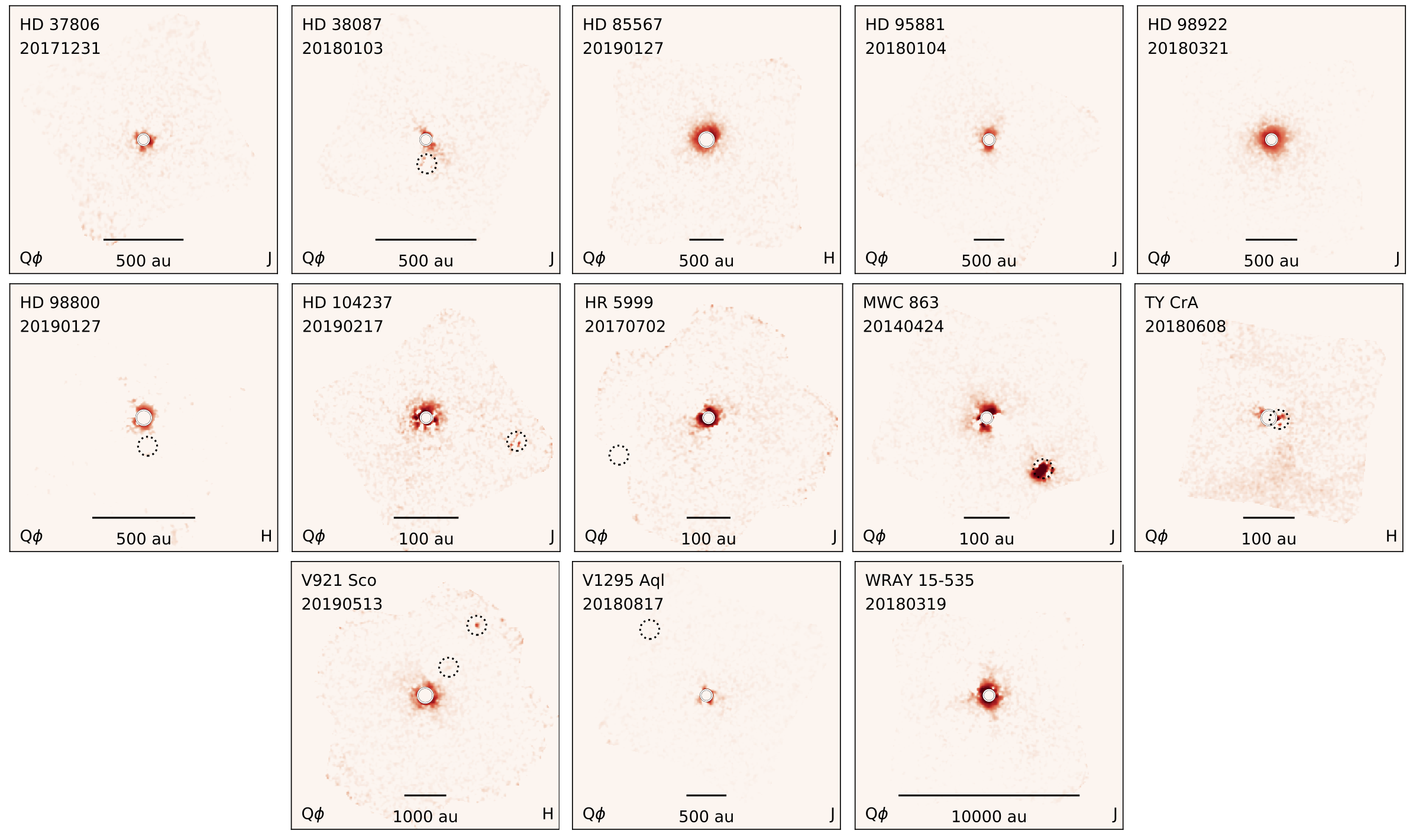}
    \caption{Undetermined Disk Structures: \Qphi images of disks which are insufficiently resolved to determine their morphology. Targets include: HD 37806, HD38087, HD 85567, HD 95881, HD 98922, HD98800, HD 104237, HR 5999, MWC 863, TY CrA, V921 Sco, V1295 Aql, and WRAY 15-535. See online figure sets (e.g. Figure \ref{fig:images_example} for target specific color bars. See Figure \ref{fig:giant_spirals} caption for figure details.
    \label{fig:unresolved}}
\end{figure*}

There are some targets that have a significant amount of polarized light detected (see subsection \ref{sec:nondetections}), however the polarized light in the image does not extend far enough away to reliably categorize the disk into one of the above categories. Objects that fit this description include: HD 37806, HD38087, HD 85567, HD 95881, HD 98922, HD98800, HD 104237, HR 5999, MWC 863, TY CrA, V921 Sco, V1295 Aql, and WRAY 15-535 as shown in Figure \ref{fig:unresolved}. We note this is a similar to the ``small disk'' category utilized by \citet{Garufi2018}. However, our sample is not strictly distance limited thus some objects have disks that do appear to be small (e.g. HD 104237 disk radius $<$ 21 au) while other objects such as V921 Sco appear to be distant and have rather large disks ($<$ 300 au). Thus we choose to classify these objects as undetermined.

\subsubsection{Categorization Summary}

\begin{figure*}[!ht]
    \centering
    \includegraphics[width=0.51\linewidth,trim=9 1 1 5 ,clip]{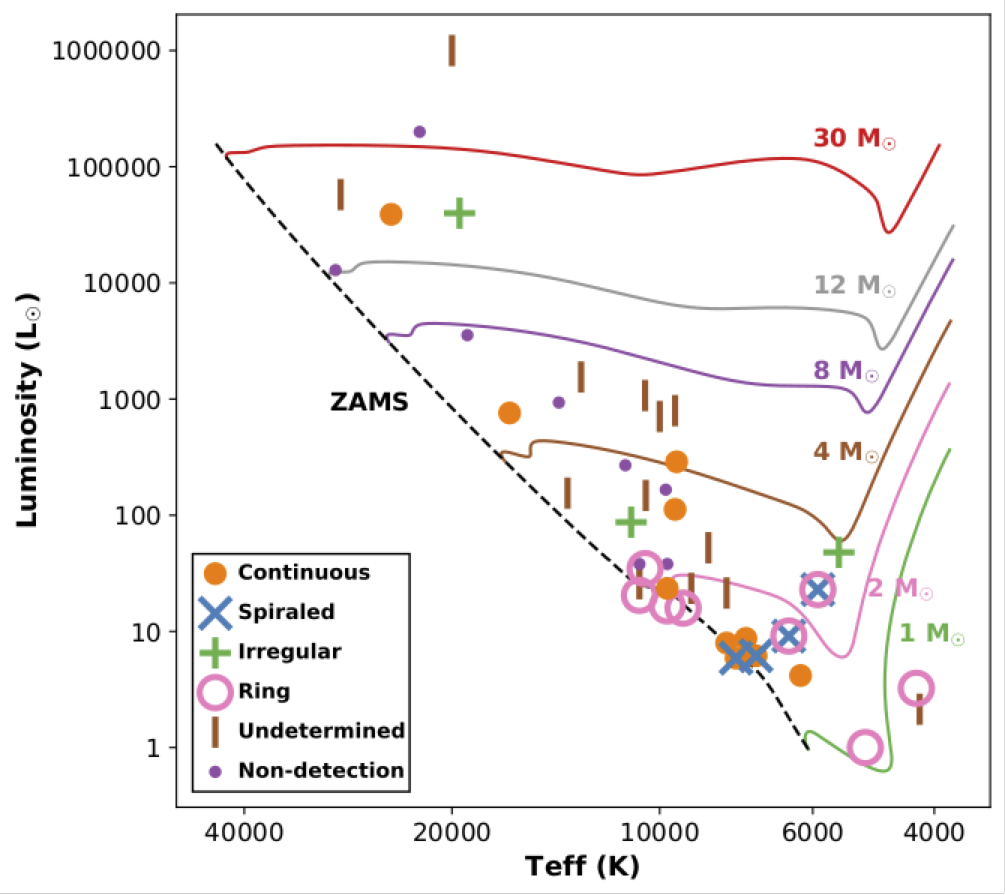}
    \includegraphics[width=0.48\linewidth,trim=1 1 1 1 ,clip]{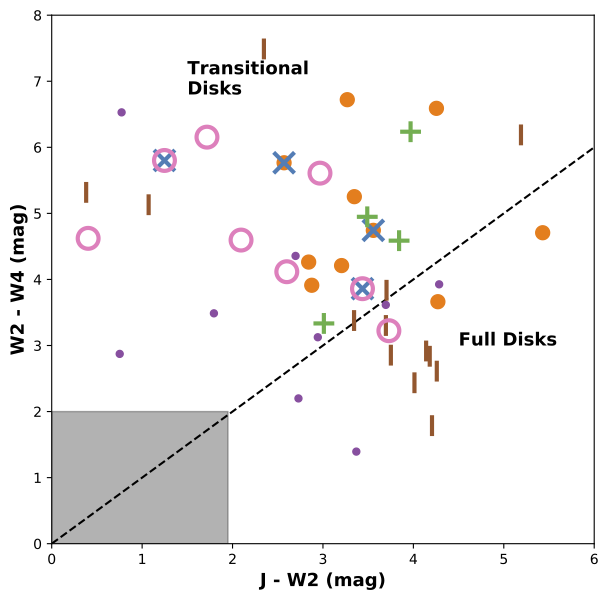}
    \caption{HR diagram (left) and color-color diagram (right) of this work's entire sample. Each target is marked by its apparent polarized light disk structure described in section \ref{sec:catagories}. HR diagram includes pre-main-sequence mass tracks (colored lines) and the zero age main sequence (ZAMS) assuming solar metallicity mass tracks from \citet{Bressan2012}. Note that FU Ori is not plotted on the HR diagram as the T$_{\textrm{eff}}$ temperature is unknown.
    \label{fig:Entire_sample}
    }
\end{figure*}

Utilizing the polarized disk morphology described above, we now investigate if we see any trends with the disk morphology. We plotted all 44 targets on an HR diagram and an infrared color-color diagram in Figure \ref{fig:Entire_sample}. We note that targets can be in multiple categories. 
The first trend is that ringed systems only occur in systems with masses $<$ 3 \Msun. This could partially be due to a distance effect where the ringed structure is within the IWA of the observations, though less likely since there are 6 systems with resolved disks but do not exhibit any ringed structures. Second, irregular systems appear to be younger; however, this is not a definitive trend as only three objects on the HR diagram are classified as irregular. 

Next, we plot the complete sample on a color-color diagram as shown in Figure \ref{fig:Entire_sample}. We generally find that full disks are unresolved, undetected, or continuous disks. Pre-transitional disks are mostly continuous disks with some irregulars and ring disks. Finally, transitional disks are dominated by ringed disks, as expected. 

\subsection{FS CMa stars}\label{sec:FSCMas}
\begin{figure*}[!ht]
    \centering
    \includegraphics[width=0.99\linewidth]{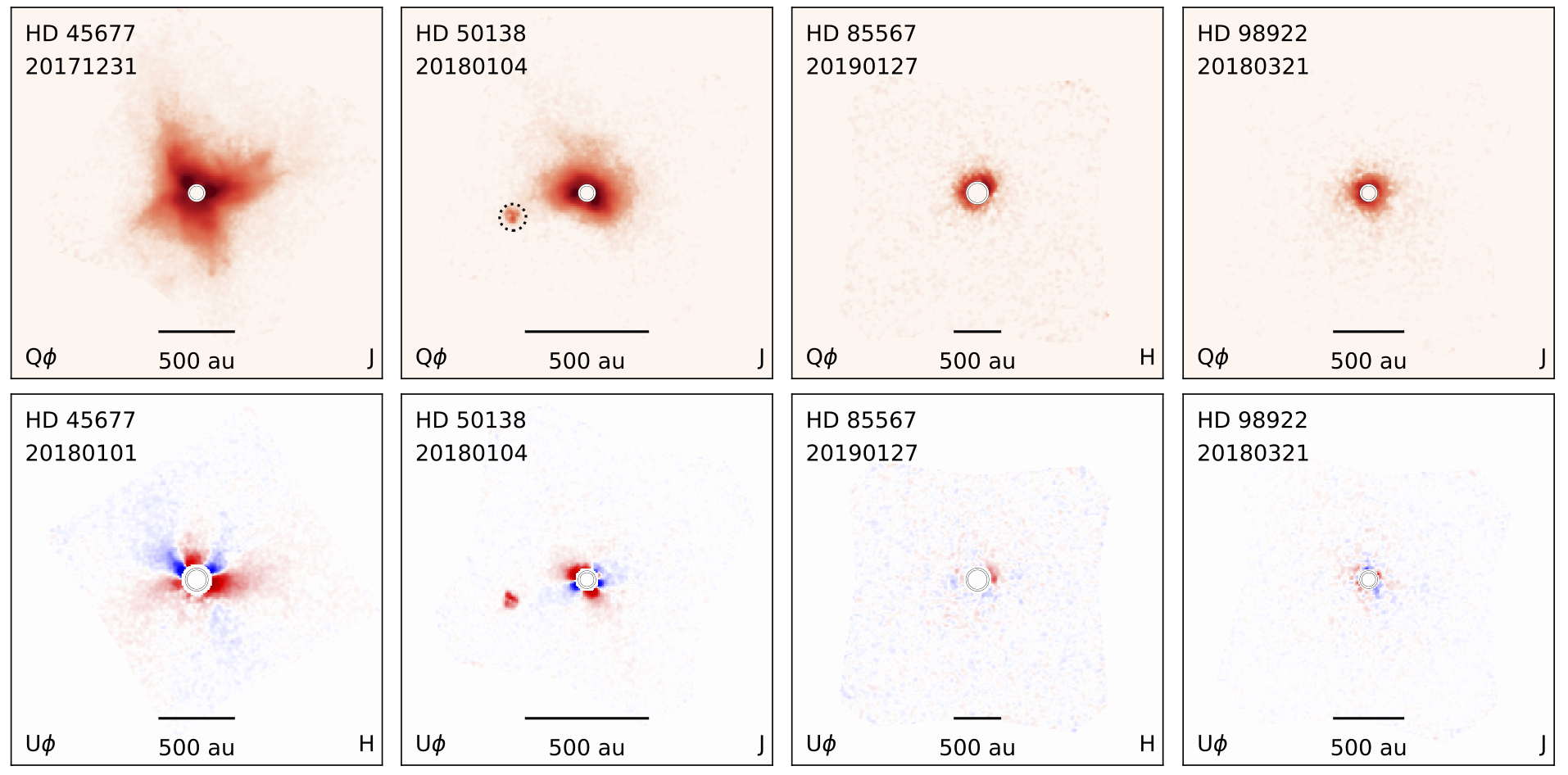}
    \caption{\Qphi and \Uphi images of targets that have previously been classified as FS CMa. Image scale is log and chosen to show target structure. See figure set appearing on the online version of this work (Figure \ref{fig:images_example}) for target specific color bars. Point sources in the FOV are labeled with a dashed black circle in \Qphi images. The name of the target and the epoch in YYYYMMDD format can be found in the upper left of each image. The type of polarized image (\Qphi or \Uphi), the 1" scale and the size in au, and the photometry band (J or H) can be found along the bottom of each sub-image. In \Uphi images, red is positive and blue is negative flux. 
    \label{fig:FSCMa}}
\end{figure*}

Four of our targets have been identified as FS CMa targets which are a sub-type of B[e] stars: HD 45677, HD 50138, and HD 85567, and HD 98922 \citep{Miroshnichenko2007,vioque2020}. These stars are potentially post-main sequence stars.  We note that HD 45677 is also known as FS CMa, the prototype for this classification.
We have the first resolved images of the dust around HD 45677, HD 50138, HD 85567 and HD 98922 as shown in Figure \ref{fig:FSCMa}. 
All objects have significant \Qphi flux that is detected out to sizable distances of $\sim$250 au (HD 50138), $\sim$700 au (HD 45677), $\sim$ 300 au (HD 85567), and $\sim$ 180 au (HD 98922). Additionally, each of the objects also show significant \Uphi flux which is suggestive of multiple scatterings due to optical depth effects or multiple illumination sources due to a binary. The \Uphi flux for both HD 45677 and HD 50138 is likely to be due to optical depth effects as the pattern is within the location of high \Qphi flux similarly seen in other optically thick sources such as HD 34700 A, GW Ori, and HD 100546. 
HD 50138 does have a point source located within 1$\farcs$0 of the central star but is not co-located with the concentric \Qphi flux around the central star. There is excess polarized flux north of the main disk that could be faint additional structure (e.g., back of the disk, spiral arm) but we are unable to definitively determine it's origin. HD 85567 is very distant (1047 pc) thus only the very outer portion of the system is imaged. Finally, HD 98922 exhibits a centro-symmetric \Qphi pattern around the inner working angle. 

\subsection{Point Source Detections in the Field of View}

We  found 24 point sources within the FOV of the PDI images with 21 targets hosting these point sources.
Point sources are identified in \Qphi and \Uphi images in the online figure sets (e.g. Figure \ref{fig:images_example}) with a dotted black circle marking the companions locations. The values are also tabulated in Table \ref{tbl:companion_results} located in the appendix. We assessed the likelihood that a given point source was a background star that the measured separation and brightness from the target star. We measured the local star density by using number of stars per half magnitude within 1 deg of the target via the 2MASS catalogue.
We then calculated the probability that a background star would be located within the projected distance of the target.
We found that 22 of the 24 point sources are likely to not be background stars within 3$\sigma$. Two point sources found around V921 Sco are less than a 2$\sigma$ probability not being background stars. Based on the likelihood that all of the point source companions are gravitationally bound, for the purposes of this work we will assume all of the 24 point sources are binaries. However, future follow-up observations, especially V921 Sco, are needed.

We estimated the masses of each of the companions using the age of the system and the companions measured flux and extrapolated the mass from evolutionary models from \citet{Baraffe2015} for masses $>$ 0.01 \Msun{} and \citet{Phillips2020} for masses $<$ 0.01 \Msun. We estimated the mass uncertainties using the uncertainty in the systems age and flux. We note that some of our targets have ages less than 0.5 Myr which is younger than the youngest age by \citet{Baraffe2015}.
In these cases, we estimated their mass based on the youngest age in the \citet{Baraffe2015} models to give an idea of their mass but note these masses should not be used for future analysis as they are not reliable. Estimated masses are tabulated in Table \ref{tbl:companion_results} located in the appendix.

\begin{figure*}[!ht]
    \centering
    \includegraphics[width=0.49\linewidth,trim=10 10 10 10 ,clip]{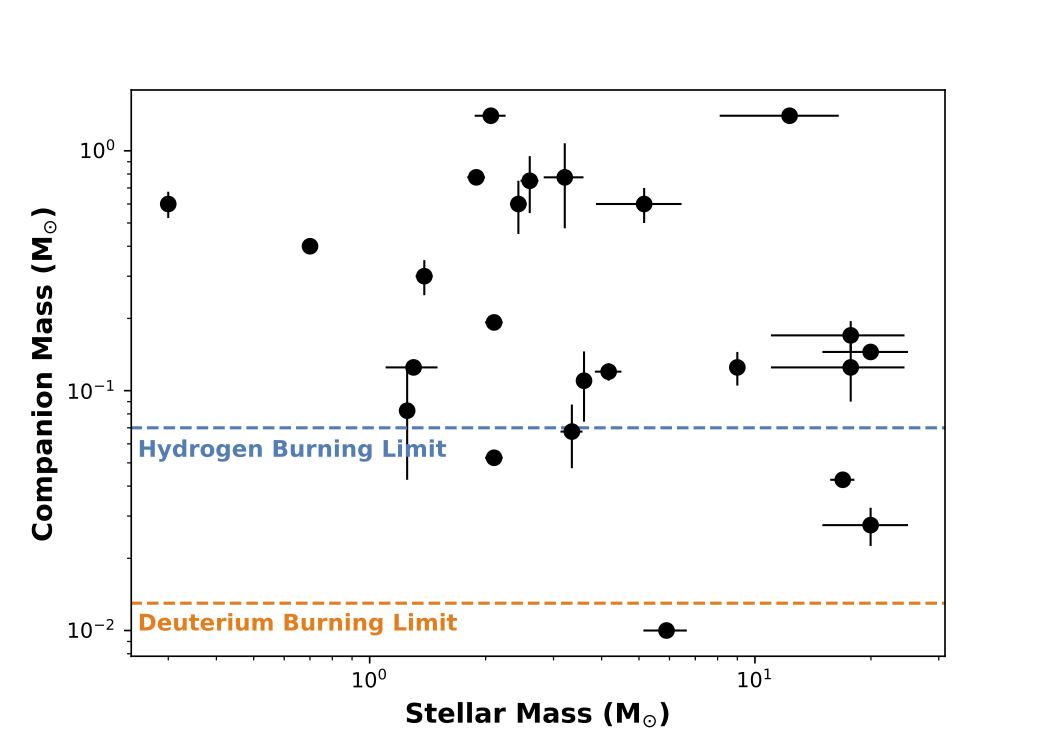}
    \includegraphics[width=0.49\linewidth,trim=10 10 10 10 ,clip]{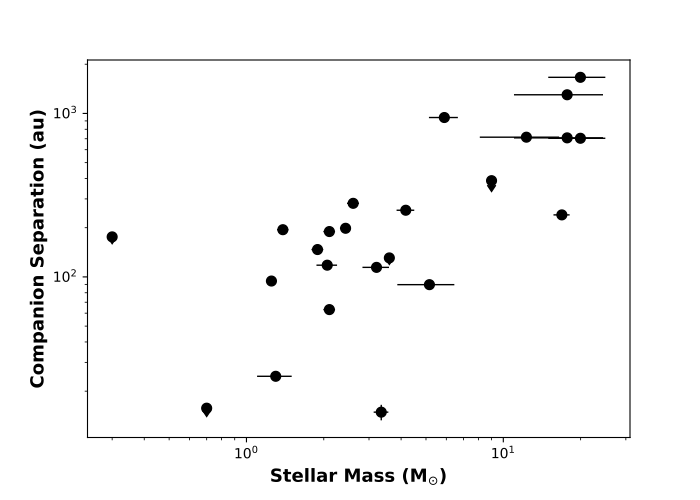}
    \caption{Comparison of stellar mass and estimated companion mass (left) and the stellar mass and the projected companion separation (right). For the companion masses, we denote the boundary between brown dwarfs and stars (hydrogen burning limit: blue) and denote the boundary between planets and brown dwarfs (deuterium burning limit: orange).
    \label{fig:point_source_trends}}
\end{figure*}

We compare the estimated companion mass and the projected companion separation to the stellar mass as shown in Figure \ref{fig:point_source_trends}. Of our 24 point sources, one point source around V1295 Aql has a mass consistent with a super-Jupiter, 4 have masses consistent with brown dwarf masses (HD 101412, HD 158643, MWC 297, and V921 Sco), 12 systems have companions with masses consistent with M-dwarfs (0.08 to 0.57 \Msun{}), and 7 have masses $>$ 0.57 \Msun{}. Further, we note that more massive stellar objects ($>$ 6 \Msun{}) have larger projected separations due to these objects being more distant. For these objects with separations $>$ 1000 au, future observations are necessary to determine if these objects are co-moving and if they are gravitationally bound.

\subsection{Possible Non-Concentric Reflection Nebulae}

\begin{figure*}[!ht]
    \centering
    \includegraphics[width=0.7\linewidth,clip]{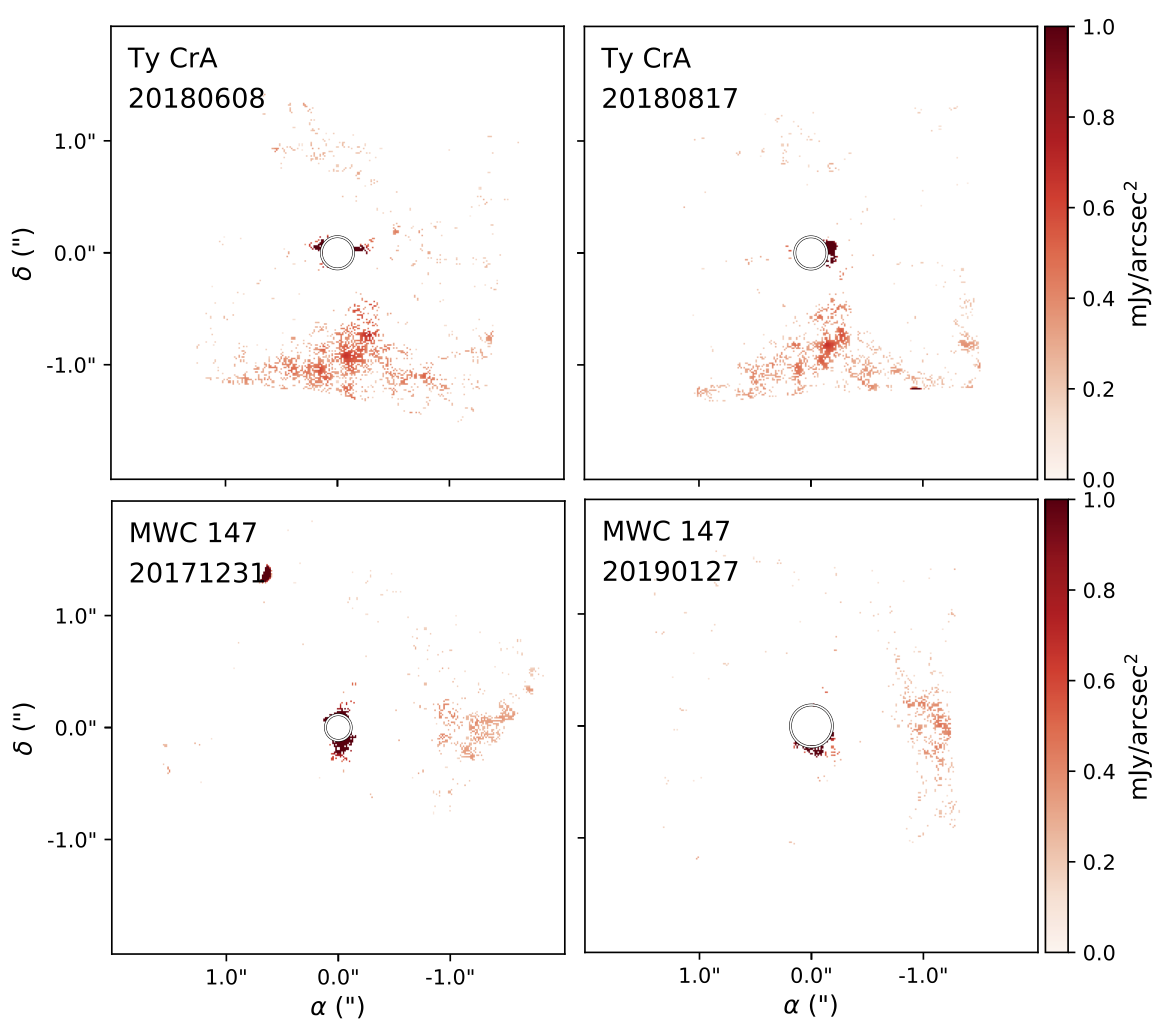}
    \caption{Target adjacent polarized flux of TY CrA (top row) and MWC 147 (bottom row)  for two different epochs. Both targets have point sources (dashed circles) close to the inner working angle (solid circle). Companion and flux $<$ 1-sigma masked in the images. Flux is in mJy ''$^{-2}$.
    \label{fig:outer_excess}}
\end{figure*}

Two targets, MWC 147 and TY CrA, have polarized flux that is adjacent to the target as shown in Figure \ref{fig:outer_excess}. The flux plotted is flux that is 1-sigma per pixel above the noise in the image. Both targets are known to be associated with reflection nebulae thus the flux is likely material part of the larger reflection nebulae. We also note that the flux may be an instrumental effect due to known issues with the flat fielding (see Section \ref{sec:flatfield}). However, this explanation is unlikely as the features appear in the same location over multiple epochs and multiple bands and no other sources show these features taken on the same night.

\subsection{Systems with Stellar Masses $>$ 8\Msun}

Six of our systems have stellar masses $>$ 8 \Msun{} including Hen 3-365, Hen 3-1330, MWC 166, MWC 297, V921 Sco, and WRAY 15-535. We do not detect polarized light around 2 of the 6 systems (Hen 3-1330 and MWC 166), and we classified the morphology of the rest as one irregular (Hen 3-365), one continuous (MWC 297), and two undetermined (V921 Sco, WRAY 15-535). Several of these objects have conflicting classifications between YSO and post main-sequence stars. Hen 3-365 is discussed by \citet{Laws2020} who noted that it is inconsistent with being a massive supergiant due to the large H I column density, but parallax distances are consistent with being an evolved star \citep{Oudmaijer1998,Maravelias2018}. Hen 3-1330 has previously been classified as a WR+O binary system \citep{Richardson2011}. V921 Sco has been classified as both a supergiant and as a Herbig Be star \citep{Kreplin2020}. WRAY 15-535 has been classified as a supergiant B[e] star \citep{Domiciano2007}.
Finally, previous work investigating MWC 166 and MWC 297 has shown that they are consistent with being Herbig Ae/Be stars \citep{Manoj2007,Wichittanakom2020}. We leave the polarized light investigations of these individual targets to future work.

\subsection{Scattered Light Diagnostics}

We calculate the total amount of \Qphi flux within the image relative to the stellar flux Fstar. To mitigate noise being added into the images, we sum all of the \Qphi light between the IWA of the target and where the radial profile of the \Qphi is within 3-$\sigma$ of zero flux using error propagation from the bootstrapped images discussed in section \ref{sec:bootstrap}. The summed \Qphi flux is then divided by the flux of the star using the 2MASS J- and H-band flux. The \Qphip/Fstar ratio can be found in appendix in Table \ref{tbl:epoch_parameters}. We note that this simplified methodology does not account for the r$^2$ flux loss. However, this method can easily be applied to our entire sample and is not dependent on knowing the disk geometry (i.e. disk inclination) which is not possible for some systems (i.e. unclassified, irregular: see Section \ref{sec:catagories}). Since there are multiple epochs of some targets, we will use the weighted average of \Qphip/Fstar ratio for all analysis.

\begin{figure*}[!ht]
    \centering
    \includegraphics[width=0.99\linewidth,trim=0 0 0 0 ,clip]{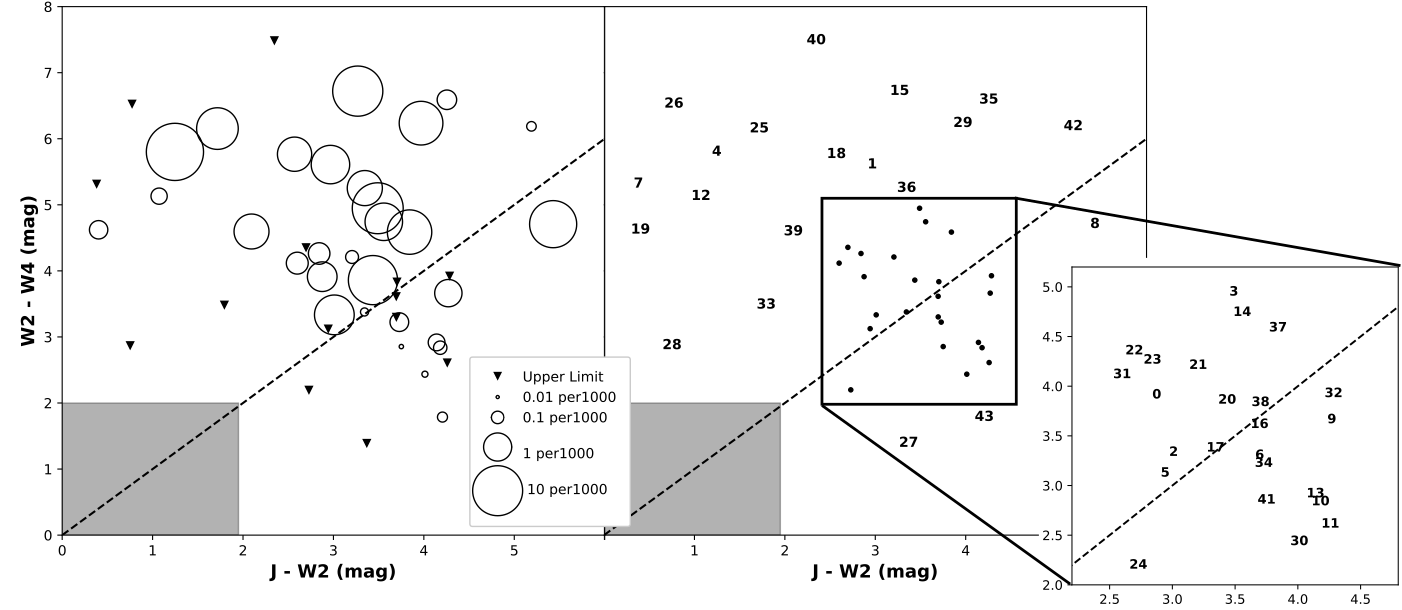}
    \includegraphics[width=0.65\linewidth,trim=5 5 5 5 ,clip]{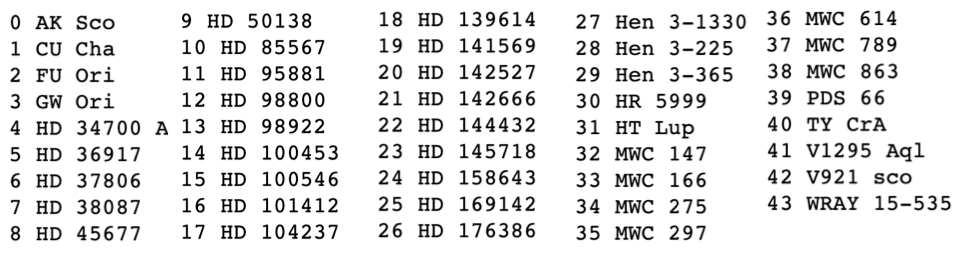}
    \caption{Infrared color-color plot (left) of the targets where the size of the circle corresponds to the magnitude of the \Qphip/Fstar ratio. Note that the size of the circles is logarithmic.  Color-color diagram is the same as shown in Figure \ref{fig:sample_GLITHS} where colors are using 2MASS J-band magnitudes and WISE 2 (4.6 $\mu m$) and 4 (22 $\mu m$) bands representing near-infrared excess (J-W2) and mid-infrared excess (W2-W4). The dashed line represents a flat spectrum SED. The grey shaded region are objects with no near-infrared or mid-infrared excess. Color-color diagram (right) labels each target with the corresponding number and target name on the bottom of the figure. \Qphip/Fstar ratios plotted are the weighted average for each target. The individual \Qphip/Fstar values can be found in Table \ref{tbl:target_properties}
    \label{fig:color_color_qphi}}
\end{figure*}

We first compare the \Qphip/Fstar ratio to the color-color diagram shown in Figure \ref{fig:color_color_qphi}, which we utilized for our target selection described in section \ref{sec:sample}. The size of the circle corresponds to the \Qphip/Fstar ratio on a logarithmic scale. We find that the targets with the largest \Qphip/Fstar ratio are commonly found in the middle of the color-color plot between 3$<$ W2-W4 $<$ 6 mag and 2 $<$ J-W2 $<$ 4 mag. 
This region coincides to the transitional disk region as outlined in Figure \ref{fig:sample_GLITHS}. Low and non-detected \Qphip/Fstar ratios are scattered throughout the diagram but have a higher concentration at the bottom with W2-W4 $<$ 3 mag. This lower right portion of the diagram coincides with the location of the full disks shown in Figure \ref{fig:sample_GLITHS}. Interpretation of these results will be discussed in section \ref{sec:discussion}.

\section{Scattered Light Trends}\label{sec:trends}

We next compare the total amount of summed scattered light per stellar flux, \Qphip/Fstar, to disk and stellar parameters to search for which parameters affect the amount of scattered light flux. Due to the diverse nature of our sample, there are some very high mass and very low mass stars. Here, we restrict our sample to only include stars between A-type stars ($>$1.4 \Msun) up to stars that will produce supernova ($<$8\Msun). When we have this restriction, we have a sample size of 33 targets that are also all in the \citet{Vioque2018} sample which provides a common source of temperatures, ages, and masses. In this section we will only analyze these 33 targets.

We will compare our sub-sample of 33 targets to their binarity, defined as either an outer binary where the stellar/sub-stellar companion can be directly imaged in our sample (separation $>$ 0$\farcs$12) or an inner binary (separation $<$ 0$\farcs$12). Binarity is determined using high-contrast imaging (see section \ref{sec:binaries}) or previously determined in the literature using spectroscopy, imaging, or interferometry. For the 33 targets in this sample,  we are able to detect binaries with masses down to brown dwarfs (0.075 \Msun) at a separation of 0$\farcs$2. There are three exceptions, HD 142527, HD 142666, and HD 158643 which didn't have sufficient field rotation to determine contrast upper limits at 0$\farcs$2. However, given the depth and that all of our binary targets identifiable in the intensity frame without the aid of ADI, we expect a similar sensitivity for these targets as well.
We note that the statistics of inner binaries is not complete as we are dependent on previous studies and it can be notoriously difficult to detect doppler shift of binaries in YSOs. Additionally, our analysis is not complete for outer binaries with separations larger than the FOV (separation $>$ 1$\farcs$9) and would not be detected by our survey.

Lastly, we investigate the commonly-used Meeus Group I and Group II categorization \citep{Meeus2001} of Herbig Ae/Be SEDs.
Group I disks have much larger far-infrared fluxes than Group II disks, and this classification is a version of the W2 - W4 color but with only two categories. Here, we calibrated our WISE-based Group I and Group II classification (see Table \ref{tbl:target_properties})  using the previous classification by \citet{Guzman2021} which has 29 of our targets in their Herbig Ae/Be sample. 

\subsection{Infrared colors}
\begin{figure*}[!ht]
    \centering
    \includegraphics[width=0.8\linewidth]{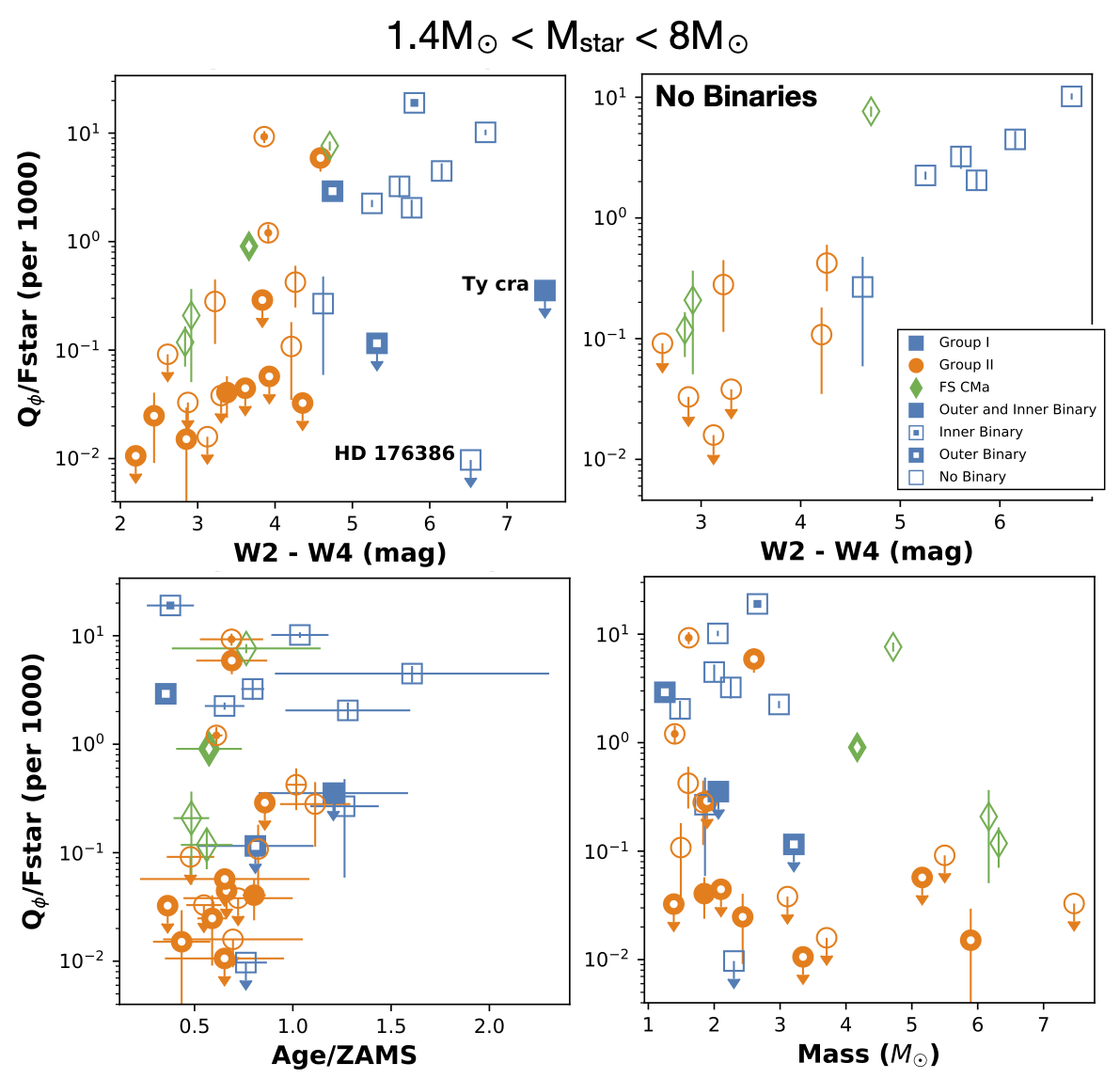}
    \caption{33 targets with 1.4\Msun$<$ Stellar Mass $<$8\Msun comparing the \Qphip/Fstar ratio to system parameters of WISE 2 (W2, 4.6 $\mu m$) - WISE 4 (W4, 22 $\mu m$) color (upper right), W2  - W4 color of only non-binary systems (upper left), 
    system age/zero age main-sequence (ZAMS) age (bottom left), and stellar mass (bottom right). Targets shapes and colors refer to their classification (Group I-blue squares, Group II-orange circles, FS CMa-green diamonds) and the shapes filling refer to its binarity (filled shapes- inner and outer binary,edge filled-outer binary, inner filled-inner binary, no filling-no binary). Outer binaries are exterior to the IWA and in the FOV as discussed in section \ref{sec:binaries}. Inner binaries are spectroscopic binaries from the literature. Error bars are plotted under the symbols with down pointing errors indicating 2$\sigma$ upper limit uncertainty.
    \label{fig:w2w4}}
\end{figure*}

We first compare the \Qphip/Fstar ratio to the W2-W4 color shown in Figure \ref{fig:w2w4}. We find that bluer targets are more likely to have lower \Qphip/Fstar ratio than redder targets. By our Group definition above, Group II targets are more likely to have lower \Qphip/Fstar ratio while Group I targets are more likely to have higher \Qphip/Fstar ratio. While there is a positive trend between \Qphip/Fstar ratio and W2-W4 color, there is a large spread of values in the trend especially noting the logarithmic plots in Figure \ref{fig:w2w4}. We attribute this large spread due to the presence of binaries. When the binaries are removed from the sample as shown in the upper right in Figure \ref{fig:w2w4}, there is not as significant of a spread in values. This suggests that binarity complicates the correlation between polarized flux from the disk vs infrared color. Very bright binaries can make it difficult to remove the SIP as discussed in section \ref{sec:pipeline} possibly resulting in a larger spread in \Qphip/Fstar ratio values. However, this is unlikely to be the main cause of the spread as the majority of binaries in these systems are not bright enough to leave the quadrupole residual in \Qphi and \Uphi.

There are two clear exceptions to this trend, TY CrA and HD 176386. Both of these targets have no detected \Qphip/Fstar ratio yet both are very red (W2-W4 $>$ 6 mag).  TY CrA and HD 176386 are both located in an extended reflection nebulae. Images from unWISE of TY CrA exhibit extended emission offset from TY CrA \citep{Lang2016}. Also, HD 176386 is most likely blended with CrA-54 in WISE bands, which CrA-54 is a K7 star which is known to host a disk \citep{Cazzoletti2019}. Thus we conclude that the W2-W4 color for TY CrA and HD 176386 are likely contaminated from their surrounding environment (reflection nebulae, close-by stars, dust clumps), not originating from the protoplanetary disk dust around these stars. However, we lack observations of adequate spatial resolution to provide accurate WISE-band fluxes thus we will keep these targets in our analysis but label them for suspicious WISE fluxes.

\subsection{Age}

We next compare the summed \Qphip/Fstar flux to ages of the systems. 
Stars will evolve towards the zero age main sequence (ZAMS) at different rates with  8\Msun stars evolving much quicker than 1.4\Msun stars. Thus we plot the ratio of the stars current age divided by the targets ZAMS for a its given estimated stellar mass as shown in the bottom left of Figure \ref{fig:w2w4}. For reference, the ZAMS is also plotted on an HR diagram for the sample in Figure \ref{fig:Entire_sample}. There are no systems with \Qphip/Fstar ratio $<$ 0.2 that are on the main sequence ($>$ 1.0 Age/ZAMS). We do not see a correlation between the system Age/ZAMS and the \Qphip/Fstar ratio.

\subsection{Stellar Mass}

Finally, we compare the stellar mass to the ratio of \Qphip/Fstar shown in the bottom right of Figure \ref{fig:w2w4}. We find that the less massive stars have the largest \Qphip/Fstar. The one key result is looking at the systems with stellar masses between  4$<$ M$_{\textrm{star}}<$8\Msun{} where only the five systems have significantly detected \Qphip/Fstar. 4 of these 5 systems are also classified as FS CMa stars. We will discuss this point further in section \ref{sec:FSCMA_reclassification}.

\section{Discussion}\label{sec:discussion}

\subsection{Ringed Systems and Stellar Mass Dependence}
We found that, in our sample, protoplanetary disks hosting ringed systems only occur in systems with stellar masses $<$ 3 \Msun{} (see Section \ref{sec:results}).
There are several potential causes that could explain this trend. First, we note that dust ringed systems have largely been theorized to exist due to the formation of exoplanets creating gaps in-between these dusty rings. Recent evidence for this theory has been provided in the case of PDS 70b,c \citep{Keppler2018}. One explanation is that the same physical scenarios that make it possible for exoplanets to form visible rings at lower stellar masses($<$ 3 \Msun{}) while the same physical scenarios do not exist for higher stellar mass systems.
One possibility is the increased binarity $>$ 3 \Msun{} stellar masses disrupts the formation of rings and gaps which was similarly observed for millimeter dust grains as part of the DSHARP sample \citep{Kurtovic2018}. However, in our sample we see the same fraction of binaries around more massive stars (42\% for 3\Msun{} $<$ Stellar Mass $<$ 8\Msun{}) than we do lower mass stars (40\% for Stellar Mass  $<$ 3\Msun{}) though we are not complete to inner and outer binaries as noted in section \ref{sec:trends}.
Second, the temperature of the higher mass disks may prevent them from forming rings and gaps. Disks around higher mass stars may be hotter, which can prevent the most volatile ice species, like CO, from freezing out onto dust grains, as found recently in a comparison of carbon depletion in T Tauri versus Herbig AeBe disks (\citealt{Sturm2022}, \citealt{Marel2021}). If the dust rings are created by the accumulation of millimeter sized grains at the edges of gaps, which may in turn be carved by planets, reduced amounts of CO or CO$_2$ ice in the warmer disks could inhibit planetesimal formation or grain growth. This could inhibit outer disk ring formation for the higher mass disks.
\citet{Garufi2018} found polarized light rings only appeared around older stars ($>$ 5 Myr) thus what we maybe observing that stellar systems $>$ 3\Msun{} do not have the time to develop polarized light rings before the dust has been removed from the system. Also, theoretical predictions suggest that the mass of a planet needed to open a gap in a disk is proportional to the mass of the star (see equation 1 in \citealt{Matsumura2003}). Thus the lack of gaps could be representative of more massive planets being needed to open up gaps around more massive stars.

Finally, the lack of detected ringed systems for stellar masses $<$ 3 \Msun{} could be an observational bias. In our sample, more massive targets are more likely to be more distant. Thus these more distant objects may host ringed structures but are located inside the IWA of our observations. Of the 20 systems with stellar masses $>$ 3 \Msun{} (see Figure \ref{fig:Entire_sample}), we find that 2 of these systems could detect all of the rings in systems with stellar masses $<$ 3 \Msun{} and an additional 11 systems are close enough to detect at least one or more of the rings. Our sample has a dust ring occurrence rate of 30\% for systems with $<$ 3 \Msun{} stars, thus we would expect to detect at least $\sim$ 1 system hosting rings with stellar masses $>$ 3 \Msun{}. However, the detectability of rings around more massive systems is complicated because the central stars are also more luminous, increasing the outer disk's brightness. Therefore ringed structures around systems $<$ 3 \Msun{} that are too dim to be detected by our imaging surveys, such as those found around MWC 275 (330 au; \citealt{Rich2019}) and CU Cha (341 au; \citealt{Ginski2016}, would be brighter and detectable around more luminous systems. If we assume the distance of the furthest known ring (341 au), we could detect such a ring around 15 of 20 systems with stellar masses $>$ 3 \Msun{}.
Our survey is the first survey to search for trends in protoplanetary disks with stellar masses $>$ 3 \Msun{} thus future work needs to verify these findings and search for massive systems that host ringed and gaped structures.

\subsection{Polarized Flux and Infrared Color   }

We find a correlation between the polarized flux and the infrared colors as shown for the entire sample in Figure \ref{fig:color_color_qphi}. This trend replicates previous studies by \citet{Garufi2017,Garufi2020}, which found a similar correlation. For the entire sample, we find that objects associated with Full Disks are more likely to have a lower \Qphip/Fstar ratio, as shown in Figure \ref{fig:color_color_qphi}. The lower \Qphip/Fstar ratio is expected as full disks are thought to self-shadow, causing the outer portions of the disk imaged with GPI to be dimmer in polarized light as compared to Transitional disks where shelf-shadowing is not occurring.

A similar trend can be seen for our limited Herbig Ae/Be sample discussed in Section \ref{sec:trends}.
The strong infrared flux of Group I objects has often been interpreted as due to strong disk flaring but could also be due to disk cavity, in either case leading to easily detected scattered light disk flux \citep{Maaskant2013,Garufi2017,Garufi2018}. The bluer Group II objects can be explained either by strong self-shadowing by the inner disk or a flatter geometry due to dust growth/settling \citep[e.g.,][]{Muro-Arena2018}. This would broadly correspond with larger scattered light flux from Group I objects and less from Group II objects. Our observations shown in Figure \ref{fig:w2w4} broadly replicates this trend though there is a large spread in values. This is primarily due to the presence of binaries. Further, we observe very few binaries that are in redder objects (Group I for Herbig's) and most binaries in our sample are associated with bluer objects as shown in Figure \ref{fig:w2w4}. This strongly suggests that a system hosting a binary plays a significant role in the Group I vs Group II classification. One potential cause is that the presence of binaries truncate the outer disks causing systems with binaries to have less polarized flux or not be resolved. This would match with previous studies suggesting that Group I and Group II systems are two distinct evolutionary pathways \citep{Maaskant2013, Garufi2018}.

\subsection{Polarized Flux and Age}
We observe a lack of systems with ages in the main sequence ($>$ 1.0 Age/ZAMS) and low \Qphip/Fstar ratios as discussed in Section \ref{sec:trends}. This could be an indicator that systems with low \Qphip/Fstar ratios and older ages ($>$ 1.0 Age/ZAMS) do not have sufficient infrared excess and are thus not in our sample. However, the brightest systems with large \Qphip/Fstar ratios may have long-lived disks. This is similar to the conclusion found by \citet{Garufi2018}, where the most massive disks are long-lived and easily observed in polarized light to older ages. Next, we do not find the correlation between  \Qphip/Fstar and age which had previously been observed by \citet{Garufi2018}. However, our sample contains a larger range in stellar mass 0.3-20\Msun{} where the \citet{Garufi2018} study only contained targets with stellar masses $<$ 3\Msun{.} Thus, the correlation between polarized light and age may not be present for higher mass stars. Finally, age results are difficult to interpret for more massive stars due to the lack of intermediate mass T-tauri stars in our sample (see Section \ref{sec:sample}). Future work is needed to include intermediate mass T-Tauri stars in polarized light imaging surveys to assess age trends in Herbig Ae/Be systems.

\subsection{Polarized Flux and Binarity}
We find that only 5 of the 12 systems with stellar masses between 4$<$ M$_{\textrm{star}}<$8\Msun{} have significantly detected \Qphip/Fstar ratios. There are four possible explanations for this trend. First, our more massive stars tend to be further away; thus, we may not be resolving as much of the polarization flux compared to closer systems. Second, these more massive stars tend to have more outer binaries (5 of 12) than compared to less massive systems ($<$4 \Msun). These binaries could be truncating or stripping the disks from the central star. The lack of polarized light may be caused by stellar evolution, where the quickly evolving bright star photoevaporates the disks so they are no longer visible. Finally, the systems may host large disks that our imaging would resolve. However, no polarized light is detected because the inner disk shadows the outer disk. This mechanism has previously been invoked to explain non-gaped protoplanetary disks having little polarized light detected \citep{Garufi2017,Muro-Arena2018}.

\subsection{Point Sources in Literature}

We compare our point source findings to those found in the literature. We limit our discussion to the sub-stellar mass companions and notable stellar companions.
Our survey has imaged five sub-stellar point sources (V921 Sco, HD 158643, MWC 297, HD 101412, and V1295 Aql). We found two companions around V921 Sco at 0$\farcs$5 (0.03 \Msun{}) and 1$\farcs$1 (0.15 \Msun{}). The first companion with a mass consistent with a brown dwarf at 0$\farcs$5 was previously discovered by \citet{Gabellini2019} with a similar mass estimate (0.06 \Msun) but has not previously been confirmed until this work. They did not observe the second companion at 1$\farcs$1, however this is expected as the second companion is located outside of the FOV of \citet{Gabellini2019} images. The brown dwarf object (0.07 \Msun{}) HD 158643 has not previously been directly imaged. However, \citet{2019Kervella} measured a variation in the proper motion of the system consistent with an object with a mass of 0.05 \Msun normalized to a separation of 1 au. These values are potentially consistent but further analysis is necessary. \citet{Ubeira2020ApJ} previously confirmed that MWC 297 B was co-moving and measured the companion mass to be 0.25 $^{+0.25}_{-0.15}$\Msun. This is much larger than our estimated mass of 0.03 \Msun. However, they were also able to measure a local high extinction of A$_V$ $\sim$ 11.9 mag from the spectral slope of the planet which is sufficient to match our findings. Finally, there is no known previous discovery of the planetary mass companion around V1295 Aql or a brown dwarf companion around HD 101412. Follow-up observations to confirm that these objects are co-moving is necessary. 

Ty CrA is a known quadrupal system with \citet{Chauvin2003} observing the 4th stellar companion around Ty CrA. The \citet{Chauvin2003} work was unable to verify if the object was co-moving. Given the 16 year difference between our observations and \citet{Chauvin2003} observations and assuming the proper motion of Ty CrA we find that if the 4th component was not co-moving the object would have moved by 0$\farcs$52 while we measure a positional difference of 0$\farcs$21. Given the large timescale and that the 4th component of Ty CrA is at a projected separation of $\sim$ 18 au, it is likely we are observing the orbital motion of the 4th component of Ty CrA. However, confirmation of the object being bound is necessary given the systems complex orbital dynamics.

\subsection{Classification of FS CMa stars}\label{sec:FSCMA_reclassification}

For the four systems that are FS CMa candidates in our sample, HD 45677, HD 50138, HD 85567, and HD 98922, we observe polarized flux at large projected distances ($>$ 200 au) as shown in Figure \ref{fig:FSCMa}. Additionally, these polarized signals are  bright and in the case of HD 45677 and HD 50138 there is significant \Uphi polarized flux suggesting multiple scattering events due to the optically thick nature of the material.

Previous studies have detected extended structures around B[e] stars via H-alpha \citep{Marston2008,Liimets2022}. These structures are typically shell-like or indicative of bi-polar outflows. We do not observe similar structures around our four FS CMa candidates which appear to be closer to typical protoplanetary disk structures. However, previous studies were at a larger spatial scale ($>$ 1') and emission while our observations are at smaller spatial scales and from a polarized source. Finally, our targets have a lower mass than those studied by \citet{Marston2008} and \citet{Liimets2022}.

Stars that exhibit B[e] phenomenon are a heterogeneous group which include pre-main sequence, main sequence, and evolved systems. Additionally, many of these stars exhibit similar features making classification difficult. 
FS CMa as defined by \citet{Miroshnichenko2007} hosts emission-line spectra containing hydrogen lines, large infrared excess that peaks at 10-30$\mu m$, located outside of a star-forming region, and if it has a secondary companion it is either fainter and cooler than the primary or degenerate. However, even given these parameters, distinguishing between B[e] classifications can be difficult. For example, a thorough investigation by \citet{Varga2019} which studied HD 50138 concluded that the evolutionary state of HD 50138 could not be unambiguously determined through mid-IR spectroscopy. We conclude that the structures observed around the FS CMa candidate stars in our \Qphi and \Uphi images are more similar to the expected structures of protoplanetary disks rather than outflow from FS CMa or another type of evolved star. Future investigations into the resolved circumstellar material around stars are necessary including resolved ALMA observations to test the kinematics of the gas in the systems. The classification of these systems to be protoplanetary is extremely important as they represent four of the five systems with 4\Msun{} $<$ M$_{\textrm{star}}$ $<$ 8\Msun{} in our sample that have detected scattered light flux.

\section{Summary}

We have presented the Gemini-LIGHTS survey which observed 44 bright Herbig Ae/Be stars and T-Tauri stars with the GPI instrument at Gemini South.
We constructed our sample based on their near- and mid-infrared colors selecting a number of transitional, pre-transitional, and full disks in order to create an unbaised sample. Importantly, we did not select against unequal mass binaries with moderate separation. 
Observations of these 44 targets were taken in J- and H-band utilizing high-contrast polarized imagery.
Our selection criteria facilitates an unbiased approach to studying bright Herbig Ae/Be and T-Tauri stars that does not favor famous targets with large disks ($>$ 100 au).
We discussed how we uniformly reduced and analyzed the 70 epochs of observations that are part of the sample using the GPI DRP pipeline along with our own python wrapper. We discussed improvements to finding centers of images with bright companions and improvements in removing SIP. 

Within our sample of 44 targets, we found several significant trends:
\begin{enumerate}
    \item  We detect scattered light signatures around 80\% of our 44 targets.
    \item Systems with large separation binaries are more likely to have bluer mid infrared colors (Group II) and these systems are more likely to have no detectable \Qphip/Fstar flux.
    \item We find that all ringed classified systems have stellar masses $<$ 3 \Msun{}, potentially indicating that if the disk rings arise from planet formation the planet formation process or disk evolution is different around more massive stars.
    \item We find a similar infrared (Wise2 - Wise4) correlation with \Qphip/Fstar identified by \citet{Garufi2017}. We find a large spread in values primarily due to binary systems as the trend is much tighter when binaries are removed.
    \item Four of five of our targets with 4\Msun{} $<$ M$_{\textrm{star}}$ $<$ 8\Msun{} have detected scattered light flux are also classified as FS CMa (HD 45677, HD 50138, HD 85567, and HD 98922). Due to the large radial extent of the polarized flux in the images ($>$ 200 au), we conclude that these objects are likely young systems.
    \item We detect 24 point sources consistent with being star companions. We find 1 point source around V1295 Aql which is consistent with a super-Jupiter mass and 3 point sources consistent with brown dwarf masses. We confirm the existence of two brown dwarf candidates (V921 Sco, HD 158643) from previous direct imaging and proper motion discoveries. We find one brown dwarf candidate HD 101412 which had previously not been observed before.
\end{enumerate}

\subsection{Acknowledgements}
Thanks to Christian Ginski for their collaboration on defining the FITS header standard for high contrast scattered light imaging. We would also like to thank Bruce Macintosh, Fredrik Rantakyro, Marshall Perrin, Max Millar-Blanchaer, Tom Esposito, Robert De Rosa, Jeffrey Chilcote, and  Ren\'{e} Rutten for their help on this survey.
E.A.R. and J.D.M. acknowledges support from NSF AST 1830728. A.A. acknowledges support from NSF AST-1311698. 
S.K. acknowledges support from an ERC Consolidator Grant (Grant Agreement ID 101003096). 
L.P. gratefully acknowledges support by the ANID BASAL projects ACE210002 and FB210003, and by ANID, -- Millennium Science Initiative Program -- NCN19\_171.
We thank the anonymous referee for feedback that helped to improve this paper.
This work has made use of data from the European Space Agency (ESA) mission
{\it Gaia} (\url{https://www.cosmos.esa.int/gaia}), processed by the {\it Gaia}
Data Processing and Analysis Consortium (DPAC,
\url{https://www.cosmos.esa.int/web/gaia/dpac/consortium}). Funding for the DPAC
has been provided by national institutions, in particular the institutions
participating in the {\it Gaia} Multilateral Agreement.

Based on observations obtained at the international Gemini Observatory, a program of NSF’s NOIRLab, which is managed by the Association of Universities for Research in Astronomy (AURA) under a cooperative agreement with the National Science Foundation. on behalf of the Gemini Observatory partnership: the National Science Foundation (United States), National Research Council (Canada), Agencia Nacional de Investigaci\'{o}n y Desarrollo (Chile), Ministerio de Ciencia, Tecnolog\'{i}a e Innovaci\'{o}n (Argentina), Minist\'{e}rio da Ci\^{e}ncia, Tecnologia, Inova\c{c}\~{o}es e Comunica\c{c}\~{o}es (Brazil), and Korea Astronomy and Space Science Institute (Republic of Korea). List of program ID's where the data were obtained: GS-2017A-LP-12, GS-2017B-LP-12, GS-2018A-LP-12, GS-2018B-LP-12, GS-2015A-C-1, GS-2016B-DD-5, GS-2015B-Q-501 ,GS-2014B-Q-503 ,GS-2018A-FT-101-17 ,GS-2018A-FT-101-15, GS-2018A-FT-101-5, GS-2014A-SV-414-14, GS-2014A-SV-414-8, GS-2014A-SV-406-5, GS-2017B-Q-500-16, GS-2014A-SV-412, and GS-2015A-Q-49.

\bibliography{bibliography}{}
\bibliographystyle{aasjournal}

\appendix

\section{Target Properties}
In this section, we provide the target parameters, observation log, and calculated values used in this work. Table \ref{tbl:target_coords} lists all of the targets, HD names, their coordinates, and their GAIA distances.
Table \ref{tbl:target_properties} lists the targets of Gemini-LIGHTS sample along with target system properties such as stellar mass, age, effective temperature, and luminosity. The photometry used in this work is listed in Table \ref{tbl:target_SED_table}.

Table \ref{tbl:obs_log} lists each of the 70 observational epochs observed as part of the Gemini-LIGHTs sample. The table lists each target, filter, date of observation, length of image exposure, and the number of frames. Additionally, the table lists the average \% polarization that was removed during the reduction. Note that this value is a combination of SIP and for small values of \% polarization, the polarization will be dominated by the instrumental polarization.  

Table \ref{tbl:epoch_parameters} lists the measured \Qphip/Fstar ratio for every observed epoch. A description of how \Qphip/Fstar is calculated can be found in section \ref{sec:trends}. Additionally, the 5-$\sigma$ upper-limit of potential point sources detected at a separation of 0$\farcs$2 is listed in table \ref{tbl:epoch_parameters}. Details on the ADI reduction and analysis can be found in section \ref{sec:binaries}.

\clearpage
\startlongtable
\begin{deluxetable*}{lccccc}
\tablecaption{Gemini-LIGHTS Target Sample}
\small
\tablehead{
\colhead{Target} & \colhead{HD Name} & \colhead{2MASS Name} & \colhead{RA} & \colhead{Dec} & \colhead{Distance} \\
\colhead{} & \colhead{} & \colhead{} & \colhead{($^{\circ}$)} & \colhead{Dec ($^{\circ}$)} & \colhead{(pc)}}
\label{tbl:target_coords}
\startdata
AK Sco & HD 152404 & 16544485-3653185 & 253.6868708 & -36.8887086 & 139.8$\pm$ 0.6  \\ 
CU Cha & HD 97048 & 11080329-7739174 & 167.013825 & -77.6548583 & 184.4$\pm$ 0.8  \\ 
FU Ori & \nodata & 05452235+0904123 & 86.3431986 & 9.0700691 & 407.5$\pm$ 3.0  \\ 
GW Ori & HD 244138 & 05290838+1152126 & 82.2849625 & 11.8701944 & 408.0$\pm$ 10.4  \\ 
HD 34700 A & HD 34700 & 05194140+0538428 & 79.9225333 & 5.6452111 & 350.5$\pm$ 2.5  \\ 
HD 36917 & HD 36917 & 05344698-0534145 & 83.6957719 & -5.5707229 & 450.4$\pm$ 11.3  \\ 
HD 37806 & HD 37806 & 05410229-0243006 & 85.2595665 & -2.7168744 & 401.6$\pm$ 4.4  \\ 
HD 38087 & HD 38087 & 05430057-0218454 & 85.752413 & -2.3125911 & 377.0$\pm$ 5.4  \\ 
HD 45677 & HD 45677 & 06281742-1303109 & 97.0726019 & -13.0530934 & 572.1$\pm$ 14.6  \\ 
HD 50138 & HD 50138 & 06513340-0657592 & 102.8891461 & -6.9664934 & 351.0$\pm$ 5.7  \\ 
HD 85567 & HD 85567 & 09502853-6058029 & 147.6188579 & -60.9674551 & 1047.4$\pm$ 18.0  \\ 
HD 95881 & HD 95881 & 11015764-7130484 & 165.4900393 & -71.5134176 & 1109.9$\pm$ 24.3  \\ 
HD 98800 & HD 98800 & 11220530-2446393 & 170.5215893 & -24.7778864 & 42.1$\pm$ 1.0  \\ 
HD 98922 & HD 98922 & 11223166-5322114 & 170.631975 & -53.3698472 & 650.9$\pm$ 8.8  \\ 
HD 100453 & HD100453 & 11330559-5419285 & 173.2730745 & -54.3246148 & 103.8$\pm$ 0.2  \\ 
HD 100546 & HD100546 & 11332542-7011412 & 173.3558188 & -70.1947875 & 108.1$\pm$ 0.4  \\ 
HD 101412 & HD 101412 & 11394445-6010278 & 174.9352141 & -60.1743876 & 412.2$\pm$ 2.5  \\ 
HD 104237 & HD 104237 & 12000511-7811346 & 180.0211875 & -78.1929333 & 106.6$\pm$ 0.5  \\ 
HD 139614 & HD 139614 & 15404638-4229536 & 235.1931708 & -42.4983306 & 133.6$\pm$ 0.5  \\ 
HD 141569 & HD 141569 & 15495775-0355162 & 237.4905224 & -3.9213039 & 111.6$\pm$ 0.4  \\ 
HD 142527 & HD 142527 & 15564188-4219232 & 239.1744971 & -42.323228 & 159.3$\pm$ 0.7  \\ 
HD 142666 & HD 142666 & 15564002-2201400 & 239.1667625 & -22.0277806 & 146.3$\pm$ 0.5  \\ 
HD 144432 & HD 144432 & 16065795-2743094 & 241.7414345 & -27.7195047 & 154.8$\pm$ 0.6  \\ 
HD 145718 & HD 145718 & 16131158-2229066 & 243.2982588 & -22.4853188 & 154.7$\pm$ 0.5  \\ 
HD 158643 & HD 158643 & 17312497-2357453 & 262.8540017 & -23.9627733 & 125.7$\pm$ 1.7  \\ 
HD 169142 & HD 169142 & 18242978-2946492 & 276.1240709 & -29.7805404 & 114.9$\pm$ 0.4  \\ 
HD 176386 & HD 176386 & 19013892-3653264 & 285.4122394 & -36.890856 & 155.1$\pm$ 0.7  \\ 
Hen 3-1330 & HD 326823 & 17065390-4236397 & 256.7246196 & -42.6110631 & 1445.2$\pm$ 40.0  \\ 
Hen 3-225 & HD 76534 & 08550867-4327596 & 133.7862435 & -43.4666156 & 885.0$\pm$ 23.1  \\ 
Hen 3-365 & HD 87643 & 10043028-5839521 & 151.1265 & -58.6645278 & 1589.1$\pm$ 309.7  \\ 
HR 5999 & HD 144668 & 16083427-3906181 & 242.142875 & -39.1050278 & 158.6$\pm$ 0.9  \\ 
HT Lup & \nodata & 15451286-3417305 & 236.3036208 & -34.2918389 & 153.5$\pm$ 1.3  \\ 
MWC 147 & HD 259431 & 06330519+1019199 & 98.2716174 & 10.3222059 & 653.4$\pm$ 11.6  \\ 
MWC 166 & HD 53367 & 07042551-1027156 & 106.1063792 & -10.4543722 & 1219.6$\pm$ 314.4  \\ 
MWC 275 & HD163296 & 17562128-2157218 & 269.0886685 & -21.9562308 & 101.0$\pm$ 0.4  \\ 
MWC 297 & \nodata & 18273952-0349520 & 276.9146958 & -3.831125 & 417.8$\pm$ 5.3  \\ 
MWC 614 & HD 179218 & 19111124+1547155 & 287.7969193 & 15.7875669 & 260.1$\pm$ 2.2  \\ 
MWC 789 & HD 250550 & 06015998+1630567 & 90.4999527 & 16.5157295 & 760.3$\pm$ 28.5  \\ 
MWC 863 & HD150193 & 16401792-2353452 & 250.0746503 & -23.8959517 & 150.8$\pm$ 0.5  \\ 
PDS 66 & \nodata & 13220753-6938121 & 200.5314458 & -69.6367194 & 97.9$\pm$ 0.1  \\ 
Ty CrA & \nodata & 19014081-3652337 & 285.4201238 & -36.8762242 & 159.1$\pm$ 4.4  \\ 
V1295 Aql & HD 190073 & 20030250+0544166 & 300.7604637 & 5.7379176 & 847.9$\pm$ 22.5  \\ 
V921 Sco & \nodata & 16590677-4242083 & 254.7782539 & -42.7023413 & 1482.4$\pm$ 77.4  \\ 
WRAY 15-535 & \nodata & 10152198-5751427 & 153.841516 & -57.8618381 & 4989.9$\pm$ 478.1  \\ 
\enddata
\tablecomments{ Distances from Gaia DR3 early release  \citet{Gaia2021}.  }
\end{deluxetable*}
\clearpage
\startlongtable
\begin{deluxetable*}{lccccc}
\tablecaption{Gemini-LIGHTS Target Characteristics}
\tablehead{
\colhead{Target} & \colhead{Mass} & \colhead{Age} & \colhead{Teff} & \colhead{log$_{10}$(L)} & \colhead{Disk Classification} \\
\colhead{} & \colhead{(M\textsubscript{\(\odot\)})} & \colhead{(Myr)} & \colhead{(K)} & \colhead{log$_{10}$((L\textsubscript{\(\odot\)}))} & \colhead{} }
\label{tbl:target_properties}
\startdata
AK Sco & 1.4 $^{+0.07}_{-0.07}$ (b) & 8.38 $^{+1.72}_{-0.42}$ (b) & 6250 $^{+250}_{-250}$ (b) & 0.62 $^{+0.03}_{-0.01}$ (b) & Cont  \\ 
CU Cha & 2.25 $^{+0.11}_{-0.14}$ (b) & 4.37 $^{+1.11}_{-0.32}$ (b) & 10500 $^{+500}_{-500}$ (b) & 1.54 $^{+0.07}_{-0.06}$ (b) & Ring  \\ 
FU Ori & 0.3  (f) & 2.0   (i) & \nodata & \nodata & Irr  \\ 
GW Ori & 2.8   (ab) & 5.0   (j) & 5500.0   (ag) & 1.68 $^{+0.1}_{-0.08}$ (ah) & Irr  \\ 
HD 100453 & 1.25 $^{+0.06}_{-0.06}$ (b) & 6.53 $^{+0.45}_{-0.49}$ (b) & 7250 $^{+250}_{-250}$ (b) & 0.79 $^{+0.02}_{-0.0}$ (b) & Spiral, Cont  \\ 
HD 100546 & 2.06 $^{+0.1}_{-0.12}$ (b) & 5.48 $^{+1.41}_{-0.77}$ (b) & 9750 $^{+500}_{-500}$ (b) & 1.37 $^{+0.07}_{-0.05}$ (b) & Cont  \\ 
HD 101412 & 2.1 $^{+0.11}_{-0.11}$ (b) & 4.37 $^{+0.22}_{-0.32}$ (b) & 9750 $^{+250}_{-250}$ (b) & 1.58 $^{+0.05}_{-0.04}$ (b) & Nondet  \\ 
HD 104237 & 1.85 $^{+0.09}_{-0.09}$ (b) & 5.48 $^{+0.27}_{-0.4}$ (b) & 8000 $^{+250}_{-250}$ (b) & 1.33 $^{+0.04}_{-0.01}$ (b) & Undet  \\ 
HD 139614 & 1.48 $^{+0.07}_{-0.07}$ (b) & 14.49 $^{+1.41}_{-3.6}$ (b) & 7750 $^{+250}_{-250}$ (b) & 0.77 $^{+0.03}_{-0.01}$ (b) & Cont  \\ 
HD 141569 & 1.86 $^{+0.09}_{-0.09}$ (b) & 8.62 $^{+11.38}_{-1.19}$ (b) & 9750 $^{+250}_{-250}$ (b) & 1.22 $^{+0.03}_{-0.03}$ (b) & Ring  \\ 
HD 142527 & 1.61 $^{+0.12}_{-0.08}$ (b) & 6.63 $^{+0.33}_{-1.55}$ (b) & 6500 $^{+250}_{-250}$ (b) & 0.96 $^{+0.03}_{-0.0}$ (b) & Spiral, Ring  \\ 
HD 142666 & 1.49 $^{+0.08}_{-0.08}$ (b) & 9.33 $^{+0.77}_{-0.47}$ (b) & 7500 $^{+250}_{-250}$ (b) & 0.94 $^{+0.04}_{-0.05}$ (b) & Cont  \\ 
HD 144432 & 1.39 $^{+0.07}_{-0.07}$ (b) & 4.98 $^{+0.25}_{-0.55}$ (b) & 7500 $^{+250}_{-250}$ (b) & 0.97 $^{+0.04}_{-0.01}$ (b) & Nondet  \\ 
HD 145718 & 1.6 $^{+0.08}_{-0.08}$ (b) & 9.8 $^{+2.8}_{-0.49}$ (b) & 8000 $^{+250}_{-250}$ (b) & 0.9 $^{+0.05}_{-0.04}$ (b) & Cont  \\ 
HD 158643 & 3.35 $^{+0.79}_{-0.22}$ (b) & 1.22 $^{+0.29}_{-0.57}$ (b) & 9800 $^{+900}_{-300}$ (b) & 2.22 $^{+0.26}_{-0.07}$ (b) & Nondet  \\ 
HD 169142 & 2.0 $^{+0.13}_{-0.13}$ (b) & 8.98 $^{+11.02}_{-3.9}$ (b) & 10700 $^{+800}_{-900}$ (b) & 1.31 $^{+0.12}_{-0.22}$ (b) & Ring  \\ 
HD 176386 & 2.3 $^{+0.14}_{-0.3}$ (b) & 4.05 $^{+15.95}_{-0.57}$ (b) & 10700 $^{+800}_{-900}$ (b) & 1.58 $^{+0.12}_{-0.22}$ (b) & Nondet  \\ 
HD 34700 A & 2.66 $^{+0.32}_{-0.13}$ (b) & 1.4 $^{+0.23}_{-0.44}$ (b) & 5900 $^{+110}_{-100}$ (b) & 1.36 $^{+0.1}_{-0.02}$ (b) & Spiral, Ring  \\ 
HD 36917 & 3.71 $^{+0.94}_{-0.75}$ (b) & 0.99 $^{+0.9}_{-0.5}$ (b) & 11215 $^{+1109}_{-1316}$ (b) & 2.43 $^{+0.24}_{-0.29}$ (b) & Nondet  \\ 
HD 37806 & 3.11 $^{+0.55}_{-0.33}$ (b) & 1.56 $^{+0.64}_{-0.6}$ (b) & 10475 $^{+1025}_{-675}$ (b) & 2.17 $^{+0.19}_{-0.14}$ (b) & Undet  \\ 
HD 38087 & 3.21 $^{+0.79}_{-0.38}$ (b) & 1.75 $^{+9.15}_{-0.64}$ (b) & 13600 $^{+2900}_{-830}$ (b) & 2.19 $^{+0.3}_{-0.22}$ (b) & Nondet  \\ 
HD 45677 & 4.72 $^{+1.19}_{-0.39}$ (b) & 0.61 $^{+3.77}_{-0.3}$ (b) & 16500 $^{+3000}_{-750}$ (b) & 2.88 $^{+0.32}_{-0.17}$ (b) & Cont  \\ 
HD 50138 & 4.17 $^{+0.46}_{-0.33}$ (b) & 0.63 $^{+0.19}_{-0.18}$ (b) & 9450 $^{+450}_{-450}$ (b) & 2.46 $^{+0.13}_{-0.09}$ (b) & Cont  \\ 
HD 85567 & 6.32 $^{+0.53}_{-0.39}$ (b) & 0.22 $^{+0.05}_{-0.05}$ (b) & 13000 $^{+500}_{-500}$ (b) & 3.19 $^{+0.1}_{-0.08}$ (b) & Undet  \\ 
HD 95881 & 5.5 $^{+0.5}_{-0.28}$ (b) & 0.28 $^{+0.05}_{-0.07}$ (b) & 10000 $^{+250}_{-250}$ (b) & 2.85 $^{+0.1}_{-0.07}$ (b) & Undet  \\ 
HD 98800 & 0.7   (ab) & 8.5 $^{+1.5}_{-1.5}$ (h) & 4200   (al) & 0.33   (al) & Undet  \\ 
HD 98922 & 6.17 $^{+0.37}_{-0.31}$ (b) & 0.2 $^{+0.01}_{-0.04}$ (b) & 10500 $^{+250}_{-250}$ (b) & 3.03 $^{+0.06}_{-0.05}$ (b) & Undet  \\ 
Hen 3-1330 & 9.0  &  \nodata &    22280   (ai) & 5.3   (ai) & Nondet  \\ 
Hen 3-225 & 7.46 $^{+0.51}_{-0.37}$ (b) & 0.17 $^{+0.02}_{-0.03}$ (b) & 19000 $^{+500}_{-500}$ (b) & 3.55 $^{+0.09}_{-0.07}$ (b) & Nondet  \\ 
Hen 3-365 & 17.72 $^{+10.87}_{-6.72}$ (b) & 0.02 $^{+0.05}_{-0.01}$ (b) & 19500 $^{+5000}_{-3000}$ (b) & 4.6 $^{+0.64}_{-0.53}$ (b) & Irr  \\ 
HR 5999 & 2.43 $^{+0.12}_{-0.12}$ (b) & 2.73 $^{+0.26}_{-0.35}$ (b) & 8500 $^{+250}_{-250}$ (b) & 1.72 $^{+0.05}_{-0.04}$ (b) & Nondet  \\ 
HT Lup & 1.3 $^{+0.2}_{-0.2}$ (d) & 0.5 $^{+0.02}_{-0.4}$ (d) & 4247 $^{+161}_{-237}$ (k) & 0.51 $^{+0.01}_{-0.01}$ (k) & Ring  \\ 
MWC 147 & 5.16 $^{+1.84}_{-1.29}$ (b) & 0.42 $^{+0.53}_{-0.28}$ (b) & 14000 $^{+2125}_{-2900}$ (b) & 2.97 $^{+0.27}_{-0.4}$ (b) & Nondet  \\ 
MWC 166 & 12.3 $^{+4.2}_{-4.2}$ (e) & 0.08 $^{+0.08}_{-0.08}$ (e) & 29500 $^{+1000}_{-1000}$ (e) & 4.11 $^{+0.37}_{-0.37}$ (e) & Nondet  \\ 
MWC 275 & 1.83 $^{+0.09}_{-0.09}$ (b) & 7.6 $^{+1.05}_{-1.22}$ (b) & 9250 $^{+250}_{-250}$ (b) & 1.2 $^{+0.06}_{-0.03}$ (b) & Ring  \\ 
MWC 297 & 16.9 $^{+1.87}_{-1.22}$ (b) & 0.03 $^{+0.01}_{-0.01}$ (b) & 24500 $^{+1500}_{-1500}$ (b) & 4.59 $^{+0.12}_{-0.12}$ (b) & Cont  \\ 
MWC 614 & 2.98 $^{+0.18}_{-0.3}$ (b) & 1.66 $^{+0.54}_{-0.26}$ (b) & 9500 $^{+200}_{-200}$ (b) & 2.05 $^{+0.09}_{-0.14}$ (b) & Cont  \\ 
MWC 789 & 2.6 $^{+0.3}_{-0.14}$ (b) & 2.56 $^{+0.43}_{-0.67}$ (b) & 11000 $^{+500}_{-500}$ (b) & 1.94 $^{+0.17}_{-0.12}$ (b) & Irr  \\ 
MWC 863 & 1.89 $^{+0.1}_{-0.1}$ (b) & 5.48 $^{+0.44}_{-0.27}$ (b) & 9000 $^{+250}_{-250}$ (b) & 1.37 $^{+0.04}_{-0.04}$ (b) & Undet  \\ 
PDS 66 & 1.2   (ak) & 6.0 $^{+1.0}_{-1.0}$ (g) & 5035 $^{+19}_{-19}$ (ak) & 0.0 $^{+0.01}_{-0.01}$ (ak) & Ring  \\ 
Ty CrA & 2.06 $^{+0.22}_{-0.19}$ (b) & 6.38 $^{+13.62}_{-2.01}$ (b) & 10700 $^{+800}_{-900}$ (b) & 1.41 $^{+0.14}_{-0.23}$ (b) & Nondet  \\ 
V1295 Aql & 5.89 $^{+0.8}_{-0.76}$ (b) & 0.22 $^{+0.11}_{-0.07}$ (b) & 9500 $^{+200}_{-200}$ (b) & 2.9 $^{+0.16}_{-0.2}$ (b) & Nondet  \\ 
V921 Sco & 19.96 $^{+6.98}_{-5.0}$ (b) & 0.02 $^{+0.03}_{-0.01}$ (b) & 29000 $^{+3882}_{-4500}$ (b) & 4.76 $^{+0.33}_{-0.34}$ (b) & Undet  \\ 
WRAY 15-535 & 17.5 $^{+2.5}_{-2.5}$ (c) & \nodata &    20000 $^{+3000}_{-3000}$ (aj) & 6.0   (aj) & Undet  \\ 
\enddata
\tablecomments{Stellar and disk characteristics of stellar mass, Age, Teff, Luminosity, and our disk classification. Disk classification described in section \ref{sec:catagories}. In the table Undet is short for Undetermined and Nondet is short for Non-detections. Citations for specific values are from the following: (b) \citet{Vioque2018}, (c) \citet{Maravelias2018}, (d) \citet{Garufi2020}, (e) \citet{Fairlamb2015}, (f) \citet{Zhu2007}, (g) \citet{Murphy2013}, (h) \citet{2018Ribas}, (i)\citet{Beck2012}, (j) \citet{Monnier2019}, (k) \citet{Gaia2018}, (l) \citet{Takami2018}}
\end{deluxetable*}
\startlongtable
\begin{deluxetable*}{lccccc}
\tablecaption{Gemini-LIGHTS Photometry and Meeus Group}
\tablehead{
\colhead{Target} & \colhead{WISE 2} & \colhead{WISE 4} & \colhead{J-band} & \colhead{H-band} & \colhead{Meeus Group} \\
\colhead{} & \colhead{(mag)} & \colhead{(mag)} & \colhead{(mag)} & \colhead{(mag)} & \colhead{}}
\label{tbl:target_SED_table}
\startdata
AK Sco & 4.8$\pm$ 0.036 & 0.888$\pm$ 0.023 & 7.676$\pm$ 0.026 & 7.059$\pm$ 0.033 & II  \\ 
CU Cha & 4.3$\pm$ 0.043 & -1.309$\pm$ 0.013 & 7.267$\pm$ 0.023 & 6.665$\pm$ 0.049 & I  \\ 
FU Ori & 3.509$\pm$ 0.065 & 0.175$\pm$ 0.021 & 6.519$\pm$ 0.023 & 5.699$\pm$ 0.033 & \nodata   \\ 
GW Ori & 4.208$\pm$ 0.045 & -0.74$\pm$ 0.011 & 7.698$\pm$ 0.03 & 7.103$\pm$ 0.029 & I  \\ 
HD 100453 & 3.388$\pm$ 0.065 & -1.355$\pm$ 0.007 & 6.945$\pm$ 0.026 & 6.39$\pm$ 0.038 & I  \\ 
HD 100546 & 3.156$\pm$ 0.049 & -3.565$\pm$ 0.001 & 6.425$\pm$ 0.02 & 5.962$\pm$ 0.031 & I  \\ 
HD 101412 & 4.94$\pm$ 0.031 & 1.325$\pm$ 0.013 & 8.635$\pm$ 0.023 & 8.217$\pm$ 0.047 & II  \\ 
HD 104237 & 2.469$\pm$ 0.071 & -0.909$\pm$ 0.009 & 5.813$\pm$ 0.023 & 5.246$\pm$ 0.059 & II  \\ 
HD 139614 & 5.099$\pm$ 0.03 & -0.667$\pm$ 0.008 & 7.669$\pm$ 0.026 & 7.333$\pm$ 0.04 & I  \\ 
HD 141569 & 6.469$\pm$ 0.02 & 1.847$\pm$ 0.012 & 6.872$\pm$ 0.027 & 6.861$\pm$ 0.04 & I  \\ 
HD 142527 & 3.066$\pm$ 0.076 & -0.794$\pm$ 0.009 & 6.503$\pm$ 0.029 & 5.715$\pm$ 0.031 & II  \\ 
HD 142666 & 4.145$\pm$ 0.049 & -0.065$\pm$ 0.023 & 7.351$\pm$ 0.026 & 6.739$\pm$ 0.027 & II  \\ 
HD 144432 & 4.398$\pm$ 0.044 & 0.042$\pm$ 0.016 & 7.095$\pm$ 0.032 & 6.538$\pm$ 0.067 & II  \\ 
HD 145718 & 4.848$\pm$ 0.04 & 0.585$\pm$ 0.015 & 7.69$\pm$ 0.024 & 7.263$\pm$ 0.029 & II  \\ 
HD 158643 & 2.171$\pm$ 0.091 & -0.028$\pm$ 0.014 & 4.9$\pm$ 0.186 & 4.712$\pm$ 0.206 & II  \\ 
HD 169142 & 5.593$\pm$ 0.023 & -0.561$\pm$ 0.008 & 7.31$\pm$ 0.021 & 6.911$\pm$ 0.038 & I  \\ 
HD 176386 & 6.074$\pm$ 0.015 & -0.455$\pm$ 0.012 & 6.847$\pm$ 0.02 & 6.809$\pm$ 0.031 & I  \\ 
HD 34700 A & 6.794$\pm$ 0.016 & 0.993$\pm$ 0.008 & 8.041$\pm$ 0.023 & 7.706$\pm$ 0.023 & I  \\ 
HD 36917 & 4.278$\pm$ 0.044 & 1.153$\pm$ 0.025 & 7.221$\pm$ 0.019 & 6.964$\pm$ 0.034 & II  \\ 
HD 37806 & 3.419$\pm$ 0.115 & 0.116$\pm$ 0.051 & 7.115$\pm$ 0.02 & 6.252$\pm$ 0.033 & II  \\ 
HD 38087 & 7.207$\pm$ 0.018 & 1.889$\pm$ 0.019 & 7.588$\pm$ 0.024 & 7.386$\pm$ 0.042 & I  \\ 
HD 45677 & 1.812$\pm$ 0.03 & -2.895$\pm$ 0.001 & 7.242$\pm$ 0.026 & 6.347$\pm$ 0.023 & I  \\ 
HD 50138 & 1.585$\pm$ 0.08 & -2.078$\pm$ 0.001 & 5.856$\pm$ 0.027 & 5.093$\pm$ 0.029 & II  \\ 
HD 85567 & 3.291$\pm$ 0.062 & 0.454$\pm$ 0.016 & 7.472$\pm$ 0.024 & 6.68$\pm$ 0.031 & II  \\ 
HD 95881 & 3.125$\pm$ 0.069 & 0.513$\pm$ 0.008 & 7.384$\pm$ 0.026 & 6.662$\pm$ 0.044 & II  \\ 
HD 98800 & 5.325$\pm$ 0.032 & 0.194$\pm$ 0.01 & 6.397$\pm$ 0.02 & 5.759$\pm$ 0.027 & \nodata   \\ 
HD 98922 & 1.863$\pm$ 0.014 & -1.054$\pm$ 0.006 & 6.004$\pm$ 0.02 & 5.226$\pm$ 0.029 & II  \\ 
Hen 3-1330 & 3.344$\pm$ 0.063 & 1.95$\pm$ 0.02 & 6.713$\pm$ 0.024 & 6.103$\pm$ 0.042 & \nodata   \\ 
Hen 3-225 & 7.066$\pm$ 0.02 & 4.193$\pm$ 0.027 & 7.818$\pm$ 0.024 & 7.858$\pm$ 0.04 & II  \\ 
Hen 3-365 & 2.248$\pm$ 0.018 & -3.988$\pm$ 0.001 & 6.217$\pm$ 0.037 & 4.756$\pm$ 0.268 & \nodata   \\ 
HR 5999 & 1.895$\pm$ 0.012 & -0.541$\pm$ 0.01 & 5.907$\pm$ 0.018 & 5.22$\pm$ 0.027 & II  \\ 
HT Lup & 4.973$\pm$ 0.032 & 0.856$\pm$ 0.016 & 7.573$\pm$ 0.021 & 6.866$\pm$ 0.029 & \nodata   \\ 
MWC 147 & 3.169$\pm$ 0.074 & -0.757$\pm$ 0.013 & 7.454$\pm$ 0.026 & 6.666$\pm$ 0.034 & II  \\ 
MWC 166 & 4.538$\pm$ 0.044 & 1.051$\pm$ 0.059 & 6.332$\pm$ 0.02 & 6.22$\pm$ 0.033 & \nodata   \\ 
MWC 275 & 2.466$\pm$ 0.061 & -0.758$\pm$ 0.007 & 6.195$\pm$ 0.021 & 5.531$\pm$ 0.036 & II  \\ 
MWC 297 & 1.872$\pm$ 0.018 & -4.718$\pm$ 0.001 & 6.127$\pm$ 0.019 & 4.387$\pm$ 0.214 & \nodata   \\ 
MWC 614 & 3.647$\pm$ 0.051 & -1.606$\pm$ 0.006 & 6.994$\pm$ 0.02 & 6.645$\pm$ 0.026 & I  \\ 
MWC 789 & 4.633$\pm$ 0.039 & 0.046$\pm$ 0.015 & 8.475$\pm$ 0.02 & 7.528$\pm$ 0.026 & II  \\ 
MWC 863 & 3.244$\pm$ 0.065 & -0.592$\pm$ 0.012 & 6.947$\pm$ 0.02 & 6.214$\pm$ 0.02 & II  \\ 
PDS 66 & 6.183$\pm$ 0.022 & 1.587$\pm$ 0.016 & 8.277$\pm$ 0.032 & 7.641$\pm$ 0.023 & \nodata   \\ 
Ty CrA & 5.139$\pm$ 0.027 & -2.35$\pm$ 0.011 & 7.486$\pm$ 0.024 & 6.97$\pm$ 0.026 & I  \\ 
V1295 Aql & 3.443$\pm$ 0.064 & 0.589$\pm$ 0.016 & 7.194$\pm$ 0.019 & 6.647$\pm$ 0.017 & II  \\ 
V921 Sco & 2.045$\pm$ 0.02 & -4.143$\pm$ 0.001 & 7.235$\pm$ 0.027 & 5.918$\pm$ 0.045 & \nodata   \\ 
WRAY 15-535 & 1.556$\pm$ 0.011 & -0.231$\pm$ 0.011 & 5.762$\pm$ 0.018 & 4.959$\pm$ 0.059 & \nodata   \\ 
\enddata
\tablecomments{WISE photometry from WISE ALL-SKY survey \citep{Cutri2012} and J- and H-band photometry from 2MASS survey \citep{Cutri2003}. Meeus Group categorized in this work as described in section \ref{sec:trends}. Notes: (a) \citep{Lieman2016}, (b)	\citep{Henning1993}, (c) \citep{Liu2018}, (d) \citep{Fang2017}, (e)	This Work, (f) \citep{Sylvester1996}, (g) \citep{Plas2019}, (h) \citep{Miley2019}, (i) \citep{Nilsson2010}, (j) \citep{Kataoka2016}, (k) \citep{Ansdell2018}, (l) \citep{Andrews2018}, (m) \citep{Cazzoletti2019}, (n) \citep{Benac2020}}
\end{deluxetable*} 
\startlongtable
\begin{deluxetable*}{lcccccccc}
\tablecaption{Observational Log and Stellar/Instrumental Polarization}
\tablehead{
\colhead{Target} & \colhead{Filter} & \colhead{Epoch} & \colhead{Exposure Time} & \colhead{\# of Frames} & \multicolumn{2}{c}{Stellar/Instrumental Polarization} \\
\colhead{} & \colhead{} & \colhead{(YYYYMMDD)} & \colhead{(s)} & \colhead{} & \colhead{\% Pol} & \colhead{PA ($^\circ$)}}
\label{tbl:obs_log}
\startdata
AK Sco & H & 20180811 & 59.6 & 40 & 1.69$\pm$ 0.06 & 138$\pm$ 1  \\ 
CU Cha & J & 20170406 & 58.2 & 32 & 1.42$\pm$ 0.12 & 117$\pm$ 3  \\ 
CU Cha & H & 20180413 & 58.2 & 36 & 1.5$\pm$ 0.12 & 112$\pm$ 2  \\ 
FU Ori & J & 20180103 & 58.2 & 24 & 0.16$\pm$ 0.03 & 87$\pm$ 6  \\ 
GW Ori & J & 20180104 & 58.2 & 32 & 0.48$\pm$ 0.1 & 96$\pm$ 6  \\ 
GW Ori & H & 20180104 & 58.2 & 36 & 0.34$\pm$ 0.11 & 110$\pm$ 10  \\ 
HD 100453 & J & 20150410 & 14.5 & 140 & 0.07$\pm$ 0.22 & 41$\pm$ 93  \\ 
HD 100546 & J & 20170220 & 43.6 & 40 & 1.16$\pm$ 0.07 & 60$\pm$ 2  \\ 
HD 101412 & H & 20180319 & 59.6 & 40 & 0.68$\pm$ 0.05 & 92$\pm$ 2  \\ 
HD 104237 & J & 20170407 & 58.2 & 32 & 0.44$\pm$ 0.05 & 98$\pm$ 3  \\ 
HD 104237 & J & 20180317 & 58.2 & 32 & 0.31$\pm$ 0.01 & 94$\pm$ 1  \\ 
HD 104237 & J & 20180520 & 58.2 & 32 & 0.38$\pm$ 0.04 & 119$\pm$ 3  \\ 
HD 104237 & J & 20190217 & 58.2 & 32 & 0.51$\pm$ 0.04 & 121$\pm$ 2  \\ 
HD 139614 & J & 20170406 & 58.2 & 64 & 0.46$\pm$ 0.09 & 179$\pm$ 5  \\ 
HD 139614 & H & 20180608 & 58.2 & 28 & 0.56$\pm$ 0.03 & 11$\pm$ 1  \\ 
HD 139614 & J & 20190513 & 58.2 & 36 & 0.17$\pm$ 0.08 & 130$\pm$ 14  \\ 
HD 141569 & J & 20180609 & 58.2 & 32 & 0.65$\pm$ 0.05 & 105$\pm$ 2  \\ 
HD 141569 & J & 20190223 & 58.2 & 32 & 0.52$\pm$ 0.06 & 114$\pm$ 3  \\ 
HD 142527 & H & 20140425 & 4.4 & 92 & 0.81$\pm$ 0.15 & 40$\pm$ 5  \\ 
HD 142666 & J & 20170703 & 58.2 & 20 & 0.49$\pm$ 0.06 & 84$\pm$ 4  \\ 
HD 142666 & J & 20180609 & 58.2 & 32 & 0.46$\pm$ 0.02 & 79$\pm$ 1  \\ 
HD 144432 & H & 20150709 & 58.2 & 32 & 1.26$\pm$ 0.09 & 175$\pm$ 2  \\ 
HD 145718 & J & 20180607 & 58.2 & 32 & 1.08$\pm$ 0.16 & 102$\pm$ 4  \\ 
HD 145718 & H & 20180608 & 58.2 & 32 & 1.51$\pm$ 0.18 & 82$\pm$ 3  \\ 
HD 158643 & J & 20180609 & 58.2 & 32 & 0.5$\pm$ 0.06 & 1$\pm$ 3  \\ 
HD 158643 & J & 20190223 & 58.2 & 32 & 0.55$\pm$ 0.02 & 3$\pm$ 1  \\ 
HD 169142 & J & 20140425 & 58.2 & 56 & 0.45$\pm$ 0.13 & 175$\pm$ 8  \\ 
HD 176386 & H & 20180607 & 58.2 & 32 & 0.28$\pm$ 0.02 & 2$\pm$ 2  \\ 
HD 176386 & H & 20180816 & 58.2 & 32 & 0.27$\pm$ 0.05 & 2$\pm$ 5  \\ 
HD 176386 & J & 20180817 & 58.2 & 64 & 0.15$\pm$ 0.14 & 40$\pm$ 27  \\ 
HD 34700 A & J & 20180103 & 58.2 & 32 & 0.65$\pm$ 0.13 & 102$\pm$ 6  \\ 
HD 34700 A & H & 20180103 & 58.2 & 32 & 0.36$\pm$ 0.05 & 104$\pm$ 4  \\ 
HD 36917 & J & 20171231 & 58.2 & 32 & 0.21$\pm$ 0.03 & 31$\pm$ 4  \\ 
HD 36917 & H & 20190126 & 58.2 & 36 & 0.79$\pm$ 0.03 & 44$\pm$ 1  \\ 
HD 37806 & J & 20171231 & 58.2 & 32 & 0.89$\pm$ 0.05 & 129$\pm$ 2  \\ 
HD 38087 & J & 20180103 & 58.2 & 32 & 1.3$\pm$ 0.06 & 109$\pm$ 1  \\ 
HD 45677 & J & 20171231 & 58.2 & 32 & 0.42$\pm$ 0.04 & 46$\pm$ 3  \\ 
HD 45677 & H & 20180101 & 58.2 & 32 & 0.46$\pm$ 0.04 & 11$\pm$ 3  \\ 
HD 50138 & J & 20180104 & 58.2 & 32 & 0.29$\pm$ 0.06 & 143$\pm$ 6  \\ 
HD 85567 & J & 20180103 & 58.2 & 32 & 0.45$\pm$ 0.05 & 71$\pm$ 3  \\ 
HD 85567 & J & 20180520 & 58.2 & 32 & 0.46$\pm$ 0.02 & 128$\pm$ 1  \\ 
HD 85567 & H & 20190127 & 58.2 & 28 & 0.48$\pm$ 0.03 & 84$\pm$ 2  \\ 
HD 95881 & J & 20180104 & 58.2 & 40 & 0.91$\pm$ 0.09 & 86$\pm$ 3  \\ 
HD 98800 & H & 20190127 & 34.9 & 28 & 0.74$\pm$ 0.1 & 154$\pm$ 4  \\ 
HD 98922 & J & 20180321 & 58.2 & 32 & 0.41$\pm$ 0.07 & 64$\pm$ 5  \\ 
Hen 3-1330 & J & 20190512 & 58.2 & 32 & 2.02$\pm$ 0.06 & 27$\pm$ 1  \\ 
Hen 3-225 & J & 20190127 & 58.2 & 32 & 0.47$\pm$ 0.05 & 147$\pm$ 3  \\ 
Hen 3-225 & H & 20190127 & 58.2 & 32 & 0.21$\pm$ 0.08 & 39$\pm$ 12  \\ 
Hen 3-365 & J & 20170406 & 58.2 & 32 & 0.79$\pm$ 0.09 & 166$\pm$ 3  \\ 
HR 5999 & J & 20170406 & 58.2 & 32 & 1.13$\pm$ 0.11 & 28$\pm$ 3  \\ 
HR 5999 & J & 20170702 & 58.2 & 36 & 0.32$\pm$ 0.19 & 19$\pm$ 17  \\ 
HT Lup & H & 20190514 & 52.4 & 64 & 0.77$\pm$ 0.12 & 36$\pm$ 4  \\ 
MWC 147 & J & 20171231 & 58.2 & 32 & 0.79$\pm$ 0.02 & 88$\pm$ 1  \\ 
MWC 147 & H & 20190127 & 58.2 & 32 & 1.0$\pm$ 0.07 & 97$\pm$ 2  \\ 
MWC 166 & J & 20170406 & 58.2 & 16 & 0.81$\pm$ 0.05 & 48$\pm$ 2  \\ 
MWC 275 & J & 20140424 & 58.2 & 32 & 0.8$\pm$ 0.34 & 34$\pm$ 12  \\ 
MWC 297 & H & 20180608 & 58.2 & 32 & 1.21$\pm$ 0.05 & 94$\pm$ 1  \\ 
MWC 297 & H & 20180817 & 43.6 & 32 & 1.15$\pm$ 0.04 & 99$\pm$ 1  \\ 
MWC 614 & J & 20180816 & 58.2 & 64 & 0.94$\pm$ 0.07 & 95$\pm$ 2  \\ 
MWC 614 & H & 20180816 & 58.2 & 64 & 1.24$\pm$ 0.11 & 101$\pm$ 2  \\ 
MWC 789 & H & 20181120 & 58.2 & 40 & 0.92$\pm$ 0.13 & 37$\pm$ 4  \\ 
MWC 863 & J & 20140424 & 58.2 & 32 & 2.95$\pm$ 0.28 & 56$\pm$ 3  \\ 
PDS 66 & J & 20160306 & 59.6 & 104 & 0.82$\pm$ 0.06 & 99$\pm$ 2  \\ 
Ty CrA & H & 20180608 & 58.2 & 32 & 0.33$\pm$ 0.04 & 148$\pm$ 3  \\ 
Ty CrA & H & 20180817 & 52.4 & 32 & 0.15$\pm$ 0.04 & 164$\pm$ 7  \\ 
V1295 Aql & H & 20180608 & 58.2 & 32 & 0.59$\pm$ 0.06 & 104$\pm$ 3  \\ 
V1295 Aql & J & 20180816 & 58.2 & 64 & 0.58$\pm$ 0.06 & 75$\pm$ 3  \\ 
V1295 Aql & H & 20180816 & 58.2 & 32 & 0.56$\pm$ 0.04 & 53$\pm$ 2  \\ 
V921 Sco & H & 20190513 & 58.2 & 32 & 0.98$\pm$ 0.21 & 138$\pm$ 6  \\ 
WRAY 15-535 & J & 20180319 & 58.2 & 32 & 2.1$\pm$ 0.09 & 53$\pm$ 1  \\ 
WRAY 15-535 & J & 20190222 & 58.2 & 32 & 2.02$\pm$ 0.1 & 58$\pm$ 1  \\ 
\enddata
\tablecomments{All epochs observed as part of the Gemini-LIGHTS survey. PA and \%Pol are the average polarization angle and \% polarization of stellar and instrumental polarization removed.}
\end{deluxetable*} 
\clearpage
\startlongtable
\begin{deluxetable*}{lccccc}
\tablecaption{Polarization Flux and Point Source Detection Upper Limits}
\small
\tablehead{
\colhead{Target} & \colhead{Band} & \colhead{Epoch} & \colhead{\Qphi/Fstar}  & \multicolumn{2}{c}{5$\sigma$ Point Source Detection Upper limit} \\
\colhead{} & \colhead{} & \colhead{(YYYYMMDD)} & \colhead{(per 1000)} & \colhead{Contrast at 0$\farcs$2} & \colhead{Mass Sensitivity (\Msun)}}
\startdata
AK Sco & H & 20180811 & 1.2 $\pm$ 0.2 & 1.6E-04 & 0.011  \\ 
CU Cha & J & 20170406 & 2.7 $\pm$ 0.9 & 3.9E-05 & 0.006  \\ 
CU Cha & H & 20180413 & 3.8 $\pm$ 1.0 & 2.5E-04 & 0.02  \\ 
FU Ori & J & 20180103 & 3.7 $\pm$ 0.6 & 1.2E-04 & 0.03  \\ 
GW Ori & J & 20180104 & 9.4 $\pm$ 0.9 & 1.6E-04 & 0.011  \\ 
GW Ori & H & 20180104 & 13.3 $\pm$ 1.1 & 5.9E-05 & 0.007  \\ 
HD 34700 A & J & 20180103 & 15.7 $\pm$ 1.8 & 1.3E-04 & 0.008  \\ 
HD 34700 A & H & 20180103 & 22.2 $\pm$ 1.8 & 5.3E-05 & 0.005  \\ 
HD 36917 & J & 20171231 & $<$0.02 &  3.6E-05 & 0.007  \\ 
HD 36917 & H & 20190126 & $<$0.03 &  8.6E-05 & 0.01  \\ 
HD 37806 & J & 20171231 & $<$0.04 &  3.8E-05 & 0.008  \\ 
HD 38087 & J & 20180103 & $<$0.12 &  6.1E-05 & 0.008  \\ 
HD 45677 & J & 20171231 & 8.4 $\pm$ 1.1 & 5.5E-05 & 0.01  \\ 
HD 45677 & H & 20180101 & 6.9 $\pm$ 1.0 & 3.0E-05 & 0.01  \\ 
HD 50138 & J & 20180104 & 0.9 $\pm$ 0.2 & 6.0E-05 & 0.011  \\ 
HD 85567 & J & 20180103 & 0.10 $\pm$ 0.08 & 1.0E-04 & 0.03  \\ 
HD 85567 & J & 20180520 & 0.12 $\pm$ 0.07 & 4.7E-05 & 0.02  \\ 
HD 85567 & H & 20190127 & 0.13 $\pm$ 0.11 & 2.5E-04 & 0.06  \\ 
HD 95881 & J & 20180104 & $<$0.09 &  9.6E-05 & 0.03  \\ 
HD 98800 & H & 20190127 & 0.19 $\pm$ 0.14 & \nodata & \nodata  \\ 
HD 98922 & J & 20180321 & 0.21 $\pm$ 0.16 & 6.9E-05 & 0.03  \\ 
HD 100453 & J & 20150410 & 2.9 $\pm$ 0.2 & \nodata & \nodata  \\ 
HD 100546 & J & 20170220 & 10.2 $\pm$ 0.6 & 8.7E-05 & 0.008  \\ 
HD 101412 & H & 20180319 & $<$0.04 &  1.7E-04 & 0.013  \\ 
HD 104237 & J & 20170407 & 0.04 $\pm$ 0.03 & 9.2E-05 & 0.013  \\ 
HD 104237 & J & 20180317 & 0.03 $\pm$ 0.02 & 1.6E-04 & 0.014  \\ 
HD 104237 & J & 20180520 & 0.06 $\pm$ 0.03 & 2.2E-04 & 0.02  \\ 
HD 104237 & J & 20190217 & $<$0.06 &  1.4E-04 & 0.014  \\ 
HD 139614 & J & 20170406 & 2.5 $\pm$ 0.7 & 8.9E-05 & 0.01  \\ 
HD 139614 & J & 20190513 & 1.5 $\pm$ 0.6 & 6.7E-05 & 0.009  \\ 
HD 139614 & H & 20180608 & 2.5 $\pm$ 0.7 & 3.1E-03 & 0.06  \\ 
HD 141569 & J & 20180609 & 0.29 $\pm$ 0.28 & 1.5E-04 & 0.013  \\ 
HD 141569 & J & 20190223 & $<$0.3 &  4.8E-05 & 0.007  \\ 
HD 142527 & H & 20140425 & 9.26 $\pm$ 1.17 & \nodata & \nodata  \\ 
HD 142666 & J & 20170703 & 0.19 $\pm$ 0.17 & \nodata & \nodata  \\ 
HD 142666 & J & 20180609 & 0.09 $\pm$ 0.08 & \nodata & \nodata  \\ 
HD 144432 & H & 20150709 & $<$0.03 &  \nodata & \nodata  \\ 
HD 145718 & J & 20180607 & 0.4 $\pm$ 0.3 & 4.4E-05 & 0.007  \\ 
HD 145718 & H & 20180608 & 0.4 $\pm$ 0.2 & 1.1E-04 & 0.011  \\ 
HD 158643 & J & 20180609 & $<$0.01 &  2.4E-05 & 0.005  \\ 
HD 158643 & J & 20190223 & $<$0.04 &  \nodata & \nodata  \\ 
HD 169142 & J & 20140425 & 4.5 $\pm$ 0.8 & 1.0E-04 & 0.008  \\ 
HD 176386 & J & 20180817 & $<$0.01 &  2.8E-05 & 0.005  \\ 
HD 176386 & H & 20180607 & $<$0.05 &  6.7E-05 & 0.006  \\ 
HD 176386 & H & 20180816 & $<$0.02 &  6.8E-05 & 0.006  \\ 
Hen 3-1330 & J & 20190512 & $<$0.03 &  4.8E-05 & 0.04  \\ 
Hen 3-225 & J & 20190127 & $<$0.04 &  7.3E-05 & 0.013  \\ 
Hen 3-225 & H & 20190127 & $<$0.06 &  6.7E-05 & 0.01  \\ 
Hen 3-365 & J & 20170406 & 5.5 $\pm$ 0.6 & 3.3E-05 & 0.05  \\ 
HR 5999 & J & 20170406 & $<$0.02 &  3.3E-04 & 0.03  \\ 
HR 5999 & J & 20170702 & 0.03 $\pm$ 0.02 & 7.8E-05 & 0.01  \\ 
HT Lup & H & 20190514 & 0.4 $\pm$ 0.2 & 5.6E-04 & 0.009  \\ 
MWC 147 & J & 20171231 & $<$0.07 &  8.4E-05 & 0.011  \\ 
MWC 147 & H & 20190127 & $<$0.1 &  3.3E-04 & 0.04  \\ 
MWC 166 & J & 20170406 & $<$0.06 &  6.0E-03 & 1.4  \\ 
MWC 275 & J & 20140424 & 0.3 $\pm$ 0.2 & 7.3E-05 & 0.01  \\ 
MWC 297 & H & 20180608 & 0.3 $\pm$ 0.2 & 8.5E-05 & 0.04  \\ 
MWC 297 & H & 20180817 & 0.3 $\pm$ 0.2 & 1.0E-04 & 0.04  \\ 
MWC 614 & J & 20180816 & 2.0 $\pm$ 0.3 & 1.5E-04 & 0.014  \\ 
MWC 614 & H & 20180816 & 2.4 $\pm$ 0.2 & 1.7E-04 & 0.013  \\ 
MWC 789 & H & 20181120 & 5.9 $\pm$ 1.5 & 6.3E-04 & 0.06  \\ 
MWC 863 & J & 20140424 & $<$0.3 &  1.1E-03 & 0.03  \\ 
PDS 66 & J & 20160306 & 2.2 $\pm$ 1.1 & 5.9E-05 & 0.003  \\ 
Ty CrA & H & 20180608 & $<$0.6 &  8.7E-04 & 0.3  \\ 
Ty CrA & H & 20180817 & $<$0.44 &  1.5E-04 & 0.06  \\ 
V1295 Aql & J & 20180816 & $<$0.02 &  2.5E-04 & 0.05  \\ 
V1295 Aql & H & 20180608 & $<$0.02 &  1.0E-04 & 0.03  \\ 
V1295 Aql & H & 20180816 & $<$0.04 &  2.9E-04 & 0.05  \\ 
V921 Sco & H & 20190513 & 0.06 $\pm$ 0.05 & 1.6E-04 & 0.13  \\ 
WRAY 15-535 & J & 20180319 & 0.11 $\pm$ 0.07 & 1.0E-04 & \nodata  \\ 
WRAY 15-535 & J & 20190222 & $<$0.04 &  1.5E-05 & \nodata  \\ 
\enddata
\tablecomments{Mass estimate based on measured companion flux and the system age (see Table \ref{tbl:target_properties} and use models from \citet{Baraffe2015} for masses $>$ 0.01 \Msun{} and \citet{Phillips2020} for masses $<$ 0.01 \Msun{}. \Qphi/Fstar values utilized in this work are the weighted average of the epochs listed above.}\label{tbl:epoch_parameters}
\end{deluxetable*} 
\clearpage
\startlongtable
\begin{deluxetable*}{lcccccccc}
\tablecaption{Point Sources Detected in Gemini-LIGHTS sample}
\small
\tablehead{
\colhead{Target} & \colhead{Band} & \colhead{Epoch} & \colhead{Comp. \#} & \colhead{Separation} & \colhead{PA} & \colhead{Flux} & \colhead{Saturated} & \colhead{Mass Est.} \\
\colhead{} & \colhead{} & \colhead{(YYYYMMDD)} & \colhead{}  & \colhead{(``)} & \colhead{($^{\circ}$)} & \colhead{(mJy)} & \colhead{} & \colhead{(\Msun)}}
\label{tbl:companion_results}
\startdata
FU Ori & J & 20180103 & 1 & 0.431 $\pm$ 0.003 & 163.3 $\pm$ 0.3 & 27 $\pm$ 4 & TRUE & 0.6 $_{-0.1} ^{+0.1}$ \\ 
HD 100453 & J & 20150411 & 1 & 0.909 $\pm$ 0.002 & 131.63 $\pm$ 0.09 & 15.4 $\pm$ 1.3 & FALSE & 0.08 $_{-0.04} ^{+0.01}$ \\ 
HD 101412 & H & 20180319 & 1 & 0.460 $\pm$ 0.002 & 149.9 $\pm$ 0.2 & 1.46 $\pm$ 0.11 & FALSE & 0.053 $_{-0.003} ^{+0.03}$ \\ 
HD 101412 & H & 20180319 & 2 & 0.153 $\pm$ 0.003 & 184.9 $\pm$ 1.0 & 5.0 $\pm$ 0.4 & FALSE & 0.19 $_{-0.01} ^{+0.06}$ \\ 
HD 104237 & J & 20170407 & 1 & 1.226 $\pm$ 0.004 & 255.43 $\pm$ 0.14 & 29 $\pm$ 9 & FALSE & 0.11 $_{-0.04} ^{+0.04}$ \\ 
HD 104237 & J & 20180317 & 1 & 1.214 $\pm$ 0.004 & 255.034 $\pm$ 0.13 & 19 $\pm$ 6 & FALSE & 0.065 $_{-0.02} ^{+0.02}$ \\ 
HD 104237 & J & 20180520 & 1 & 1.204 $\pm$ 0.003 & 255.1 $\pm$ 0.11 & 32 $\pm$ 6 & FALSE & 0.12 $_{-0.02} ^{+0.02}$ \\ 
HD 104237 & J & 20190217 & 1 & 1.198 $\pm$ 0.004 & 255.28 $\pm$ 0.11 & 34 $\pm$ 5 & FALSE & 0.13 $_{-0.02} ^{+0.02}$ \\ 
HD 144432 & H & 20150708 & 1 & 1.256 $\pm$ 0.004 & 6.07 $\pm$ 0.16 & 51 $\pm$ 7 & TRUE & 0.3 $_{-0.05} ^{+0.03}$ \\ 
HD 158643 & J & 20180609 & 1 & 0.118 $\pm$ 0.013 & 294 $\pm$ 5 & 30 $\pm$ 4 & FALSE & 0.07 $_{-0.02} ^{+0.01}$ \\ 
HD 38087 & J & 20180103 & 1 & 0.303 $\pm$ 0.003 & 183.7 $\pm$ 0.4 & 47 $\pm$ 8 & TRUE & 0.8 $_{-0.3} ^{+0.5}$ \\ 
HD 50138 & J & 20180104 & 1 & 0.729 $\pm$ 0.003 & 106.6 $\pm$ 0.2 & 10.1 $\pm$ 1.5 & FALSE & $^\ast$0.12    \\ 
HD 98800 & H & 20190127 & 1 & 0.374 $\pm$ 0.010 & 9.8 $\pm$ 0.6 & 633 $\pm$ 9 & TRUE & 0.4 $_{-0.03} ^{+0.03}$ \\ 
HR 5999 & J & 20170406 & 1 & 1.251 $\pm$ 0.004 & 112.55 $\pm$ 0.11 & 141 $\pm$ 16 & TRUE & 0.6 $_{-0.2} ^{+0.1}$ \\ 
HR 5999 & J & 20170702 & 1 & 1.260 $\pm$ 0.005 & 112.4 $\pm$ 0.2 & 128 $\pm$ 19 & TRUE & 0.6 $_{-0.2} ^{+0.1}$ \\ 
HT Lup & H & 20190514 & 1 & 0.161 $\pm$ 0.003 & 246.6 $\pm$ 0.7 & 70 $\pm$ 7 & TRUE & 0.13 $_{-0.01} ^{+0.01}$ \\ 
Hen 3-365 & J & 20170406 & 1 & 0.818 $\pm$ 0.003 & 101.08 $\pm$ 0.14 & 1.3 $\pm$ 0.2 & FALSE & $^\ast$0.17    \\ 
Hen 3-365 & J & 20170406 & 2 & 0.446 $\pm$ 0.008 & 133.0 $\pm$ 0.8 & 0.6 $\pm$ 0.1 & FALSE & $^\ast$0.13    \\ 
Hen 3-225 & J & 20190127 & 1 & 1.564 $\pm$ 0.019 & 303.4 $\pm$ 0.3 & 1.46 $\pm$ 0.02 & FALSE & $^\ast$0.12    \\ 
Hen 3-1330 & J & 20190512 & 1 & 0.268 $\pm$ 0.017 & 49 $\pm$ 2 & 0.8 $\pm$ 0.3 & FALSE & $^\ast$0.13    \\ 
MWC 147 & J & 20171231 & 1 & 0.137 $\pm$ 0.004 & 56 $\pm$ 1 & 30 $\pm$ 6 & TRUE & 0.6 $_{-0.1} ^{+0.4}$ \\ 
MWC 147 & J & 20190127 & 1 & 0.156 $\pm$ 0.003 & 54.1 $\pm$ 0.8 & 33 $\pm$ 5 & TRUE & 0.5 $_{-0.1} ^{+0.4}$ \\ 
MWC 166 & J & 20170406 & 1 & 0.587 $\pm$ 0.004 & 298.4 $\pm$ 0.3 & 282 $\pm$ 9 & TRUE & $^\ast >$1.4    \\ 
MWC 297 & H & 20180608 & 1 & 0.573 $\pm$ 0.003 & 86.5 $\pm$ 0.2 & 2.06 $\pm$ 0.09 & FALSE & $^\ast$0.04    \\ 
MWC 297 & H & 20180817 & 1 & 0.576 $\pm$ 0.004 & 86.6 $\pm$ 0.2 & 1.74 $\pm$ 0.14 & FALSE & $^\ast$0.04    \\ 
MWC 789 & H & 20181120 & 1 & 0.371 $\pm$ 0.002 & 216.8 $\pm$ 0.3 & 10.7 $\pm$ 0.8 & FALSE & 0.8 $_{-0.2} ^{+0.1}$ \\ 
MWC 863 & J & 20140421 & 1 & 0.975 $\pm$ 0.004 & 226.97 $\pm$ 0.14 & 154 $\pm$ 19 & TRUE & 0.8 $_{-0.1} ^{+0.1}$ \\ 
Ty CrA & H & 20180608 & 1 & 0.129 $\pm$ 0.004 & 255.2 $\pm$ 1.3 & 26 $\pm$ 4 & FALSE & $>$1.4    \\ 
Ty CrA & H & 20180817 & 1 & 0.112 $\pm$ 0.003 & 257.3 $\pm$ 1.0 & 41 $\pm$ 4 & FALSE & $>$1.4    \\ 
V921 Sco & H & 20190513 & 1 & 0.475 $\pm$ 0.013 & 319.0 $\pm$ 1.1 & 0.09 $\pm$ 0.03 & FALSE & 0.03 $_{-0.01} ^{+0.01}$ \\ 
V921 Sco & H & 20190513 & 2 & 1.122 $\pm$ 0.003 & 323.12 $\pm$ 0.09 & 0.96 $\pm$ 0.08 & FALSE & 0.15 $_{-0.01} ^{+0.01}$ \\ 
V1295 Aql & H & 20180608 & 1 & 1.113 $\pm$ 0.003 & 40.52 $\pm$ 0.11 & 0.026 $\pm$ 0.002 & FALSE & $^\ast $0.010    \\ 
V1295 Aql & H & 20180816 & 1 & 1.135 $\pm$ 0.016 & 40.1 $\pm$ 0.5 & 0.055 $\pm$ 0.009 & FALSE & $^\ast $0.013    \\ 
V1295 Aql & J & 20180817 & 1 & 1.119 $\pm$ 0.005 & 40.6 $\pm$ 0.2 & 0.015 $\pm$ 0.002 & FALSE & $^\ast $0.010    \\ 
\enddata
\tablecomments{Point sources with their PSF cores saturated are noted in the table. Mass estimate based on measured companion flux and the system age (see Table \ref{tbl:target_properties}). Mass estimates use models from \citet{Baraffe2015} for masses $>$ 0.01 \Msun{} and \citet{Phillips2020} for masses $<$ 0.01 \Msun{}. A $\ast $ denotes any target with ages $<$ 0.5 Myr which are younger than the youngest model in \citet{Baraffe2015} or \citet{Phillips2020}.}
\end{deluxetable*}

\clearpage

\section{Polarized Light FITS header Definition}\label{sec:header}

Here we define a standard FITS file for high-contrast polarization data. This is motivated by large numbers of observations of protoplanetary disks from multiple telescopes (e.g. GPI, SPHERE/IRDIS, Subaru/HiCIAO, Subaru/CHARIS). A standard FITS file will allow for better comparison of protoplanetary disk polarization imagery between different studies and instruments. This FITS standard was created in collaboration with Christian Ginski. 

The data is held in a 3 dimensional cube of 5 $\times$ x $\times$ y where x and y are the pixel dimensions of the image and 5 are the different image types. The five different image types are I, \Qphi, \Uphi, Q, U, and :LP\_I where I is Intensity image without stellar light subtracted, \Qphi, \Uphi, Q, and U as defined above in Section \ref{sec:pipeline}, and LP\_I which is the linear polarized intensity or $(Q^2 + U^2)^{1/2}$.

A sample header is shown in Table \ref{tbl:header_example} taken from GPI observations of MWC 275 (HD 163296) used in this work. Standard WCS headers are included to allow use and image overlay with programs such as DS9.
Additionally, reduction information such as star locations (X-STAR,Y-STAR), flux calibration (ZEROPT,REFMAG,CALIBFFAC), and fraction of stellar/instrumental polarization removed (FQ, FU) will help facilitate better comparisons between different epochs of observations of the same target.

\begin{deluxetable*}{lcc}
\tablecaption{Sample Header}
\tablehead{
\colhead{Keyword} & \colhead{Value} & \colhead{Comment}}
\label{tbl:header_example}
\startdata
DATE-OBS & '2014-04-24' & UT start date of exposure \\
CCDSIZE & '2048x2048' & Array dimensions \\
CREATOR & 'GPI DRP, v1.5.0,revc0cad3f' & This file created by GPI Data Reduction \\
OBSMODE & 'J\_coron ' & Currently selected observation Mode \\
RA & 269.0887 & Target Right Ascension \\
DEC & -21.956075 & Target Declination \\
DATE & '2014-04-24' & UTC Date of observation (YYYY-MM-DD) \\          
EPOCH & 2000.0 & Target Coordinate Epoch \\
GEMPRGID & 'GS-2014A-SV-412' & Gemini programme ID \\
INSTRUME & 'GPI     ' & Instrument used to acquire data \\
OBSERVAT & 'Gemini-South' & Observatory (Gemini-North|Gemini-South) \\
OBSID & 'GS-2014A-SV-412-6' & Gemini Observation ID \\
CD1\_1 & -3.92777777778E-06 & partial of first axis coordinate w.r.t. x \\
CD1\_2 & 9.84069992011E-14 & partial of first axis coordinate w.r.t. y \\
CD2\_1 & 9.8406999118E-14 & partial of second axis coordinate w.r.t. x \\
CD2\_2 & 3.92777777778E-06 & partial of second axis coordinate w.r.t. y \\
CDELT1 & 0.0014 & Coordinate increment \\
CDELT2 & 0.0014 & Coordinate increment \\
CRPIX1 & 141.0 & x-coordinate of ref pixel [note: first pixel is \\
CRPIX2 & 141.0 & y-coordinate of ref pixel [note: first pixel is \\
CRVAL1 & 269.0887 & Right ascension at ref point \\
CRVAL2 & -21.956075 & Declination at ref point \\
CTYPE1 & 'RA--TAN' & First axis is Right Ascension \\
CTYPE2 & 'DEC--TAN' & Second axis is Declination \\
CUNIT1 & 'deg     ' & Units of data \\
CUNIT2 & 'deg     ' & Units of data \\
RADESYS & 'FK5     ' & R.A DEC coordinate system reference \\
BSCALE  & 1 & Linear factor in scaling equation \\
BZERO & 0 & Zero point in scaling equation \\
WCSAXES & 3 & Number of axes in WCS system \\
CTYPE3 & 'STOKES  ' & Polarization \\
CUNIT3 & 'N/A     ' & Polarizations \\
CRVAL3 & 1 & I,Q\_phi,U\_phi,Q,U,LP\_I \\
CRPIX3 &  0 & Reference pixel location \\
CD3\_3 & 1 & Stokes axis: images 0 and 1 give orthogonal pol \\
FQ & 0.002997042052516556 & avg frac of stell/inst pol removed  \\
FU & 0.007443801664624007 & avg frac of stell/inst pol removed \\   
TARGET & 'MWC\_275 ' & Target Name \\
STOKES & 'I,Q\_phi,U\_phi,Q,U,LP\_I' & data cube Stokes components \\  
STAR\_X & 140.5 & star position axis1 (1-based coordinates) \\
STAR\_Y & 140.5 & star position axis2 (1-based coordinates) \\
FILTER & 'J-band  ' & filter band of observation \\
ZEROPT & 1594 & Jy \\
REFMAG & 6.195 & ref. mag of star for flux conversion \\
FLUXUNIT & 'mJy/arcsec\^{}2' & pixel flux units \\
CALIBFAC & 4.00356E-08 & Conversion factor mJy/arcsec\^{}2/ADU/sec/coadd \\
SCALE & 14.14 & Pixel Scale mas/pix \\
COMMENT & \nodata & Stokes componts only take linear pol. into account \\
COMMENT & \nodata & All pol. images are stellar pol. subtracted. \\
COMMENT & \nodata & 2MASS magnitudes used for flux conversion. \\
\enddata
\tablecomments{}
\end{deluxetable*}

\section{Full sample I, \Qphip, and \Uphi images} \label{sec:IQU_appendix}

Full sample I, \Qphip, and \Uphi images for all epochs for ArXiv version only. Images accepted to AJ can be found as a figure set on the online version of the manuscript. See Figure \ref{fig:images_example} for caption information.

\begin{figure*}[!ht]
    \centering
    \includegraphics[width=0.99\linewidth ,clip]{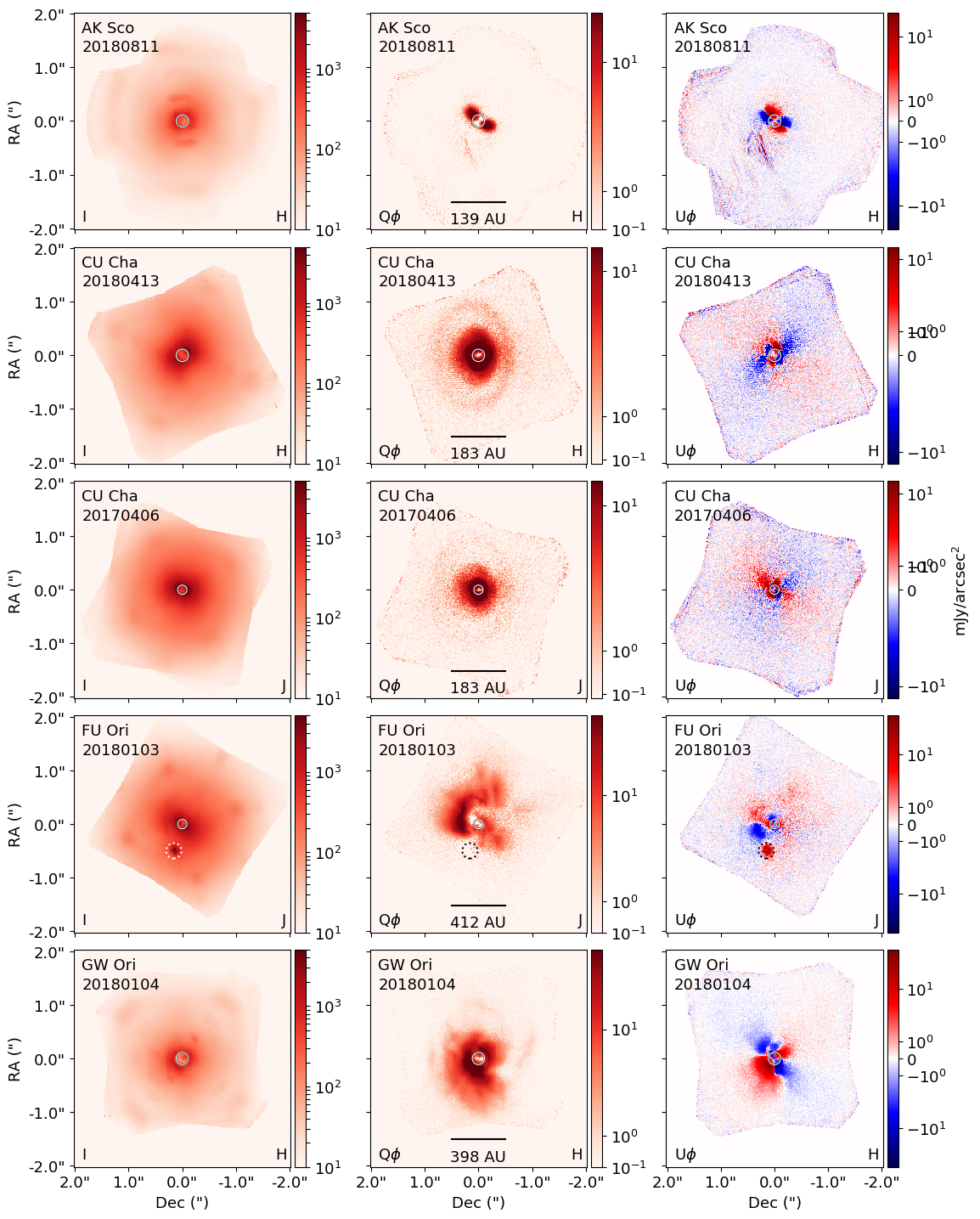}
    \caption{}
\end{figure*}

\begin{figure*}[!ht]
    \centering
    \includegraphics[width=0.99\linewidth ,clip]{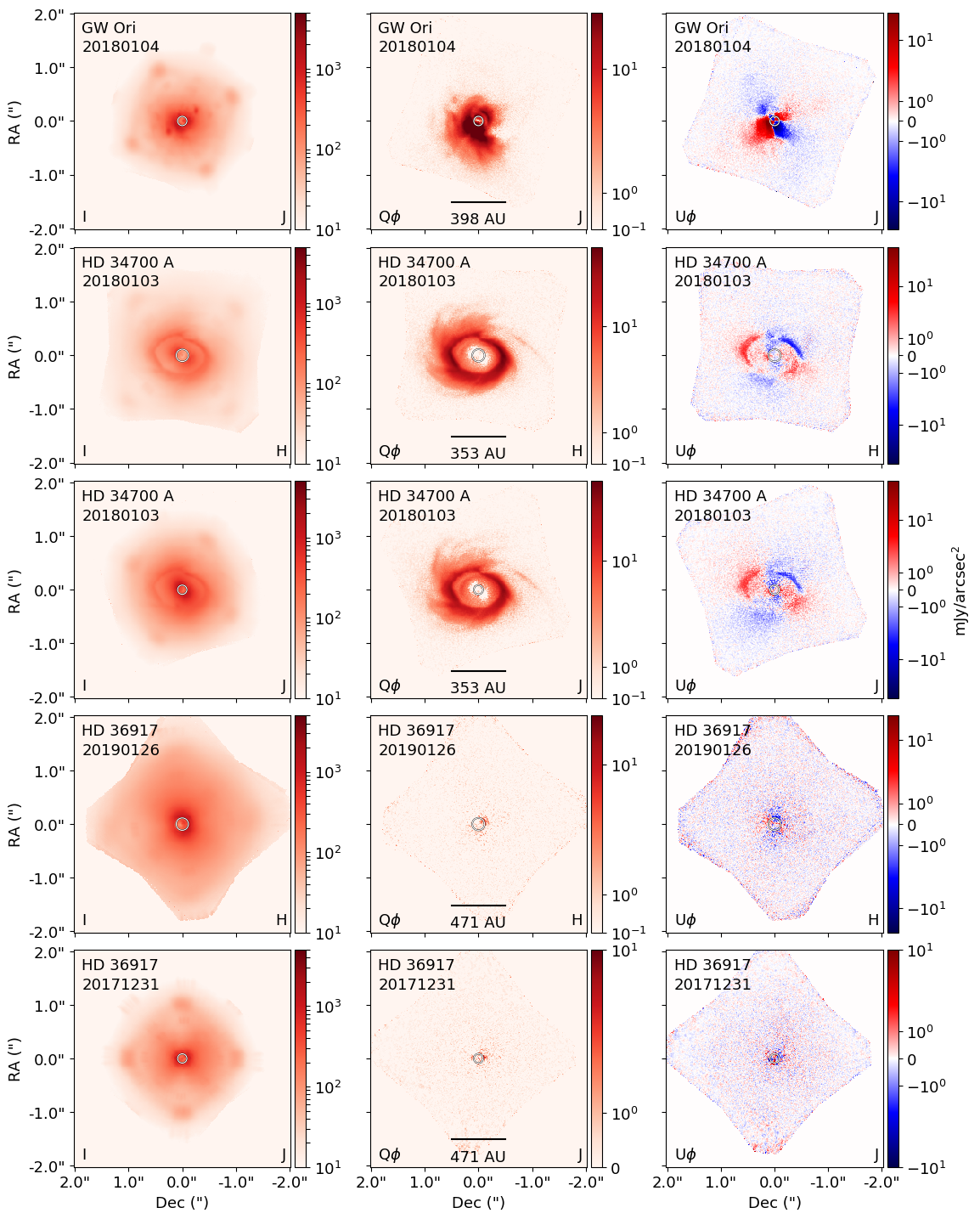}
    \caption{}
\end{figure*}

\begin{figure*}[!ht]
    \centering
    \includegraphics[width=0.99\linewidth ,clip]{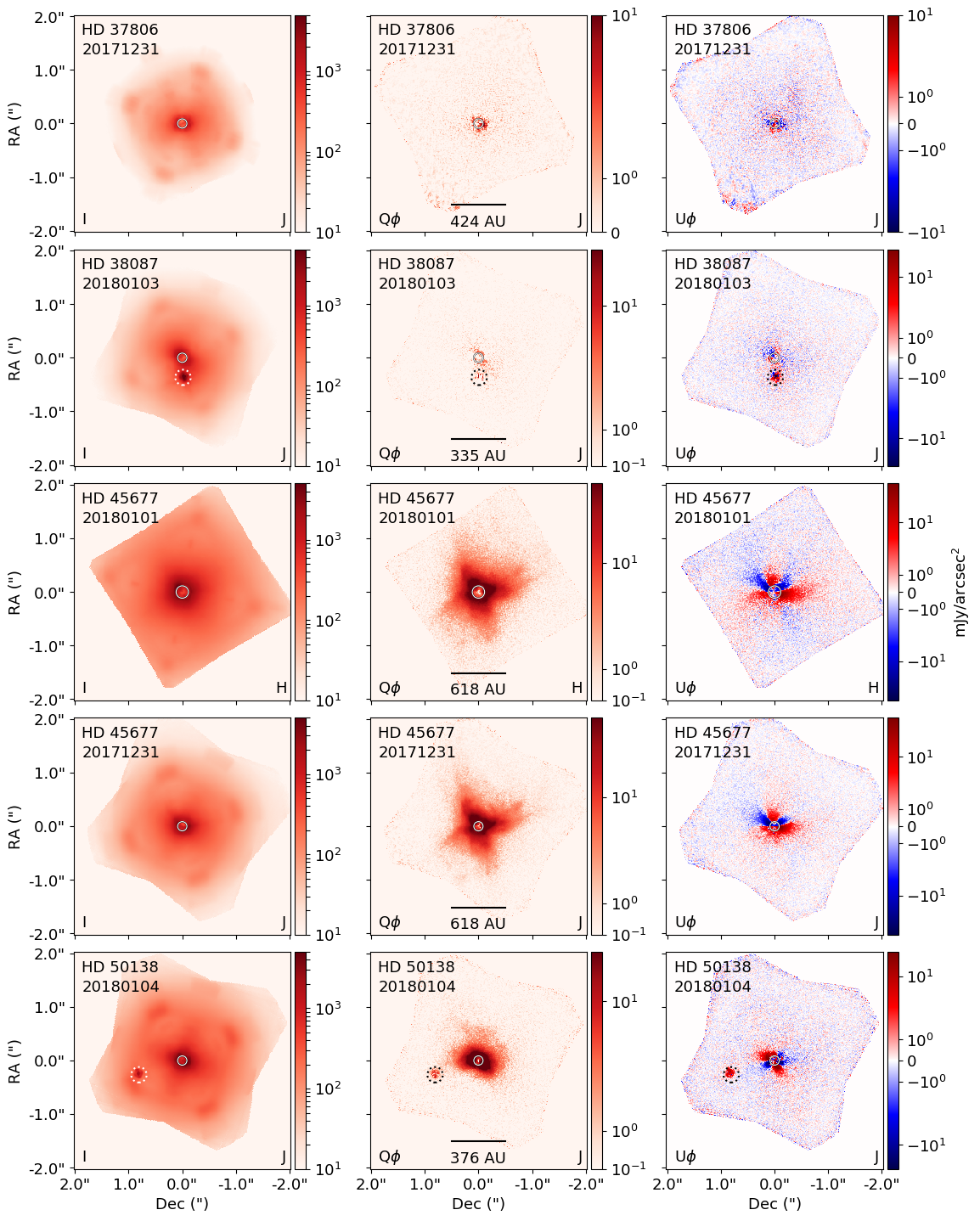}
    \caption{}
\end{figure*}

\begin{figure*}[!ht]
    \centering
    \includegraphics[width=0.99\linewidth ,clip]{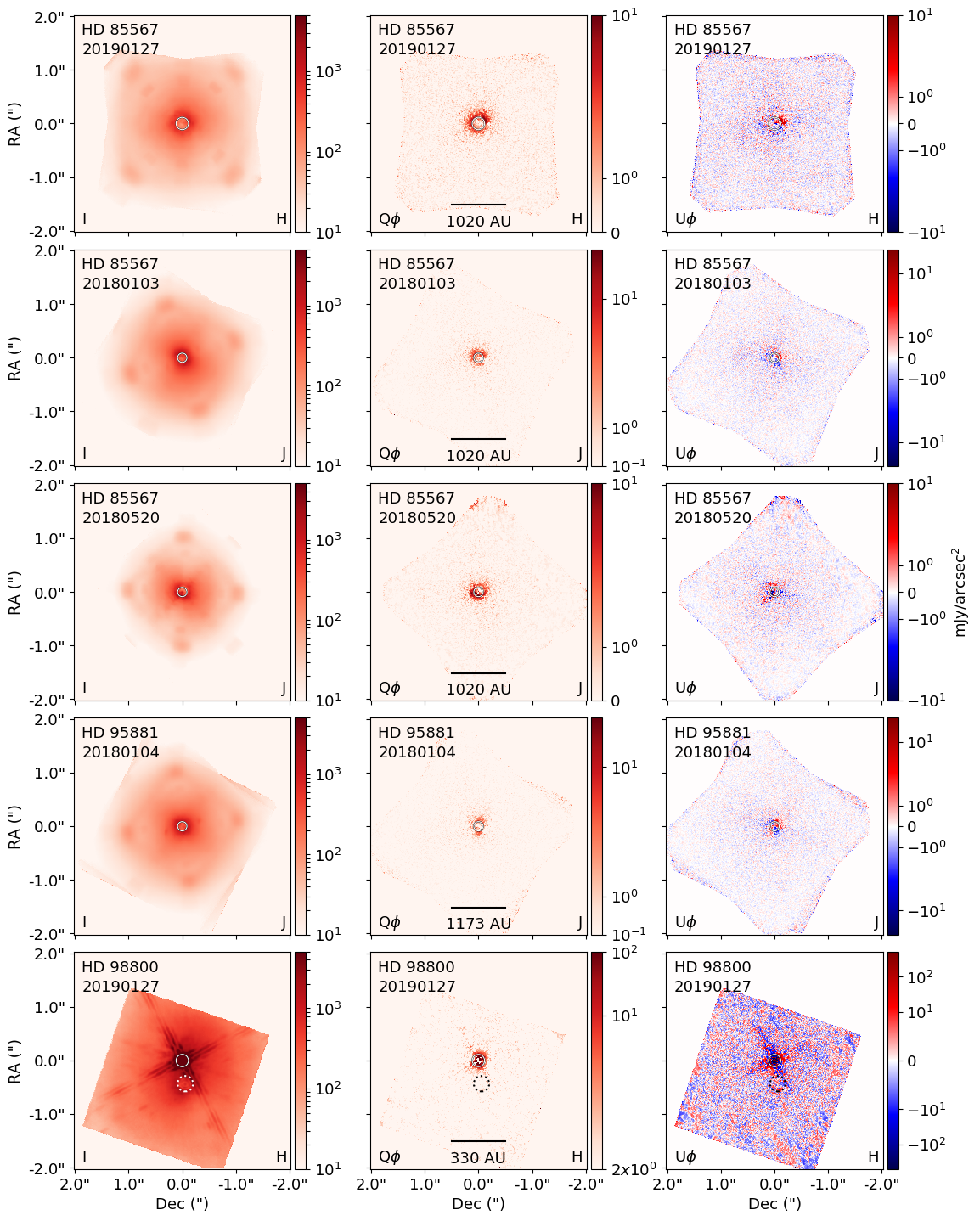}
    \caption{}
\end{figure*}

\begin{figure*}[!ht]
    \centering
    \includegraphics[width=0.99\linewidth ,clip]{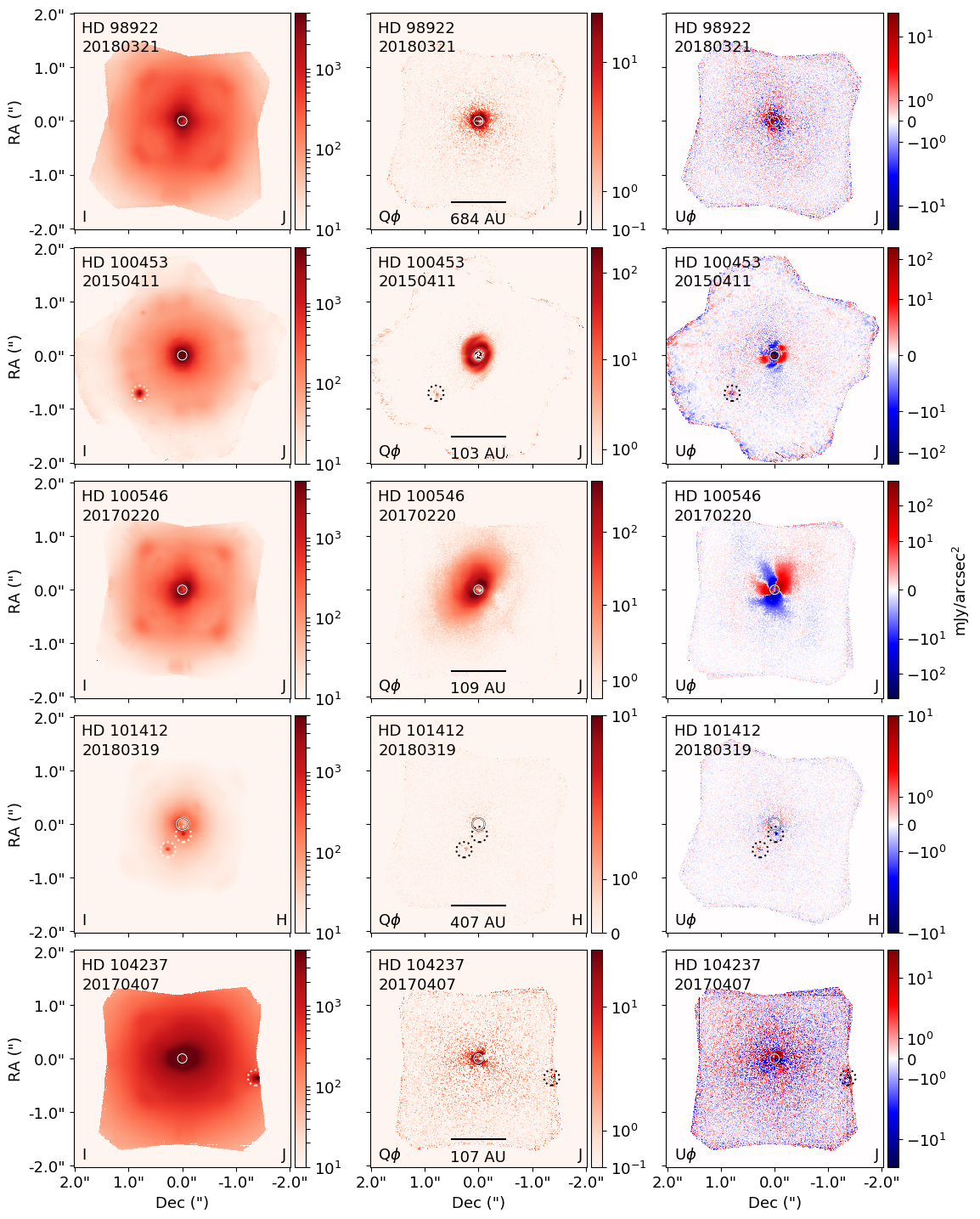}
    \caption{}
\end{figure*}

\begin{figure*}[!ht]
    \centering
    \includegraphics[width=0.99\linewidth ,clip]{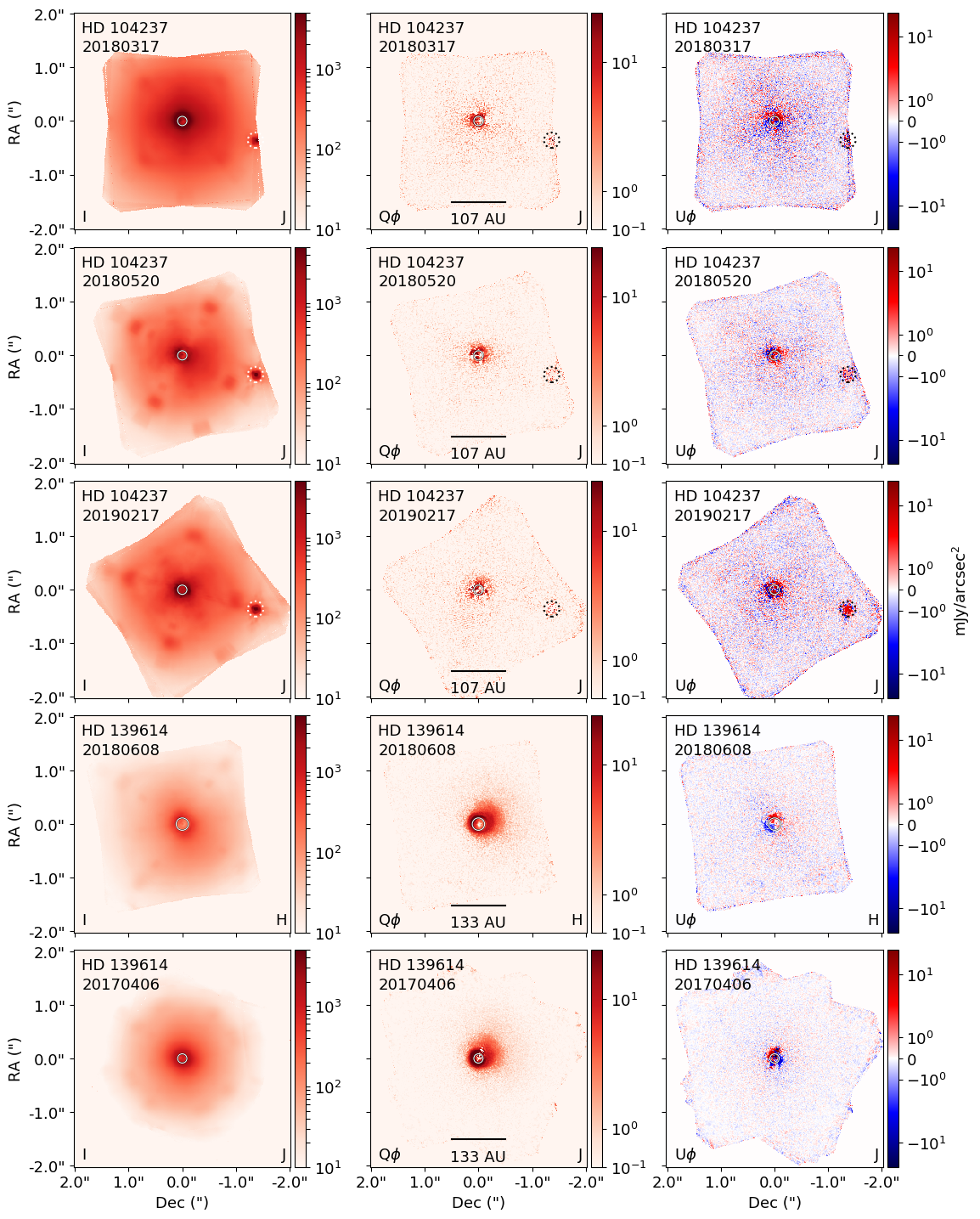}
    \caption{}
\end{figure*}

\begin{figure*}[!ht]
    \centering
    \includegraphics[width=0.99\linewidth ,clip]{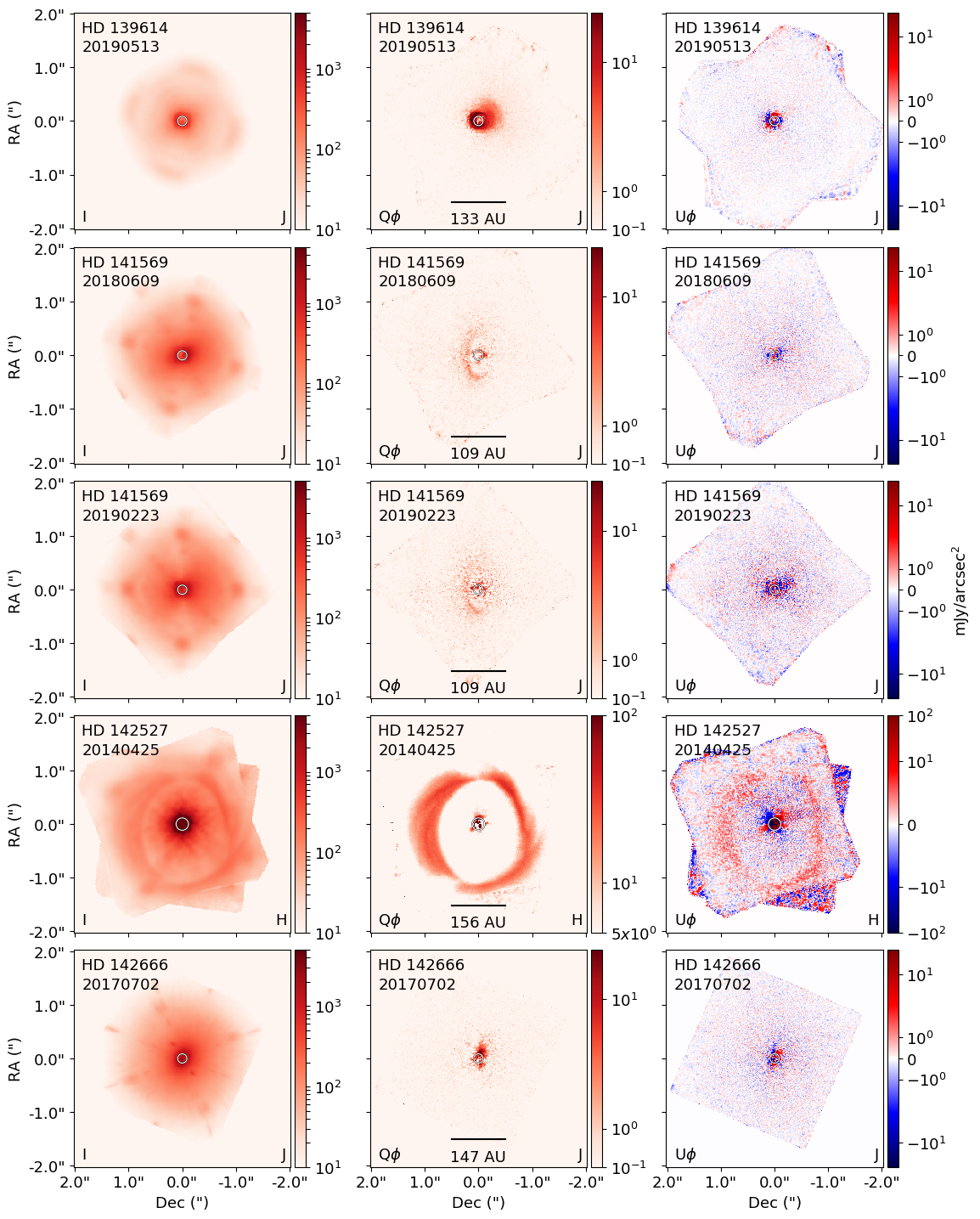}
    \caption{}
\end{figure*}

\begin{figure*}[!ht]
    \centering
    \includegraphics[width=0.99\linewidth ,clip]{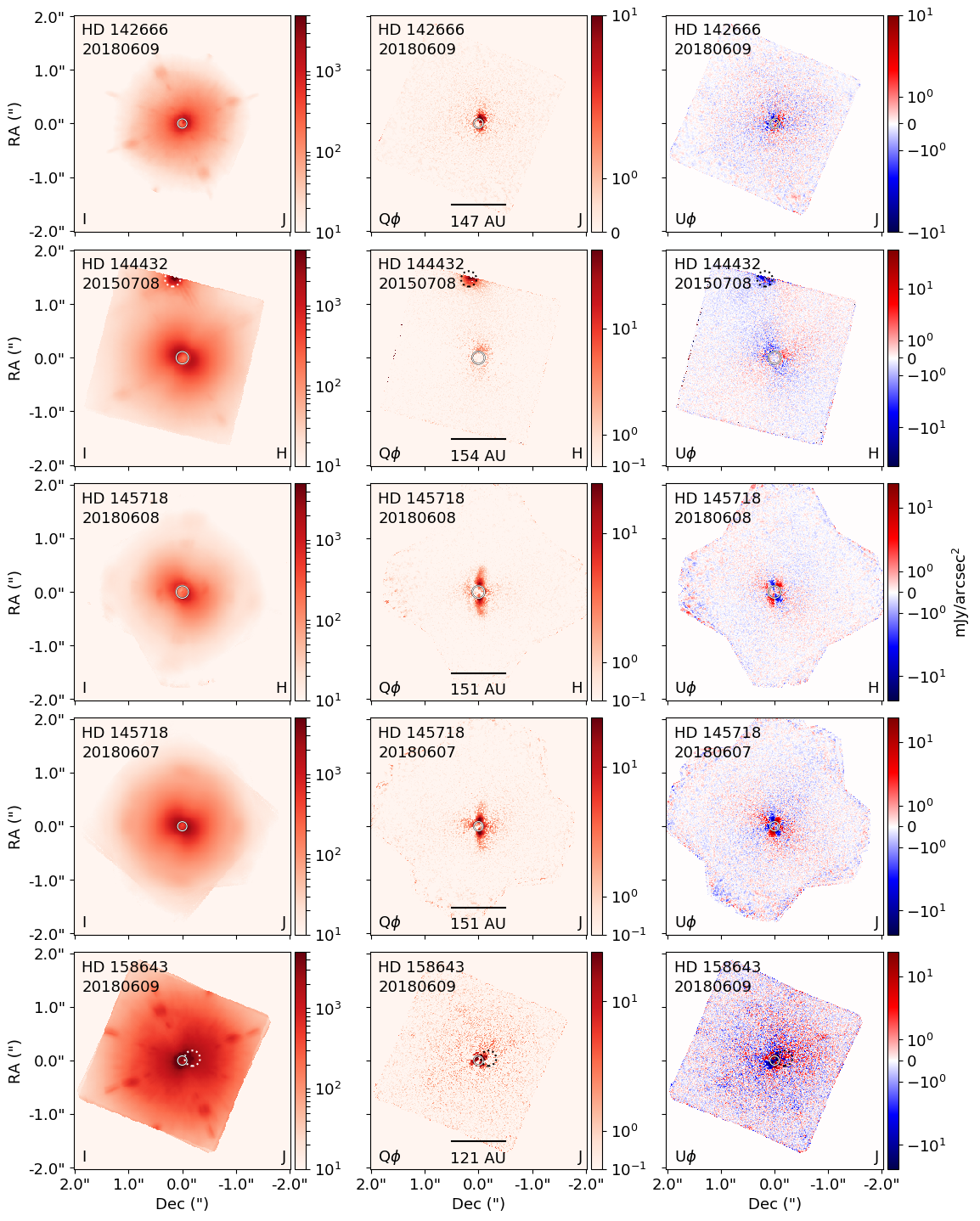}
    \caption{}
\end{figure*}

\begin{figure*}[!ht]
    \centering
    \includegraphics[width=0.99\linewidth ,clip]{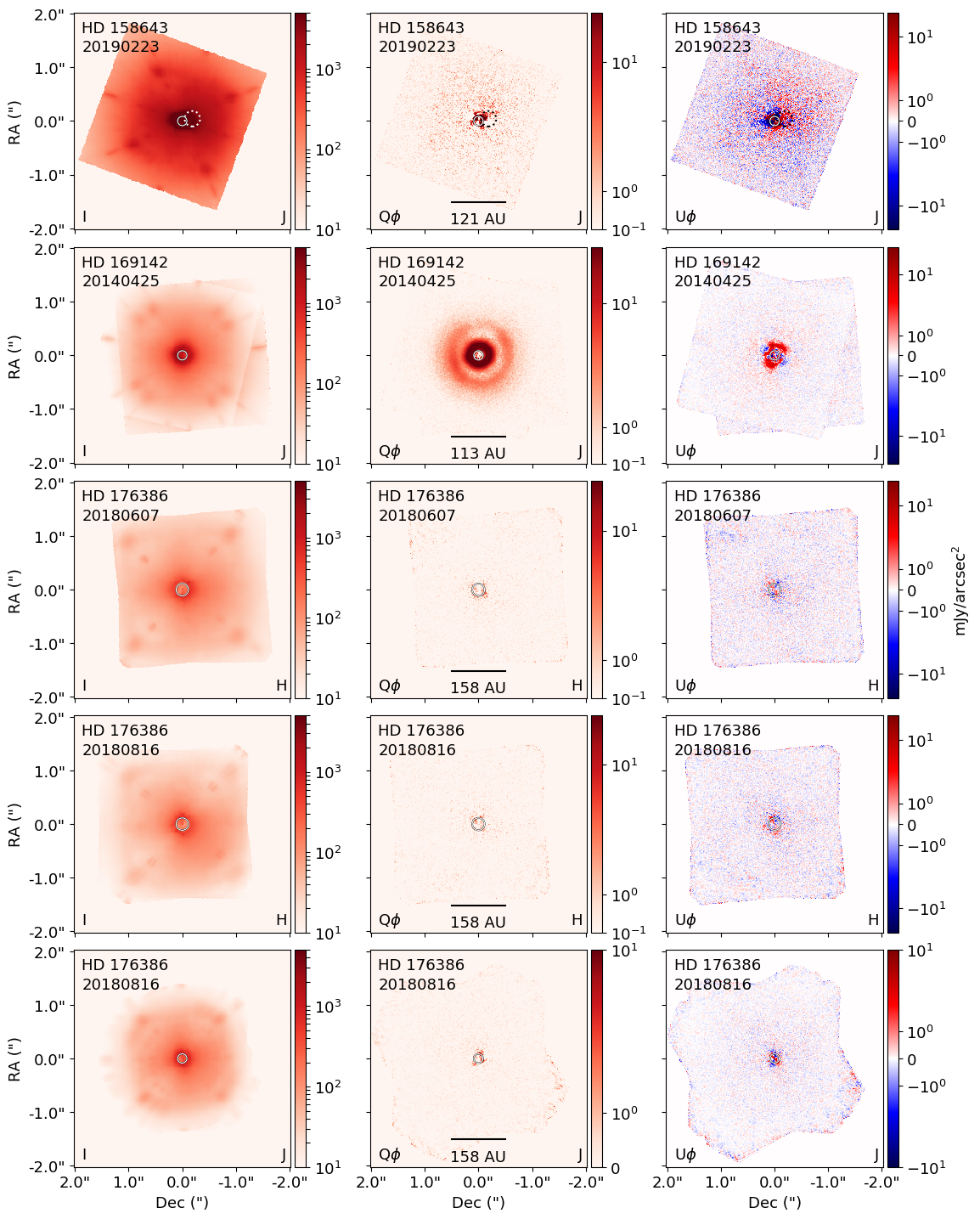}
    \caption{}
\end{figure*}

\begin{figure*}[!ht]
    \centering
    \includegraphics[width=0.99\linewidth ,clip]{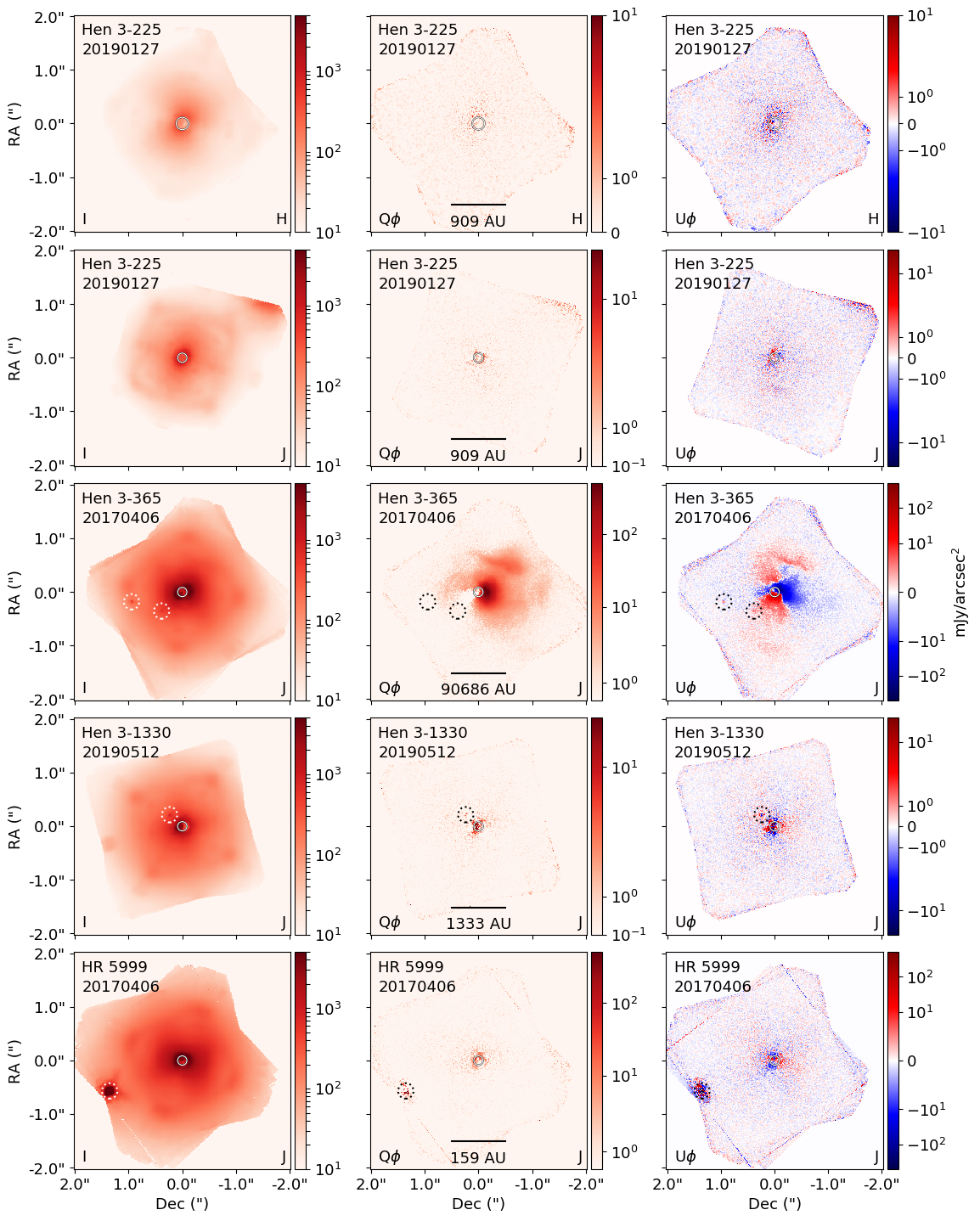}
    \caption{}
\end{figure*}

\begin{figure*}[!ht]
    \centering
    \includegraphics[width=0.99\linewidth ,clip]{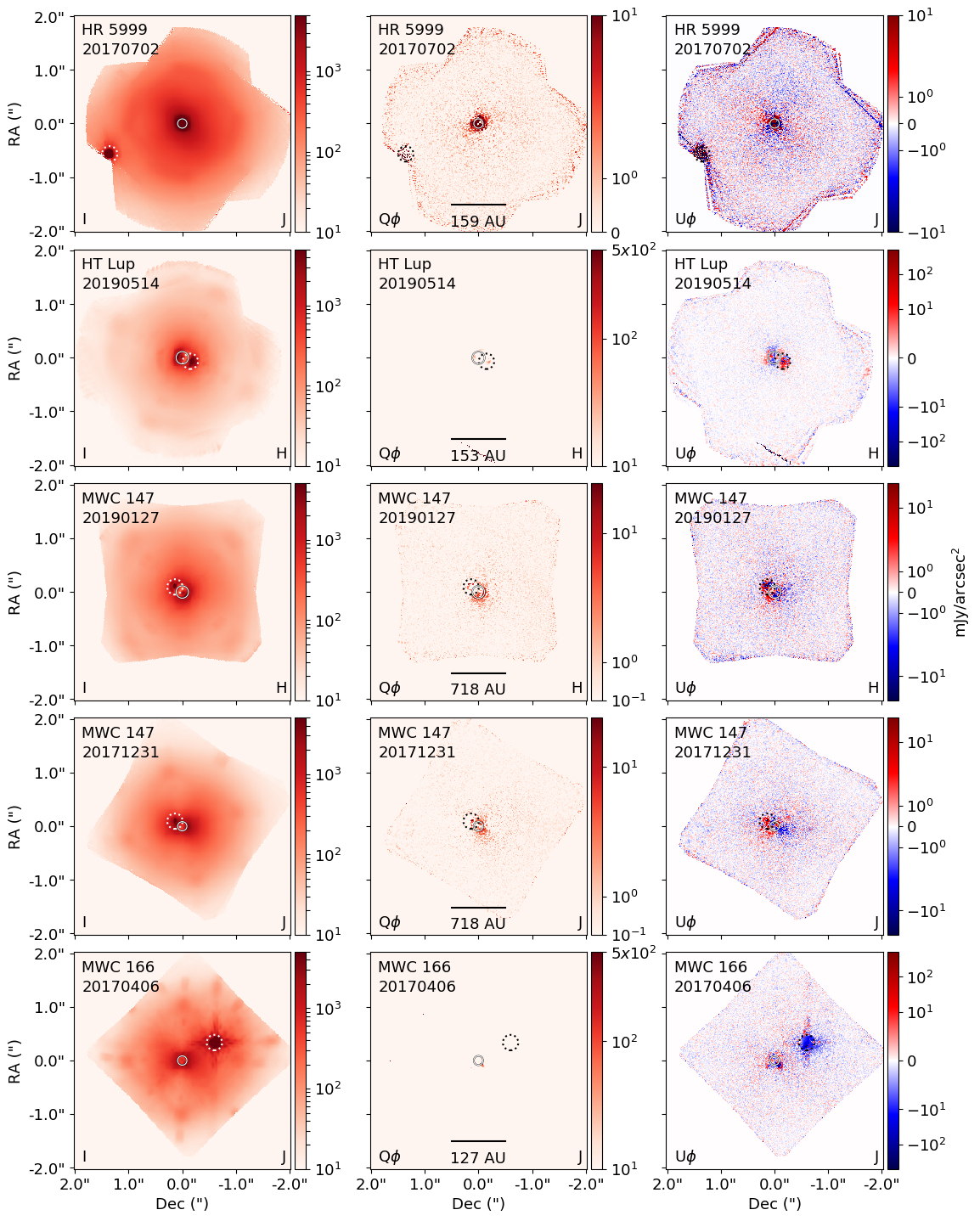}
    \caption{}
\end{figure*}

\begin{figure*}[!ht]
    \centering
    \includegraphics[width=0.99\linewidth ,clip]{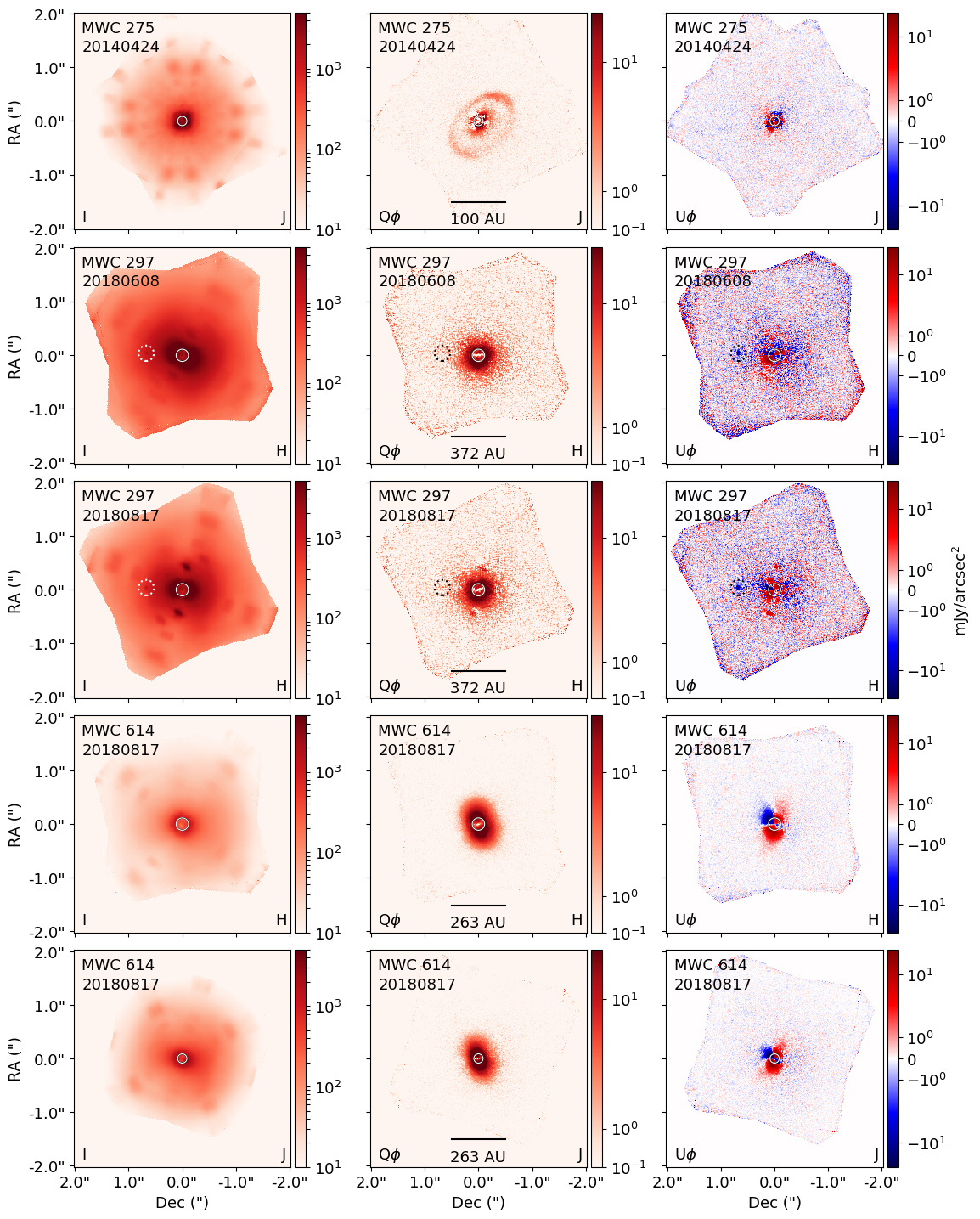}
    \caption{}
\end{figure*}

\begin{figure*}[!ht]
    \centering
    \includegraphics[width=0.99\linewidth ,clip]{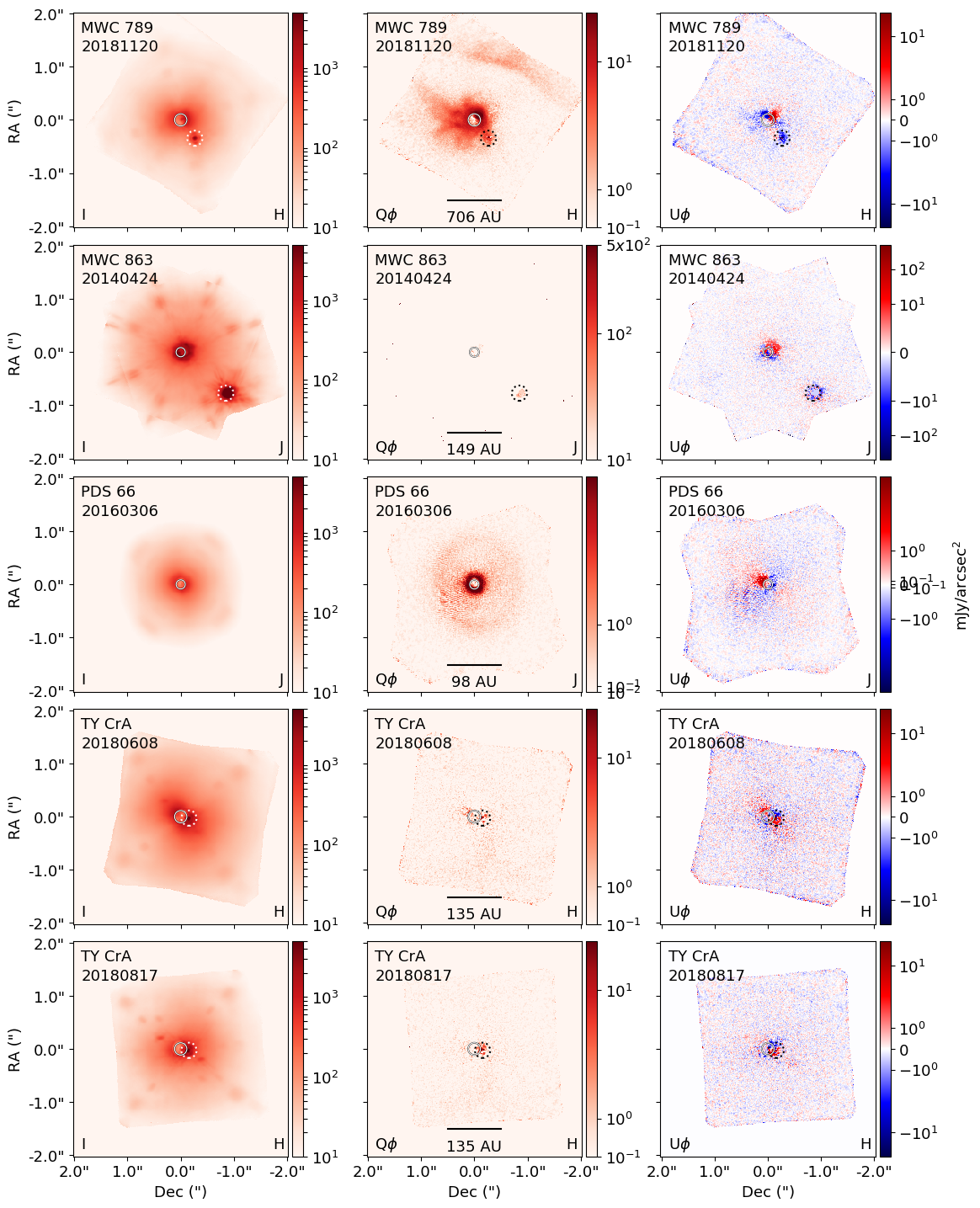}
    \caption{}
\end{figure*}

\begin{figure*}[!ht]
    \centering
    \includegraphics[width=0.99\linewidth ,clip]{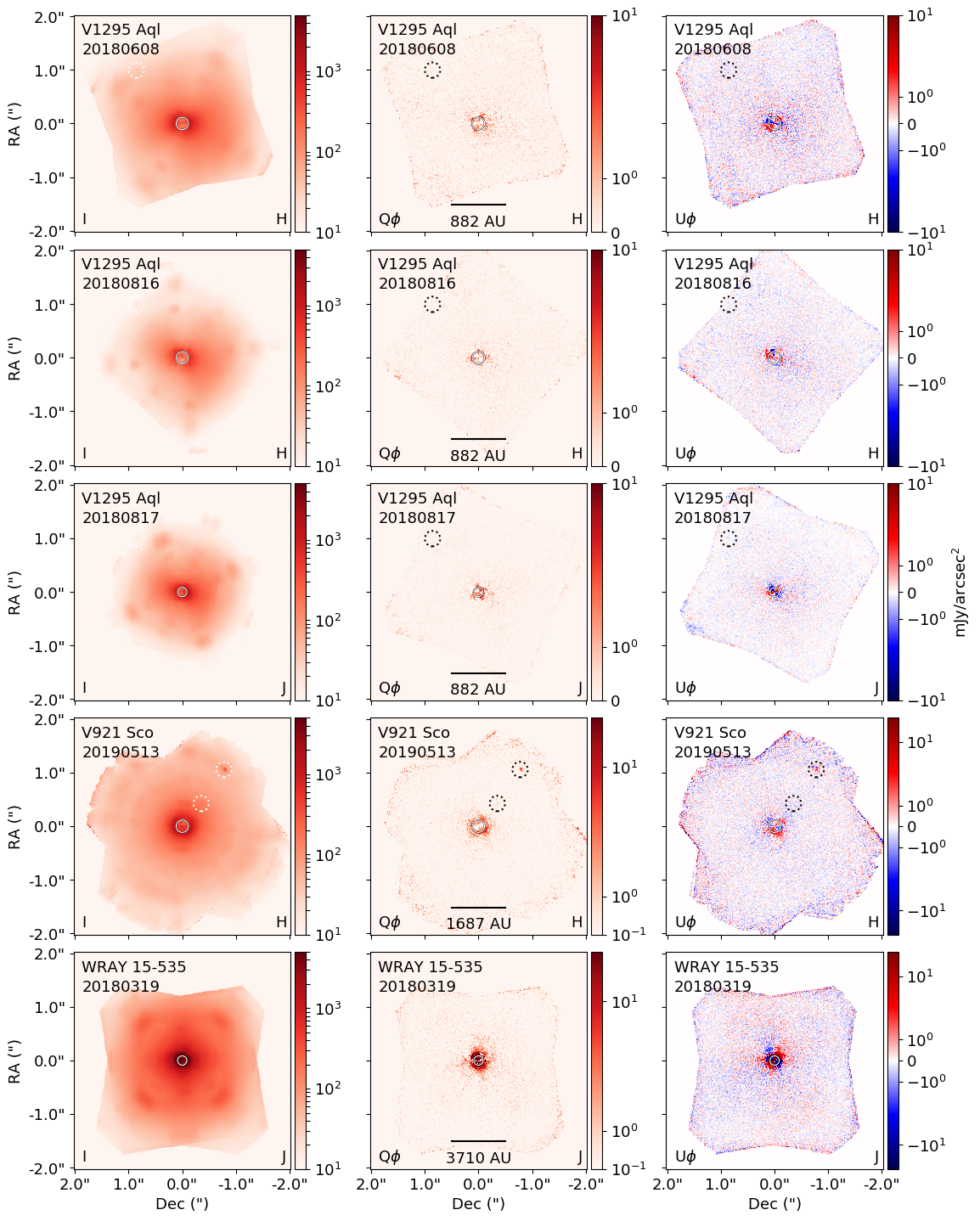}
    \caption{}
\end{figure*}

\begin{figure*}[!ht]
    \centering
    \includegraphics[width=0.99\linewidth ,clip]{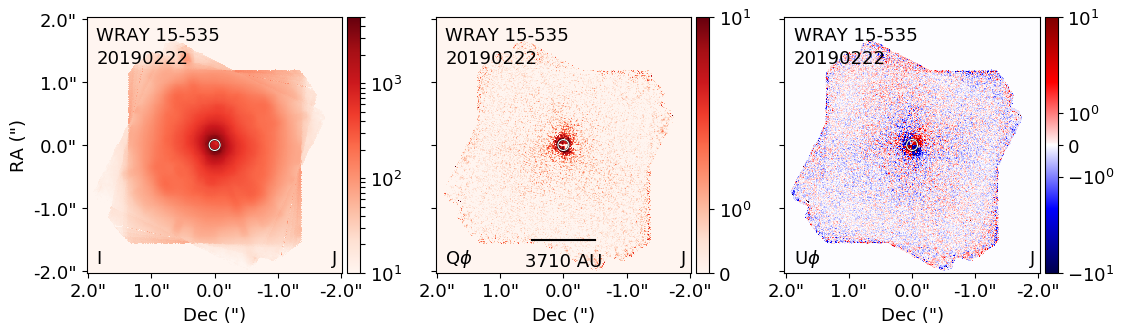}
    \caption{}
\end{figure*}

\end{document}